\documentclass[a4paper,12pt,final]{iopart}
\pdfoutput=1

\usepackage{url}
\usepackage{enumerate}
\usepackage{comment}
\usepackage{iopams}
\usepackage{setstack}
\usepackage{xcolor}
\usepackage{graphicx, caption}
\usepackage{eurosym}
\usepackage{pbox}
\usepackage{framed,color}
\definecolor{shadecolor}{rgb}{0.9,0.9,0.9}

\usepackage{iopams}

\newcommand{\comments}[1]{}
\def\P{P}					%Probability
\def\d{\mathrm{d}}
\newcommand{\beq}{\begin{equation}}
\newcommand{\eeq}{\end{equation}}
\newcommand{\bal}{\begin{eqnarray}}
\newcommand{\eal}{\end{eqnarray}}

\newcommand{\be}{\begin{equation}}
\newcommand{\ee}{\end{equation}}
\newcommand{\bd}{\begin{displaymath}}
\newcommand{\ed}{\end{displaymath}}
\newcommand{\BE}{\begin{eqnarray}}
\newcommand{\EE}{\end{eqnarray}}

\newcommand{\boldd}{\ensuremath{\mathbf{d}}}

\newcommand{\bD}{\ensuremath{\mathbf{D}}}

\newcommand{\boldSigma}{{\mbox{\boldmath $\Sigma$}}}

\usepackage[symbol]{footmisc}

\begin{document}
\title[Network Meta-Analysis: A Statistical Physics Perspective]{Network Meta-Analysis: A Statistical Physics Perspective}

\author{Annabel L. Davies${}^{\pounds}$, Tobias Galla${}^{\mbox{\scriptsize \euro},\pounds}$}

\address{$\pounds$~ Department of Physics and Astronomy, School of Natural Sciences, The University of Manchester, Manchester M13 9PL, United Kingdom}
\address{\euro~ IFISC, Instituto de F\'isica Interdisciplinar y Sistemas Complejos (CSIC-UIB),
Campus Universitat de les Illes Balears, 07122 Palma de Mallorca, Spain}
 
 \ead{annabel.davies@manchester.ac.uk, tobias.galla@ifisc.uib-csic.es}
 
\begin{abstract}
  Network meta-analysis (NMA) is a technique used in medical statistics to combine evidence from multiple medical trials. NMA defines an inference and information processing problem on a network of treatment options and trials connecting the treatments. We believe that statistical physics can offer useful ideas and tools for this area, including from the theory of complex networks, stochastic modelling and simulation techniques. The lack of a unique source that would allow physicists to learn about NMA effectively is a barrier to this. In this article we aim to present the `NMA problem' and existing approaches to it coherently and in a language accessible to statistical physicists. We also summarise existing points of contact between statistical physics and NMA, and describe our ideas of how physics might make a difference for NMA in the future. The overall goal of the article is to attract physicists to this interesting, timely and worthwhile field of research.
\end{abstract}

\tableofcontents
\markboth{Network Meta-Analysis: A Statistical Physics Perspective}{} 

 \section{Introduction} 

Physicists in general, and statistical physicists in particular, have a propensity to draw inspirations from problems across the borders of traditional disciplines. The application of ideas and methods from physics to questions in biology, economics and the social sciences is therefore well established \cite{Parisi:1993, Castellano:2009}. The following quote by the late Dietrich Stauffer encapsulates this \cite{Stauffer:2004}: {\em `The basic theorem of interdisciplinary research states: Physicists not only know everything; they know everything better.'}\footnote{The quote continues: {\em `This theorem is wrong; it is valid only for computational statistical physicists like me'.}} Arguably, not all of these invasions into the territory of other disciplines are useful, and physicists have been criticised for their, at times, ill-informed attempts to address questions outside their area of expertise \cite{Gallegati:2006}. On the other hand, it is also hard to deny that physics approaches have made useful contributions to a number of different fields.

In this perspective review we highlight network meta-analysis (NMA), a topic from medical statistics, as a field for which we think physics ideas might be useful. Meta-analysis is a statistical technique used to combine the results of multiple trials \cite{DerSimonian:1986, Smith:1995, Higgins:2009}. The aim of such trials is to establish and compare how effective different treatment options are. To do this, the different treatments are administered to groups of subjects in medical trials. Individual trials often have small sample sizes and involve subjects taken from a reduced population. Because of this, it is desirable to systematically integrate results from different trials to obtain an overall estimate of the effect of a given treatment and to compare treatment options. This is complicated by the fact that trials taking place at different locations will generally involve demographically different subject groups. The aggregation of data from different trials is not straightforward.  

\begin{figure}[t]
    \centering
    \includegraphics[width=1\linewidth]{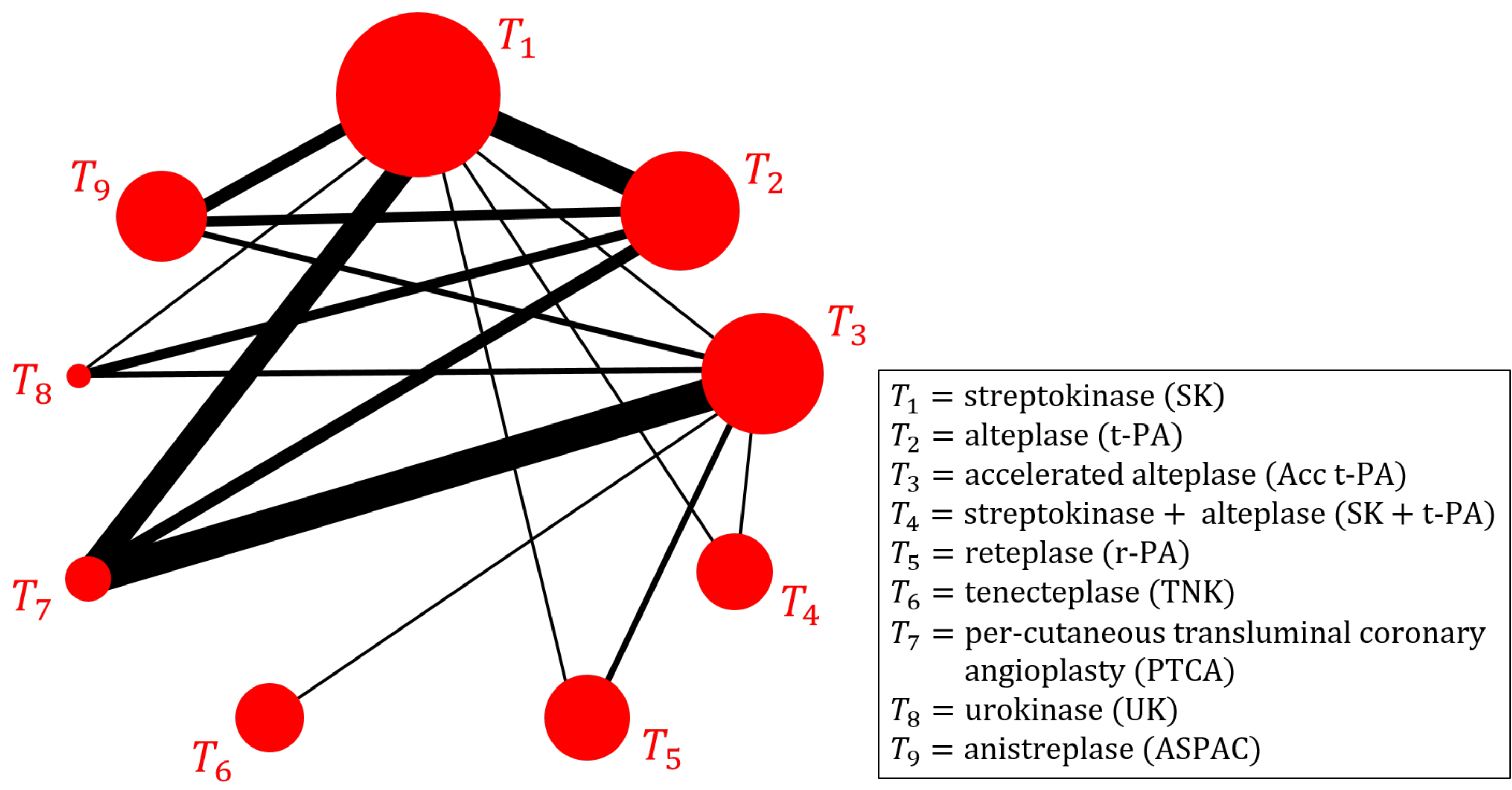}
    \caption{The network for the `thrombolytic drug data' set \cite{Boland:2003, Keeley:2003, Dias:2010} comparing nine treatments for acute myocardial infarction (heart attack). The treatments $T_1,\dots,T_9$ are labelled in the box. They consist of eight thrombolytic drugs and one angioplasty intervention ($T_7$). The thickness of the edges in the network indicate the number of trials making that comparison. The area of the node is proportional to the number of patients allocated to that treatment. The network consists of 50 trials; two $3$-arm trials (comparing $T_1,T_3,T_4$ and $T_1,T_2,T_9$, respectively) and 48 $2$-arm trials. The multi-arm trials are not explicitly indicated on the graph.
    \label{fig:RealNetEG}}
\end{figure}

\medskip
Conventional meta-analysis focuses on pairwise comparisons of treatments. More recently however, NMA (also referred to as `indirect meta-analysis', and `multiple' or `mixed treatment comparison' \cite{Tonin:2017, DIAS:2018}) has emerged as a technique for making inferences about multiple competing treatments. NMA allows one to combine data from multiple trials even when different trials test different sets of treatment options. The term `network meta-analysis' derives from a graphical representation of the treatments and trials. The nodes of the graph are the different treatment options and the connecting edges represent comparisons made between the treatments, an illustration can be found in Fig.~\ref{fig:RealNetEG}. NMA combines direct and indirect evidence for the assessment of treatments. This makes it possible to compare treatments that have not been tested together in any trial. For a textbook on NMA see \cite{DIAS:2018}.  

NMA is based on two main concepts: microscopic models for the outcomes of the different trials in the graph, and algorithms or procedures to carry out the actual NMA inference. 

\medskip

\noindent {\em Microscopic models.}  The microscopic model captures the main assumptions made on the process leading to real-world trial outcomes. Each trial tests a subset of treatments. In the so-called `random effects model' the relative treatment effects of the treatments in a particular trial are drawn from an underlying distribution. As a consequence, the effect of one treatment relative to another is not necessarily the same in two different trials. This reflects local confounding factors, for example, the fact that patient groups are chosen from different demographic subsets at different locations. In simulations of the random effects model, rates of success and failure for each trial arm are then constructed (the treatments tested in a trial are referred to as the  `arms' of the trial). A treatment success or failure occurs for a particular patient with the corresponding probability. This generates multiple layers of randomness in simulations: random treatment effects, and a binomial distribution of successes and failures for each trial arm.  
\medskip

\noindent {\em NMA inference.} The purpose of NMA is to estimate model parameters from given trial outcomes. These can either be real-world data or synthetic data from simulations. The NMA process also provides confidence levels for these parameters. This can be used to construct a `ranking' of treatments, as best, second best and so on. In more sophisticated approaches, probabilities are assigned that each treatment has a particular rank, reflecting the uncertainty on inferred treatment effects. Different ranking methods are still very much under discussion.

The NMA inference itself can either be carried out in a frequentist or a Bayesian setting. In this paper we will describe both approaches. In Bayesian NMA prior distributions are assumed for key model parameters, and posterior distributions are constructed from these and the trial outcomes. This needs to be done numerically, using Markov Chain Monte Carlo methods (in NMA specifically, the Metropolis-in-Gibbs algorithm is often used \cite{Spiegelhalter:2003}). In frequentist NMA, one defines a linear regression model dependent on the model parameters under the assumption of normally distributed random errors. The model is fitted using generalised least squares regression or, equivalently, maximum likelihood. Other procedures are also possible \cite{Efthimiou:2019, Stijnen:2010}, but are not discussed here.

We believe that NMA has a natural appeal to statistical physicists.  Those with experience in complex networks will find it interesting to connect the structure of treatment-trial networks with the outcome of NMA. Computational physicists may contribute to optimising the inference process and required sampling methods. Those interested in stochastic simulations can naturally connect with data generation methods used to obtain synthetic data for a given network of treatments and trials. NMA is an inference problem on a networked structure, and we expect that physicists working at the border to computer science and machine learning will become excited about it; for example it is conceivable that message passing methods can become a useful tool for NMA. Our own work (with collaborators) shows that random walks on the meta-analytic network and related graphs can lead to additional insights and improve methods to establish how evidence flows in NMA \cite{Davies:2021}.

One main bottleneck appears to be that there is no unique source which would allow a physicist to enter this field efficiently. While textbooks and review articles on NMA exist \cite{Tonin:2017, DIAS:2018, TSD2, TSD3, CochraneBook,  SALANTI:2012, Greco:2016, Salanti:2008, Jansen:2008, Efthimiou:2016}, these are often written for medical practitioners, or users of existing software packages. The mathematical details are frequently suppressed, or not presented in a language physicists are used to. This can make it hard to get a good grip on the actual mechanics of NMA. This perspective review is our attempt at rectifying this. Our objective is to provide a  technical introduction to NMA, accessible to physicists. We have aimed to make this self-contained, but at the same time this review is not a textbook and we have tried to keep the length to a reasonable limit. We hope we have found a sensible middle ground. We necessarily had to make a selection of topics we can cover, and attempted to choose those that are most helpful for others entering this area. We also aim to point out ideas from physics which we believe to be most promising to make a difference to NMA. We hope that this will facilitate future work by the physics community in this timely and worthwhile area of research.

The paper is organised as follows: Sec.~\ref{sec:networks_of_trials} sets the scene, defines the necessary notation and states the `NMA problem'.  In Secs.~\ref{sec:Bayes} and \ref{sec:freq} we then present the mathematical procedures used to carry out an NMA in a Bayesian and frequentist setting respectively. Sec.~\ref{report} summarises how the results of an NMA are reported. In Sec.~\ref{sec:existing} we present existing analogies connecting NMA to different systems in physics, including resistor networks and random walks. In Sec.~\ref{sec:future} we then outline some more general connections between the two fields and speculate on ways in which we think physicists may contribute to NMA in the future. Sec.~\ref{sec:summary} contains a brief summary and discussion.

\section{Networks of medical trials}\label{sec:networks_of_trials}

\subsection{General background: randomised controlled trials, meta-analysis, and network meta-analysis}
In this section we first give an informal description of the key concepts in NMA. We turn to a more formal mathematical setup in Sec.~\ref{sec:FEM_REM}.

\subsubsection{Randomised controlled trials.}

For our purposes a trial is an experiment in which a group of subjects is used to compare a given set of  treatment options. The different treatments are referred to as the {\em arms} of the trial. In particular, a `controlled' clinical trial is one with at least two arms. Typically, this involves one or more `experimental' treatment groups representing new treatments being tested. These are compared to the so-called `control' group(s) which could be alternative (existing) treatments, a placebo or no treatment \cite{NICE-gloss}. 

The allocation of subjects to the different arms is randomised to avoid any bias in treatment assignment. For example, the treatment assigned to a given subject can be chosen with equal probability from the arms of the trial. In this scenario the trial is a `randomised controlled trial' (RCT).

Once assigned to a trial arm, each subject receives the respective treatment, and undergoes follow-up. In the simplest case the outcome for each subject is binary (dichotomous), e.g. `treatment successful' vs. `treatment not successful'. We can also think of this as `an event has occurred' vs. `no event has occurred'.  For the purposes of this introductory review we focus on this  case of binary outcomes.  We exclude censoring  (e.g. patients withdrawing from the trial or otherwise not being followed up, and therefore not producing data). More complex outcomes in trial data may consist of a discrete set of more than two alternatives (e.g. `ordinal outcomes' on an $5$-point scale), or the outcome may be continuous, see e.g. \cite{TSD2}.

\subsubsection{Meta-analysis.}\label{sec:meta-analysis}
Meta-analysis in general is concerned with combining evidence from multiple trials. The simplest case is so-called pairwise or standard meta-analysis. Two treatment options are compared in a set of different trials. The purpose of meta-analysis is then to `integrate' the outcomes of the different trials, and to estimate how effective the competing treatment options are. These estimates can then be used to decide if one of the two treatments is to be preferred over the other, and which one.

In this process it is important to bear in mind that the outcomes of different trials cannot always be aggregated directly. Clinical trials taking place at different locations will draw from a local patient pool, and as a result, the general characteristics of the subjects may differ from trial to trial (e.g. age, health or economic status, level of education etc) which may affect the observed treatment effects. In order to combine evidence from multiple trials we require an underlying model -- a stochastic process with unknown parameters leading to realisations of the data observed in trials. Two of the most common modelling approaches, so-called fixed and random effects models, are discussed in Section~\ref{sec:FEM_REM}. Once a given model assumption has been made, the objective of meta-analysis is to estimate parameters of the model from the data.

\subsubsection{Networks of trials and NMA.} \label{NetTrials}
General networks of treatments and trials capture more complex situations than the one described in Section~\ref{sec:meta-analysis}. For example imagine that there are four different treatment options and several trials, each comparing a subset of treatments. Not every trial tests all four treatments, but the same pairwise comparison is perhaps made in different trials. This generates a network of treatment options and trials.

\begin{figure}
    \centering
    \includegraphics[width=0.75\linewidth]{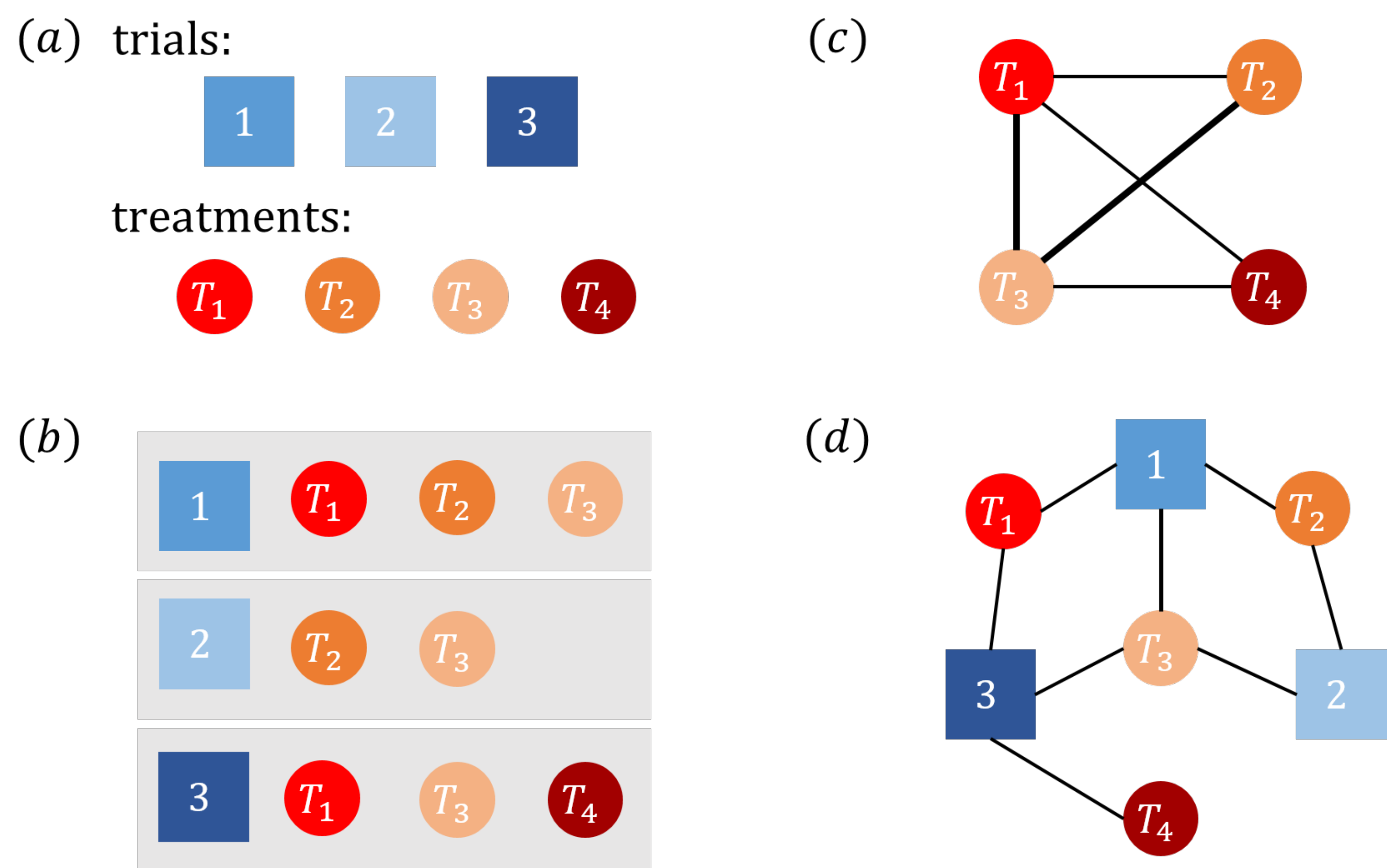}
    \caption{Illustration of a network of treatment options and trials. (a) There are three trials in the network (squares), and four treatments in total (circles). (b) Trial $1$ has three arms (treatments $T_1$, $T_2$ and $T_3$), trial $2$ is two-armed (treatments $T_2$ and $T_3$), and trial $3$ tests treatments $T_1, T_3$ and $T_4$. (c) Presentation of the network as a graph with only one type of node. Each node represents one treatment, and two treatments are connected if they have been directly compared in at least one trial. The thickness of the edge connecting two nodes is proportional to the number of trials comparing those two treatments. This representation does not contain  full information about the network of treatments and trials.  (d) Representation as a bipartite graph of treatments and trials. This can also be understood as a hypergraph (related concepts include incidence or Levi graphs \cite{Pisanski:2000}).
    \label{fig:illustration}}
\end{figure}

A possible scenario is illustrated in Fig.~\ref{fig:illustration}. The network consists of three trials (indicated by square boxes), comparing different subsets of four different treatments [panel (a)]. We label the treatments $T_1,\dots, T_4$, and indicate them as circles on the graph.   Panels (b), (c) and (d) show different graphical representations of this network; details can be found in the figure caption. Option (c) is most commonly used in practice.

Most notably in this example,  there is no pairwise comparison between two fixed treatments that is made in all three trials (i.e. no pair of treatment arms features in all three trials). The pair $T_2$--$T_3$ appears in trials $1$ and $2$, so the use of conventional (pairwise) meta-analysis would be restricted to combining  information regarding this particular pair from trials $1$ and $2$ only.

Network meta-analysis aims to integrate further information from the network. The information about the pair $T_2$--$T_3$ from trials $1$ and $2$ is referred to as {\em direct evidence} for the comparison of these two treatments. However, $T_2$ and $T_3$ are each also compared to $T_1$. For treatment option $T_2$ this happens in trial $1$, and for $T_3$ in trials $1$ and $3$. These comparisons to a common third treatment provide {\em indirect evidence} for the comparison of $T_2$ and $T_3$.

In the example, treatment options  $T_2$ and $T_4$ are not compared directly in any trial. However, each of the two are directly compared to $T_1$ and $T_3$ [see Fig.~\ref{fig:illustration} (b)]. This indirect evidence can be used to infer information about the comparison between $T_2$ and $T_4$, even though there is no direct evidence. 

\medskip

\textit{Data example.} Fig.~\ref{fig:RealNetEG} shows the network for a real NMA data set comparing nine treatments for acute myocardial infarction (heart attack) \cite{Boland:2003, Keeley:2003, Dias:2010}. The treatments are labelled $T_1,\dots,T_9$ and their names are given in the figure. They consist of eight thrombolytic drugs and one angioplasty intervention ($T_7$). The data has therefore been referred to as the `thrombolytic drug' data set \cite{Dias:2010, TSD4}. A heart attack occurs when blood flow to the heart is cut off, usually resulting from blockage of one or more of the coronary arteries. Thrombolytic drugs aim to dissolve blood clots that have blocked arteries whereas angioplasty is a procedure that tries to relieve blockage by widening the arteries. The data consists of 50 trials; two 3-arm trials (comparing $T_1,T_3,T_4$ and $T_1,T_2,T_9$, respectively) and 48 2-arm trials. Each trial records the number of deaths in each treatment arm that occur within 30 or 35 days  of a heart attack. The network is represented by a weighted graph where for each pair of treatments, the thickness of the edge link is proportional to the number of trials comparing these two treatments. The area of each node is proportional to the number of patients allocated to the treatment represented by that node. Multi-arm trials are not explicitly indicated on the graph.

\subsection{General notation for networks of trials and treatments}
We write $N$ for the total number of treatment options in the network. We label the different treatments $T_1, T_2,\dots, T_N$, and we will use the indices $a,b$ when we refer to elements of the set of treatments, i.e. $a,b\in\{T_1,\dots, T_N\}$. The number of trials in the network is denoted by $M$, and we use the indices, $i,j$ to refer to the different trials, i.e. $i,j \in\{1,\dots,M\}$. We will use the words `trial' and `study' synonymously.

Each trial compares a subset of treatments. We write $A_i\subset\{T_1,\dots, T_N\}$ for the set of treatment options compared in trial $i$. Hence $m_i\equiv |A_i|$ is the number of arms of study $i$. We number the arms of trial $i$ by   $\ell=1,\dots, m_i$, and denote the treatment given to patients in arm $\ell$ of trial $i$ by $t_{i,\ell}$. Each $t_{i,\ell}$ ($i=1,\dots,M$, $\ell=1,\dots,m_i$) is a treatment from the set $\{T_1,\dots,T_N\}$. 

In the illustration in Fig.~\ref{fig:illustration}, we have $N=4$ treatments and $M=3$ trials. For trial $3$, for example, we have $m_3=3$ (three-armed trial), $A_3=\{T_1, T_3, T_4\}$ as well as $t_{3,1}=T_1, t_{3,2}=T_3$ and $t_{3,3}=T_4$.

We write $n_{i,\ell}$ for the number of subjects receiving the $\ell$-th treatment in trial $i$, with $\ell=1,\dots,m_i$. Focusing on binary outcomes, the data available for each patient is whether an `event' has occurred by the end of the study or not. (We note that, depending on the context, an event can either be treatment success or an adverse event as in the data example in Sec.~\ref{NetTrials}.)  

For each arm $\ell$ of trial $i$, the number of resulting events is recorded. We denote this by $r_{i,\ell}$. This quantity takes integer values in the range $0,1,\dots, n_{i,\ell}$

Summarising, trial $i$ is defined by the treatments  it compares, $t_{i,1},\dots, t_{i,m_i}$, by the number of patients in each arm, $n_{i,1},\dots,n_{i,m_i}$, and by the number of events in each arm $r_{i,1},...,,r_{i,m_i}$. 

\subsection{Absolute and relative treatment effects: the logit scale}

We assume that the application of the treatment in arm $\ell$ of trial $i$ generates events with probability $p_{i,\ell}$ independently for each of the $n_{i,\ell}$ patients at the end of this trial arm \cite{Hamza:2008}. As a consequence, each $r_{i,\ell}$ is a binomial random variable,
\begin{equation}
\label{eq:Binomial}
    {\rm Prob}(r_{i,\ell}=r)=\left(\begin{array}{c} n_{i,\ell} \\ r \end{array}\right) p_{i,\ell}^r (1-p_{i,\ell})^{n_{i,\ell}-r},
\end{equation}
for $i=1,\dots,M$ and $\ell=1,\dots m_i$.

\subsubsection{Absolute treatment effects.}
The $\{p_{i,\ell}\}$ can be interpreted as `absolute treatment effects', they capture how likely it is that the different treatments produce `events'. The word `absolute' indicates that this treatment effect is not expressed with reference to any other treatment or baseline.

In the context of binary outcomes, absolute treatment effects are frequently expressed in terms of so-called `log-odds'. For a probability $p$ (i.e. a number $p\in [0,1]$) the term `odds' refers to the ratio $p/(1-p)$, and the log-odds or `logit' (`logistic unit') is defined as
\be
\label{eq:logit}
\mbox{logit}(p)=\ln\frac{p}{1-p}.
\ee
While the original probability $p$ is restricted to the range $0\leq p \leq 1$, the logit of $p$ can take values on the entire real axis, with $\lim_{p\to 0} \mbox{logit}(p)=-\infty$, and $\lim_{p\to 1}\mbox{logit}(p)=\infty$. 

Accordingly, we can express the treatment effects in terms of $\lambda_{i,\ell}\equiv \mbox{logit}(p_{i,\ell})=\ln p_{i,\ell}-\ln(1-p_{i,\ell})$. In a slight abuse of terminology we will refer to both the  $\{\lambda_{i,\ell}\}$ and to the $\{p_{i,\ell}\}$ as the {\em absolute treatment effects}. From the context it will be clear what we mean.

The logit transformation in Eq.~(\ref{eq:logit}) is an example of a so-called `link function'. Outcomes from medical trials come in many forms (e.g. time-to-event, ordered categories, continuous measurements) and are generated from a range of distributions (e.g normal, binomial, Poisson). By using a link function to transform the treatment effect associated with a particular type of data onto the continuous scale we can then use the same basic model for a range of different data types. The choice of logit in Eq.~(\ref{eq:logit}) is a practical choice for binomial data. Exchanging the definitions of event vs no event (i.e., $p\leftrightarrow 1-p$) only results in a sign reversal, i.e., it makes no difference  for the mathematical model and inference process. This is not true for some other choices of the link function \cite{Rucker:2009, Deeks:2002}.

In the following we describe the NMA model for binomial data with a logit link. The so-called likelihood function (defined further below) is a binomial distribution.
To analyse other types of data, one can use the same basic NMA model but the likelihood and link functions vary depending on the type of this data. See references \cite{TSD2, Dias:2013} for an overview of different data types and their corresponding link functions.

\subsubsection{Relative treatment effects, transitivity and common baseline.} \label{RelEffect_base}
We now  introduce the so-called {\em relative treatment effect} for treatments $a$ and $b$. If these two treatments have  absolute treatment effects $\lambda_a$ and $\lambda_b$, then we write
\be
d_{ab}\equiv \lambda_b-\lambda_a
\ee
for the relative treatment effect for this pair. This definition implies $d_{ab}=-d_{ba}$, and $d_{aa}=0$.

In formulating this setup we assume that the relative treatment effects fulfill the transitivity relation $d_{ab}=d_{ac}+d_{cb}$ for all triplets of treatments $a, b$ and $c$. Alternatively, this can be written as 
\be\label{eq:transitivity}
d_{ab}=d_{cb}-d_{ca}.
\ee
Using transitivity, the relative treatment effects of all pairs in a network of $N$ treatments, $T_1,\dots, T_N$, are fully specified by $N-1$ numbers. For example, we can designate treatment $T_1$ as the overall `global' baseline. It is then sufficient to know $d_{T_1a}$, for $a\in\{T_2,\dots, T_N\}$. These values are termed the `basic parameters' \cite{Eddy:1992, Lu:Ades:2006} and we collect them in the $(N-1)$-component vector $\boldsymbol{d}=(d_{T_1T_2},d_{T_1T_3},\dots,d_{T_1T_N})^\top$ (the notation $(\dots)^\top$ indicates transposition). The relative treatment effect $d_{ab}$ for any pair of treatments $a, b$ can then be determined from Eq.~(\ref{eq:transitivity}), using $c=T_1$. 

Transitivity describes an assumption made in setting up the model. It is applicable to all possible comparisons in the network. The `statistical manifestation' \cite{CochraneBook} of transitivity in observed data is referred to as consistency. I.e., if direct and indirect evidence exist in the data for a particular comparison, then the data is consistent if there is no discrepancy in the treatment effects obtained via the two types of evidence.

\subsection{Fixed and random effects models}\label{sec:FEM_REM}
In this section we describe the so-called random and fixed effects models used in meta-analysis. We will abbreviate these as RE and FE models respectively. The FE model is a limiting case of the RE model. 

\subsubsection{General idea: Modelling relative effects.} The usual approach to NMA is to model the \textit{relative} rather than the absolute treatment effects. Individual trials are likely to have different characteristics in terms of population demographics (e.g. age, socio-economic status, baseline health) and trial procedures (such as treatment dosage or administration). It is therefore likely that absolute treatment effect outcomes, such as the number of patients who experience an event, will vary substantially depending on these characteristics. For example, we are likely to observe a higher proportion of deaths in a trial with an older population compared to a younger population. It is then unrealistic to model the absolute effects as being comparable across all trials. A less restrictive model is to assume that \textit{differences} between the effects of treatments are similar across the trials. For example, if treatment $T_1$ is more effective than treatment $T_2$ then it is likely that in each trial we will observe fewer deaths in arm $T_1$ than in arm $T_2$ even if the overall number of deaths in each trial is very different.  

With this in mind, we assume that each trial comparing treatments $a$ and $b$ is associated with some unknown relative treatment effect $\Delta_{i,ab}$ which represents the `true' difference in effectiveness between $a$ and $b$ \textit{in trial} $i$. Trial $i$ then provides information about $\Delta_{i,ab}$ subject to some sampling error (due to the finite number of participants in that trial). This information is said to be `observed'. The FE and RE models differ in the assumptions placed on these `trial-specific' relative effects. 

The FE model assumes that each trial has the same underlying relative treatment effects, i.e. $\Delta_{i,ab}=d_{ab}$ $\forall i$. The RE model, on the other hand, is more flexible. Rather than requiring that the trial-specific relative effects are the same in every trial (as in the FE model), they are instead assumed to be `exchangeable' \cite{DerSimonian:1986, Hong:2015, Dias:2016}. In other words, the true relative treatment effects in each trial are treated as random variables, drawn from an underlying distribution \cite{Higgins:2009}. This reflects the fact that differences in trial characteristics may mean that a particular treatment option is comparatively more effective in one trial than in another. For example, treatment $T_1$ may be the most effective treatment for participants of all ages, but it could yield even better results for younger patients. Trials with a younger demographic may then observe a larger relative effect between $T_1$ and $T_2$ compared with a trial of older participants. More specifically, the RE model assumes that the relative effect of two treatments $a$ and $b$ is drawn from the {\em same} distribution for any trial involving these two treatments, i.e., this distribution is the same for all $i$. We are interested in the mean of this distribution. This indicates the typical relative effect between the treatments.

The RE model therefore consists of two levels of randomness; one due to variations \textit{between} trials and the other due to sampling \textit{within} a given trial. In the FE model, we only allow for the latter. 

\subsubsection{Transitivity of trial-specific relative effects.}
We assume that the trial-specific relative treatment effects fulfill the transitivity relations in Eq.~(\ref{eq:transitivity}). For a trial with $m$ arms it is therefore sufficient to designate a trial-specific baseline treatment and its absolute treatment effect, and the treatment effects of the $m-1$ remaining arms in the trial relative to this baseline. This will fully determine the true absolute effects of all treatments in the trial.

In our model there are $m_i$ arms in trial $i$, labelled $\ell=1,\dots,m_i$. Without loss of generality we use $\ell=1$ as the trial-specific baseline treatment. We then write 
\be
\label{eq:logit-ratio}
\Delta_{i,1\ell}=\ln \frac{p_{i,\ell}}{1-p_{i,\ell}}-\ln\frac{p_{i,1}}{1-p_{i,1}}
\ee
for the  effect of treatment $t_{i,\ell}$ in the trial relative to this baseline. 

The absolute treatment effects $p_{i,\ell}$ of all $m_i$ arms in the trial can then be obtained from the true absolute effect of the baseline, $p_{i,1}$, and  $\Delta_{i,12},\dots,\Delta_{i,1\ell_i}$.

\subsubsection{Model definitions.}
\label{ModelDef} We now formalise our model. As we have seen, the RE model consists of two level of randomness. One is between-trial variation and the other is due to sampling in a given trial. We describe these in turn\footnote{In our description the first level of randomness is the between-trial variation, whereas the second level is the sampling randomness within a trial. This reflects the `mechanistic' view a physicist might take, and focuses on how one would generate synthetic trial data in a simulation. A statistician might take a reverse view and see the trials as the starting point (hence sampling noise is the first level of randomness). The synthesis of several trials then follows later, and the between-trial randomness therefore comes second for the statistician.}. 

{\em Between-trial variation.} We assume that the relative treatment effects for a given trial $i$ are drawn from a multivariate normal distribution, 
\begin{equation}
\label{eq:multi_norm}
  \left(\matrix{
    \Delta_{i,12}\cr
    \vdots \cr
    \Delta_{i,1m_i}
    }\right) \sim \mathcal{N}\left(\left(\matrix{
    d_{t_{i,1}t_{i,2}}\cr
    \vdots \cr
    d_{t_{i,1}t_{i,m_i}}
    }\right),\boldsymbol{\Sigma_i}
    \right).
\end{equation}
 The first argument is the mean, the second argument is the covariance matrix. We will now describe these first and second moments in more detail.

Given Eq.~(\ref{eq:multi_norm}) the relative effect between two treatments $a$ and $b$ in a given trial is a Gaussian random variable. In particular, the relative effect between these treatments varies across different trials involving $a$ and $b$. The model parameters $d_{ab}$ are the averages of these random numbers. More precisely, for a given pair $a$ and $b$ the parameters $d_{ab}$ can be interpreted as the `typical' relative effect one should expect to see in a trial involving these treatments.

Using the transitivity relation in Eq.~(\ref{eq:transitivity}), we can write the vector of mean treatment effects in trial $i$, $\boldsymbol{d}_i=(d_{t_{i,1}t_{i,2}},\dots,   d_{t_{i,1}t_{i,m_i}})^\top$ in terms of the vector of basic parameters via 
\be
\boldsymbol{d}_i=\mathbf{X}_i \boldsymbol{d}.\label{eq:linear_X}
\ee
We note that $\boldsymbol{d}$ is a vector with $N-1$ entries and that $\boldsymbol{d}_i$ has $m_i-1$ components. The $(m_i-1)\times (N-1)$  matrix $\mathbf{X}_i$ describes which treatments are compared in trial $i$ and is called the `design matrix' of the trial. Each of the $N-1$ columns of $\mathbf{X}_i$ represents a treatment $a\in \{T_2,\dots,T_N\}$. The $m_i-1$ rows represent the comparisons of treatments $t_{i,\ell}$ $(\ell=2,\dots,m_i)$ to the trial-specific baseline $t_{i,1}$. In a given row $\ell$ all entries of $\mathbf{X}_i$ are zero, except those corresponding to the treatments that are being compared. More precisely, we distinguish two cases: (i) the trial-specific baseline treatment is not the global baseline ($t_{i,1}\neq T_1$), and (ii) the trial-specific baseline is the global baseline ($t_{i,1}= T_1$). 

We focus first on case (i). In row $\ell$ the matrix entry for the treatment $t_{i,\ell}$ that is being compared against the trial-specific baseline is set to $+1$. The entry in the column corresponding to the trial-specific baseline [which is among the $\{T_2,\dots, T_N\}$ in case (i)] is set to $-1$. All other $N-3$ entries in row $\ell$ are zero. In situation (ii), we again set the entry for the treatment $t_{i,\ell}$ that is being compared against the trial-specific baseline to $+1$. All other $N-2$ entries in row $\ell$ are zero.

\begin{figure}
   \centering
    \includegraphics[width=0.85\linewidth]{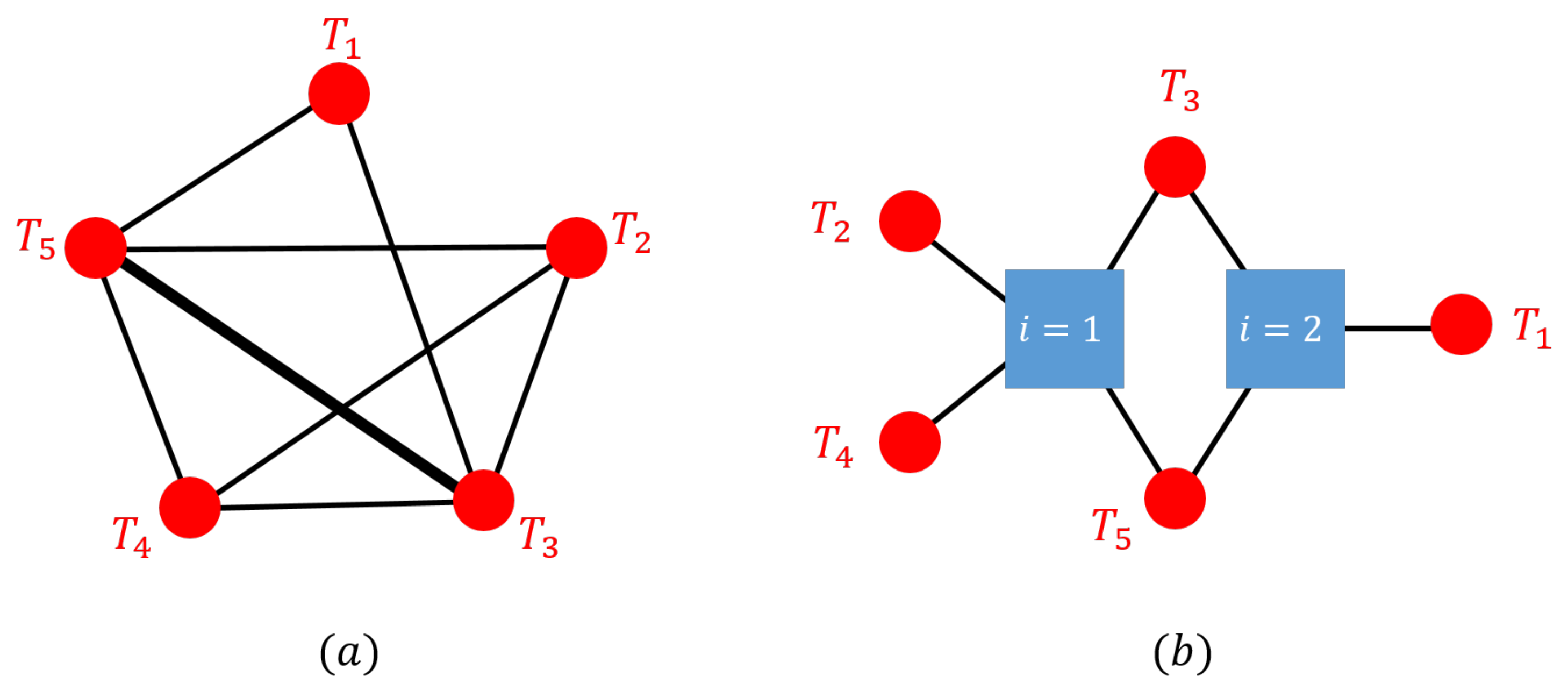}
    \caption{An fictional network with $N=5$ treatments and $M=2$ trials. Trial $i=1$ compares treatments $A_1=\{T_2,T_3,T_4,T_5\}$ and trial $i=2$ compares $A_2=\{T_1,T_3,T_5\}$. (a) Standard network representation where the thickness of each edge relates to the number of trials that make that comparison. Here, only the pair $\{T_3, T_5\}$ appears in both trials. (b) Network representation as a bipartite graph. } \label{fig:X_eg}
\end{figure}

To illustrate this, consider the example network in Fig.~\ref{fig:X_eg}. This network consists of $N=5$ treatments and $M=2$ trials. The global baseline treatment is $T_1$. The vector of basic parameters is $\boldsymbol{d}=(d_{T_1T_2},d_{T_1T_3},d_{T_1T_4},d_{T_1T_5})^\top$. Trial $i=1$ compares treatments $A_1=\{T_2,T_3,T_4,T_5\}$ with trial-specific baseline $t_{1,1}=T_2$. This is an example of option (i) and its design matrix is
\begin{equation}
    \mathbf{X}_1 = \left(\matrix{-1 & 1 & 0 & 0 \cr
    -1 & 0 & 1 & 0 \cr
    -1 & 0 & 0 & 1 
    }\right).
\end{equation}
Trial $i=2$ compares treatments $A_2=\{T_1,T_3,T_5\}$ and its trial-specific baseline is $t_{2,1}=T_1$. This is an example of option (ii) and its design matrix is 
\begin{equation}
    \mathbf{X}_2 = \left(\matrix{0 & 1 & 0 & 0 \cr
    0 & 0 & 0 & 1 
    }\right).
\end{equation}

The $(m_i-1)\times (m_i-1)$  matrix $\boldsymbol{\Sigma_i}$ in Eq.~(\ref{eq:multi_norm}) describes the variance of the relative treatment effects, $\Delta_{i,1\ell}$ ($\ell=2,\dots,m_i$), and their correlations. Following References \cite{Hig:White:1996, Lumley:2002, Lu:Ades:2004}, we will assume that its diagonal elements are all identical. We write $\tau^2$ for their common value. This is the variance of each $\Delta_{i,1\ell}$. We will further assume that the covariance between any two treatment effects is $\tau^2/2$ (these are the off-diagonal elements of $\boldsymbol{\Sigma_i}$). This ensures that the relative effect $\Delta_{i,1\ell}-\Delta_{i,1\ell'}$ between \textit{any two} treatments $\ell\neq\ell'$ in trial $i$ has variance $\tau^2$\footnote{This can be seen from $\mathrm{Var}(\Delta_{i,1\ell}-\Delta_{i,1\ell'})=\mathrm{Var}(\Delta_{i,1\ell})+\mathrm{Var}(\Delta_{i,1\ell}')-2\mathrm{Cov}(\Delta_{i,1\ell}, \Delta_{i,1\ell'})$.}. 

The between-trial variance $\tau^2$ is termed the \textit{heterogeneity} variance. We will refer to its square root, $\tau$, as the heterogeneity parameter. Occasionally we will simply use `heterogeneity' to refer to either the parameter or the variance but it should be clear from the context what we mean. Usually this distinction in not important.

The aim of network meta-analysis is to estimate the mean relative treatment effects $d_{ab}$ for all pairs $a\neq b$, and the heterogeneity parameter, $\tau$. Given the transitivity assumption in Eq.~(\ref{eq:transitivity}) not all $d_{ab}$ are independent. As explained earlier, we can use treatment $a=T_1$ as the overall global baseline treatment, and it is sufficient to estimate $d_{T_1a}$ for $a=T_2,\dots,T_N$ \cite{Lu:Ades:2006}.

\medskip

{\em Sampling noise in a given trial.} In a second level of randomness the model assumes that the application of the treatment in arm $\ell$ of trial $i$ generates events with probability $p_{i,\ell}$ independently for each of the $n_{i,\ell}$ patients at the end of this trial arm \cite{Hamza:2008}.  Each $r_{i,\ell}$ is then a binomial random variable, as described in Eq.~(\ref{eq:Binomial}).

\begin{figure}[t]
   \centering
    \includegraphics[width=1\linewidth]{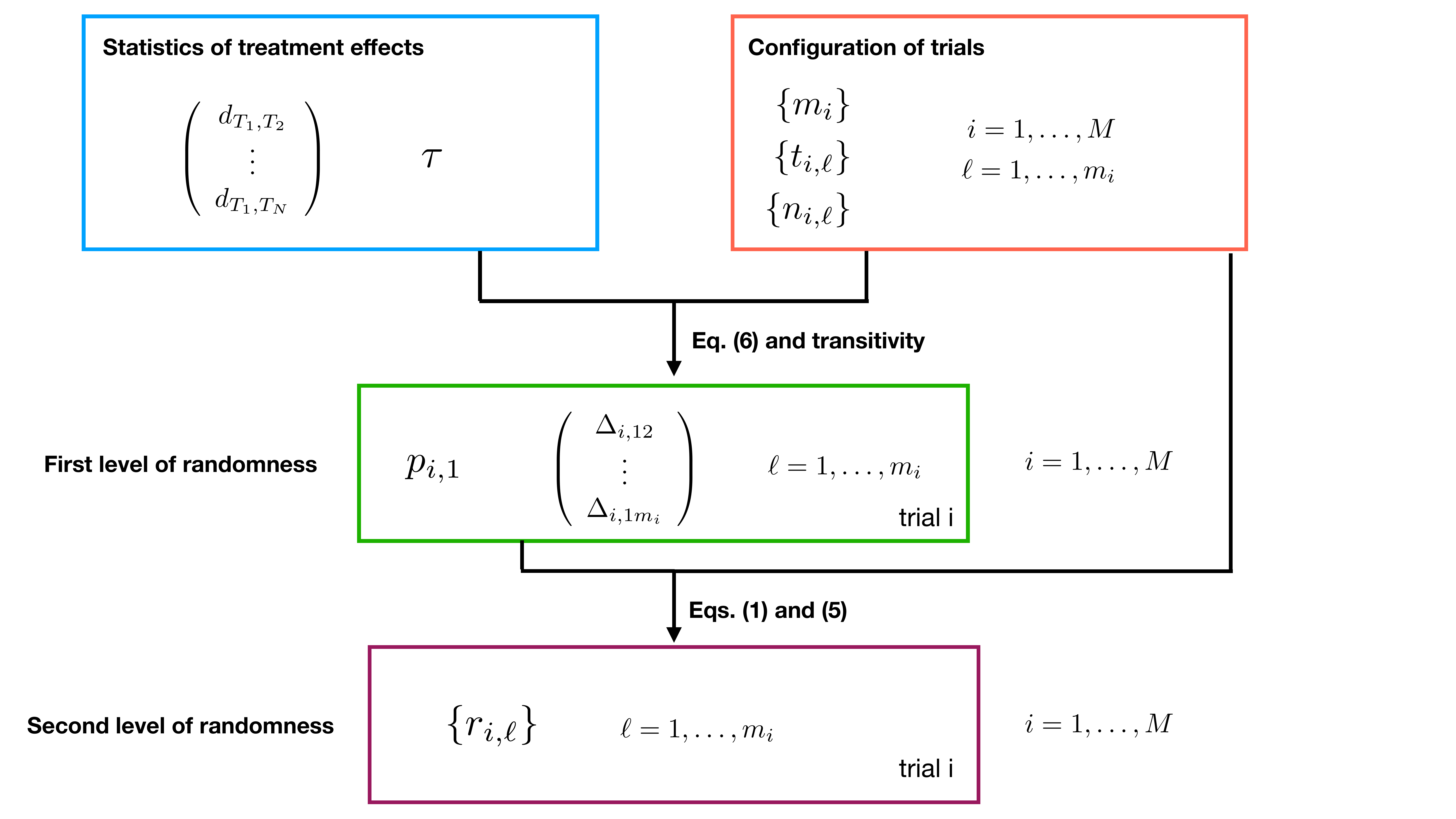}
    \caption{Diagram summarising the random effects model of NMA. The main input parameters are the configuration of trials and the statistics of treatment effects. The trial configuration is set by the number of trials $M$ in the network, the number of arms in each trial ($m_i$), the treatment options used in these arms ($t_{i,\ell}$) and the number of patients in each arm $(n_{i,\ell})$. The statistics of the treatment effects are parameterised by the mean effect of each treatment $T_2,\dots, T_N$ relative to the overall baseline treatment $T_1$, and the heterogeneity parameter $\tau$. In a first step of randomness realisations of the random variables describing the treatment effects in the different trials ($p_{i,1}$ and $\Delta_{i,12},\dots,\Delta_{i,1m_i}$) are drawn for each trial from the distribution in Eq.~(\ref{eq:multi_norm}), supplemented by a distribution for each $p_{i,1}$. These are then used along with Eq.~(\ref{eq:logit-ratio}) to construct the absolute effects of the treatments in each trial.  From these, and using the number of participants  (the $\{n_{i,\ell}\}$), the number of events in each arm (the $\{r_{i,\ell}\}$) are then drawn from the binomial distributions in Eq.~(\ref{eq:Binomial}). The fixed effect model is the special case $\tau=0$. The distribution in Eq.~(\ref{eq:multi_norm}) then turns into a delta-distribution. In this scenario, the true relative effect between two treatments $a$ and $b$ does not vary between trials and is given by $d_{ab}$.  } \label{fig:nma_diagram}
\end{figure}

The RE model is summarised and illustrated in Fig.~\ref{fig:nma_diagram}. The FE model is simply a special case of the RE model where $\tau=0$. As previously explained, for any fixed pair of treatments, there is then no variation in the relative treatment effects between trials.

\subsection{Generation of synthetic data in simulations}
\label{DataGen}
The different levels of randomness in the model can be understood by thinking about how one would simulate synthetic trial data in line with the model assumptions. 

We begin such a process by defining the fixed parameters of the network. First, we pick the network configuration; the number of treatments $N$, the total number of studies, $M$, the number of arms in each trial $\{m_i\}$, the set of treatments these arms relate to $\{t_{i,\ell}\}$, and the number of participants in each arm $\{n_{i, \ell}\}$. We then assign the `true' values of the model parameters, i.e. of  $\boldsymbol{d}=(d_{T_1T_2},d_{T_1T_3},\dots,d_{T_1T_N})^\top$ and $\tau$.

Following this set-up, we generate independent realisations $\nu=1,2,...,\Omega$ of synthetic trial outcomes. Specifically, for each $\nu$:

\begin{enumerate}[(1)]
\item For all trials $i$, randomly sample the parameters $\Delta_{i,1\ell}$, $\ell=2,\dots, m_i$, from the multivariate normal distribution in Eq.~(\ref{eq:multi_norm}). 
\item Using the $\Delta_{i,1\ell}$, $\ell=2,\dots, m_i$, and a so-called `data generating model' still to be defined (see below), construct the probabilities $p_{i,\ell}$, $\ell=1,\dots,m_i$, for all trials $i$ in the network.
\item For each trial arm, generate random event data (`observations'), $r_{i,\ell}$, from the binomial distribution in Eq.~(\ref{eq:Binomial}).
\end{enumerate}
In step (2) of the simulation procedure, the relative treatment effects $\Delta_{i,1\ell}$ ($\ell=2,\dots, m_i$) in any one trial $i$ do not uniquely define the absolute treatment effects $p_{i,\ell}$ ($\ell=1,\dots, m_i$) required for step (3). Eq.~(\ref{eq:logit-ratio}) can be re-arranged to give
\begin{equation}\label{eq:p_i_ell}
    p_{i,\ell}=p_{i,\ell}[p_{i,1},\Delta_{i,1\ell}] = \frac{p_{i,1}\mathrm{e}^{\Delta_{i,1\ell}}}{1+p_{i,1}\left(\mathrm{e}^{\Delta_{i,1\ell}}-1\right)},
\end{equation}
so that $p_{i,1}$ together with the $\Delta_{i,1\ell}$ ($\ell=2,\dots, m_i$) specifies all absolute treatment effects in trial $i$.

To fully define step (2) in the above algorithm it is therefore sufficient to specify the construction of $p_{i,1}$. A discussion of possible data generating models for $p_{i,1}$ is given in Seide et al (2019) \cite{Seide:2019}. We briefly describe two possible methods, highlighting the resulting symmetry or asymmetry introduced.  

One simple procedure involves sampling $p_{i,1}$ for each trial from some specified distribution. For example one could choose a uniform distribution between two limits (perhaps zero and one to sample the full range of effects) or from a normal distribution truncated at zero at the lower end, and at one at the upper end. One then obtains the other absolute treatment effects via Eq.~(\ref{eq:p_i_ell}). By using a different method for generating the absolute effect of the trial-specific baseline ($p_{i,1}$) compared to the other absolute treatment effects in the trial ($p_{i,\ell\neq 1}$), this method introduces asymmetry into the generation procedure. A simple way of reducing the effect of this asymmetry on the synthetic data is to randomly select the trial-specific baseline treatments at each iteration $\nu$. 

An alternative data generation model was proposed by Seide et al (2019) \cite{Seide:2019}. Here, one chooses the treatment effect for the baseline treatment to be the value that minimises the Euclidean distance of the vector $(p_{i,1},\dots,p_{i,m_i})$ from the vector $(1/2, \dots, 1/2)$, i.e.,
\begin{eqnarray}
    p_{i,1} = \min\limits_{q} \Biggl[\left(q-\frac{1}{2}\right)^2+ \sum_{\ell=2}^{m_i}\bigg(p_{i,\ell}\left[q,\Delta_{i,1\ell}\right]-\frac{1}{2}\bigg)^2 \Biggr],
\end{eqnarray}
where $p_{i,\ell}[\cdot,\cdot]$ is the expression given in Eq.~(\ref{eq:p_i_ell}). This method maintains symmetry since all $m_i$ absolute treatment effects in trial $i$ are determined simultaneously.

\subsection{The process of carrying out a network meta-analysis -- Brief overview} \label{sec:ma_and_nma}

\subsubsection{Frequentist versus Bayesian network meta-analysis.}
\label{FreqBayesNMA}
In Sections~\ref{sec:Bayes} and \ref{sec:freq} below we describe the two main approaches to carrying out an NMA, i.e., the steps that are used to infer parameters such as relative treatment effects from the observed trial data.  The two approaches correspond to the two main branches of inference in general, frequentist and Bayesian inference. 

Much has been written about the difference between Bayesian and frequentist approaches to inference (e.g. \cite{Cox:2006, Bartholomew:1965, Wagenmakers2008, Samaniego:2010, Hespanhol:2019}). One central point distinguishing the two is the conception of probability. Frequentist inference defines the probability of some event in terms of how frequently the event occurs if we repeat some process (e.g. an experiment) many times \cite{Cox:2006}. The Bayesian approach instead uses probability to describe the degree of belief in a statement \cite{Glickman:2007, Greenland:2006}. In the Bayesian framework, parameters such as treatment effects are considered random variables, where the randomness reflects the remaining uncertainty after the inference process. If the distribution for a parameter is very sharp, then this indicates that we can be fairly certain that the inferred parameter is in a given range around the mode of that distribution. If the distribution is wide then the strength of our beliefs is weak. In the Bayesian approach probability therefore becomes subjective. It is not a property of the system only, but also of the prior beliefs, and the information available to the observer. Given that probability in Bayesian statistics reflects the degree of belief in the value of parameters, we can make statements such as `Given our prior beliefs and the data we have observed, we think that treatment A is more effective than treatment B with probability 70\%'. In frequentist methodology, probability is not an expression of our beliefs and therefore an equivalent statement would be, for example, `Based on hypothesised repetitions of the experiment, treatment A would be estimated to be more effective than treatment B 70\% of the time'.

A concept used in both Bayesian and frequentist inference is the so-called `likelihood function'. For observed data $\mathbf{{D}}$, the likelihood function is the conditional probability (or probability density) of observing this data given a specific set of model parameters $\boldsymbol{{\theta}}$, $P(\mathbf{{D}}|\boldsymbol{{\theta}})$\footnote{One way of thinking about this is as follows: Consider the map $P:(\mathbf{{D}}, \boldsymbol{\theta}) \mapsto P(\mathbf{{D}}, \boldsymbol{{\theta}})$ as a real-valued function of the two arguments $\mathbf{{D}}$ and $\boldsymbol{{\theta}}$ (which each may be multi-dimensional). We can then look at this from two perspectives: (i) Fixing $\boldsymbol{\theta}$ one obtains a map $\mathbf{{D}} \mapsto P(\mathbf{{D}}, \boldsymbol{{\theta}})$ describing the probability distribution (or density) of the data for given fixed parameters. Eq.~(\ref{eq:Binomial}) is an example.  (ii) If we fix $\mathbf{D}$ we obtain a map $\boldsymbol{{\theta}} \mapsto P(\mathbf{{D}}, \boldsymbol{{\theta}})$. This is the likelihood for the parameter $\boldsymbol{\theta}$, given the (fixed) data $\mathbf{D}$. }. The likelihood function is so named because it describes how likely it is to observe the given data for different values of parameters. In fact, the likelihood is viewed as a function of the parameters rather than the data and -- somewhat confusingly -- is often written as $L(\boldsymbol{\theta}|\mathbf{{D}})$.

\subsubsection{Arm-based versus contrast-based data.}
In the setup so far we have treated the $\{r_{i,\ell}\}$ (the number of events in the arm of the trials) as the trial outcome or `data' in an NMA. Since each of the $r_{i,\ell}$ is associated with an arm in a trial, data of this type is referred to as `arm-level' data. 

The event probabilities associated with these measurements can also be expressed on the logit scale and used to calculate the log odds ratio (LOR) representing the relative effect between two treatments. For example, we write $y_{i,1\ell}$ for the \textit{observed} LOR between the effect of treatment $\ell$ in trial $i$ and the baseline treatment in that trial,
\be \label{eq:LOR}
y_{i,1\ell} = \ln \frac{r_{i,\ell}/n_{i,\ell}}{1-r_{i,\ell}/n_{i,\ell}}- \ln \frac{r_{i,1}/n_{i,1}}{1-r_{i,1}/n_{i,1}}.
\ee

Sometimes a trial will only report the log odds ratio of each treatment relative to the trial-specific baseline (the $\{y_{i,1\ell}\}$), and not the detailed number of events in each arm (the $\{r_{i,\ell}\}$). The log odds ratio is a so-called `summary statistic' and
data of this type is called `summary-level data'.

In NMA we can choose to model data on the arm level or the summary level. We refer to these approaches as `arm-based' (AB) and `contrast-based' (CB) models respectively \cite{Salanti:2008}. Arm-level data is modelled using the binomial distribution [Eq.~(\ref{eq:Binomial})]. The likelihood function of the data is then also based on this binomial distribution. This is sometimes referred to as the `exact-likelihood' or `AB-likelihood' \cite{White:2019} approach. In contrast-based models the LORs from each trial are modelled as following a normal distribution. This is an approximation, and the approach is also referred to as the `approximate-likelihood' or the `CB-likelihood' \cite{White:2019} model\footnote{The terms `contrast-based' and `arm-based' have also been used to distinguish between models of relative treatment effects (CB) and models of absolute effects (AB) \cite{Dias:2016}. The latter are not standard practice. All models discussed in this article are based on relative treatment effects and we use CB/AB to distinguish between the `level' of data that is used in constructing the model.}. 

Both frequentist and Bayesian inference methods can be used to evaluate both AB and CB models. In practice, frequentist models \cite{Lumley:2002, White:2012, Rucker:2012} are usually based on contrast-level summaries while Bayesian models \cite{Lu:Ades:2004, DIAS:2018, TSD2} tend to use arm-level data \cite{Franchini:2012}.  Clearly, if trials only report summary-level data then we are restricted to CB models.

In the next section we summarise the general Bayesian approach to inference, and describe how this is applied in an arm-based NMA model. In Sec.~\ref{sec:freq} we first describe the contrast-based NMA model and give an overview of the frequentist approach to inference. We then show how frequentist inference can be used to estimate relative treatment effects in a CB model.

\section{Bayesian Network Meta-Analysis}\label{sec:Bayes} 
In this section we discuss the Bayesian approach to network meta-analysis. We use an arm-based model and treat the number of events in each arm of each trial (the $r_{i,\ell}$) as the raw data in the model.

\subsection{General approach}
The process of Bayesian NMA converts prior beliefs on the distribution of model parameters into posterior distributions using the observed data. The approach is based on Bayes theorem, which in its simplest form can be stated as $P(A|B)=P(B|A)\frac{P(A)}{P(B)}$,  where $A$ and $B$ are outcomes of a probabilistic experiment. Writing $\btheta$ for the parameters, and $\bD$ for the data, this becomes
\be
P(\btheta|\bD)=P(\bD|\btheta)\frac{P(\btheta)}{P(\bD)}.
\ee
We are interested in the distribution of parameters given the observed data. In this context, we notice that the term $P(\bD)$ is not a function of $\btheta$, and so we can write
\be\label{eq:bayes}
P(\btheta|\bD)=\mbox{const}\times P(\bD|\btheta)P(\btheta),
\ee
where the constant on the right is to be determined from normalisation. This is the fundamental equation for Bayesian NMA (and any other type of Bayesian inference). The object $P(\btheta)$ on the right is known as the {\em prior} distribution of parameters. It reflects our beliefs about what the parameters might be, before we have taken into account the data $\bD$. The expression on the left is the {\em posterior distribution} of the parameters, it represents our updated beliefs having observed and used the data $\bD$. The factor that connects the two is the conditional probability, or likelihood function, $P(\bD|\btheta)$ (see Sec.~\ref{FreqBayesNMA}). 

\subsection{Hierarchical structure of the random effects model}
In NMA the parameters $\btheta$ are the true relative treatment effects $\boldsymbol{d}$ and the heterogeneity parameter $\tau$. These are the `parameters of interest' \cite{Basu:1977, Cox:2006} of the model and we will refer to them simply as the `model parameters'. As described in Sec.~\ref{sec:FEM_REM} there are two levels of randomness in the RE model. The first layer generates trial-specific relative treatment effects $\mathbf{\Delta}_i$ ($i=1,\dots, M$) and absolute effects for the baseline treatment in each trial (the $p_{i,1}$ in Fig.~\ref{fig:nma_diagram}). In a second layer, binomial outcomes are then produced for each trial arm.  

The trial-specific effects $\mathbf{\Delta}_i$ and the $p_{i,1}$, are so-called `nuisance parameters' \cite{Basu:1977, Cox:2006}. For the discussion of the general Bayesian approach we will call these $\bnu$. They are random variables, and their distribution is parameterised by the model parameters $\btheta$. The nuisance parameters in turn determine the distribution of the output data $\bD$. This is captured by the following relation
\be
P(\bD|\btheta)=\int d\bnu ~ P_{\rm out}(\bD|\bnu)P_{\rm in}(\bnu|\btheta).
\ee
We write $P_{\rm in}(\bnu|\btheta)$ to describe the internal layer of the model (generation of nuisance parameters from the model parameters), and $P_{\rm out}(\bD|\bnu)$ for the `output layer' (generation of data from the nuisance parameters).

Using Eq.~(\ref{eq:bayes}) we then have
\be\label{eq:layers}
P(\btheta|\bD)=\mbox{const}\times \left(\int d\bnu ~ P_{\rm out}(\bD|\bnu)P_{\rm in}(\bnu|\btheta)\right) P(\btheta),
\ee
where $P(\btheta)$ is the prior distribution of the parameters $\btheta$. 

In Sec.~\ref{sec:construct_u} below, we focus on the construction of 
\be\label{eq:q}
U(\bD,\bnu,\btheta)\equiv P_{\rm out}(\bD|\bnu)P_{\rm in}(\bnu|\btheta)P(\btheta).
\ee
This is the joint distribution of the model parameters $\btheta$, the nuisance parameters $\bnu$ and the data $\bD$. 

To obtain the posterior distribution $P(\btheta|\bD)$ one fixes $\bD$ to be the observed data. The next step then is to integrate out the nuisance parameters in Eq.~(\ref{eq:layers}). The normalisation constant in this equation can be determined at the end.

\medskip

We have now reduced the problem of carrying out an NMA to two tasks: 
\begin{enumerate}
    \item[1.] We need to construct explicit forms for the factors on the right-hand side of Eq.~(\ref{eq:q});
    \item[2.] We need a method with which to integrate out the nuisance parameters, and to extract the posterior distribution for $\btheta$.
\end{enumerate}
We will first discuss step $1$. Numerical methods for step $2$ are described in Sec.~\ref{sec:computational_techniques}.

\subsection{Construction of the joint distribution of model parameters, nuisance parameters and the data.}\label{sec:construct_u}
\subsubsection{Choice of priors for the model parameters.}
The parameters of interest in the model are the heterogeneity $\tau$, and the true treatment effects of treatments $a\in\{T_2,\dots,T_N\}$ relative to the overall baseline treatment $T_1$. The method requires distributions capturing prior beliefs on the values these parameters might take. 

It is common to choose a Gaussian distribution as the prior for the relative treatment effects. The prior for $\tau$ has evoked more discussion \cite{Hig:White:1996, DuMouchel:1994, Lu:Ades:2009, Rosenberger:2021, Gelman:2006}, but the usual practice is to use a uniform prior distribution between zero and some upper limit, $\tau_{\rm max}$, which can depend on the data \cite{TSD2, Lu:Ades:2006}. 

This results in the form
\be\label{eq:tau_d_priors}
P(\btheta)=\frac{1}{\tau_{\rm max}} 1\!\!1_{[0,\tau_{\rm max}]}(\tau)\times \prod_{a\in\{T_2,\dots,T_N\}} \frac{\exp\left(-\frac{d_{T_1 a}^2}{2\sigma_d^2}\right)}{\sqrt{2\pi\sigma_d^2}},
\ee
with the indicator function $1\!\!1_{[x,y]}(\tau)=1$ for $x\leq\tau\leq y$, and $1\!\!1_{[x,y]}(\tau)=0$ otherwise. The product over $a$ has $N-1$ factors (one for each $a\in\{T_2,\dots,T_N\}$) and indicates that the prior distribution for each of the true relative treatment effects $d_{T_1 a}$ is a Gaussian distribution with mean zero, and variance $\sigma_d^2$. In using this factorised form, we have assumed that these parameters are a priori pairwise independent \cite{TSD3, Greco:2016}.   

It is common to use so-called `non-informative' priors. These are relatively broad distributions for each of the parameters, reflecting a situation in which little information about the parameters is known a priori. This can be achieved by choosing values for $\tau_{\rm max}$ and $\sigma_d^2$ that are large compared with the typical scale of the parameters $\tau$ and $\boldsymbol{d}$ respectively. What constitutes as `large' is informed by the type of data and the medical condition/clinical question of interest.  

Occasionally it may be necessary to use an informative prior for the heterogeneity parameter \cite{TSD2, Turner:2012, Rhodes:2015, Turner:2018, Hig:White:1996}. If the network contains very few trials per comparison then there is little information about the variation between trials and the estimation of $\tau$ is likely to be imprecise \cite{Turner:2018}. In particular, when a flat prior is used with data that gives little information about between-trial variance then the posterior of $\tau$ will be dominated by the prior which may lead to unrealistically high estimates of heterogeneity \cite{TSD2}.  Informative priors have been proposed for $\tau$ based on external data, for example databases of existing meta-analyses that relate to the relevant data type, medical condition and interventions \cite{Turner:2012, Rhodes:2015}. The use of such priors then allows us to incorporate external information about $\tau$ into the inference process.

\subsubsection{Distribution of nuisance parameters for given model parameters.}
The model parameters ($\boldsymbol{d} = (d_{T_1,T_2},\dots, d_{T_1,T_N})^\top$ and $\tau$) of the RE model determine the distribution of relative treatment effects $\bDelta_i$ for each of the trials $i=1,\dots,M$. As explained in more detail in Sec.~\ref{ModelDef}, the RE model assumes that each entry of $\bDelta_i$ is drawn from a Gaussian distribution centred on $\mathbf{X}_i\boldsymbol{d}$, where $\mathbf{X}_i$ is the $(m_i-1)\times (N-1)$ design matrix for trial $i$. The variance of each component of each $\bDelta_i$ is $\tau^2$.  The correlation between any two different entries of $\bDelta_i$ is assumed to be $\tau^2/2$, and there are no correlations between $\bDelta_i$ and $\bDelta_j$ for two different trials $i\neq j$.

Putting this all together we have
\begin{equation}
\label{eq:delta_cond}
    P_{\rm in}(\bnu|\btheta)=\prod_{i=1}^M \left[ \frac{\exp\left[-\frac{1}{2}(\bDelta_i-\mathbf{X}_i\boldd)^\top\boldSigma_i^{-1}(\bDelta_i-\mathbf{X}_i\boldd)\right]}{(2\pi)^{(m_i-1)/2}\det(\boldSigma_i)^{1/2}} \times P_{{\rm bl},i}(p_{i,1})\right],
\end{equation}
where for a given trial $i$, the matrix $\bSigma_i$ is of size $(m_i-1)\times (m_i-1)$, and has diagonal entries $\tau^2$, and off-diagonal entries $\tau^2/2$ (see Sec.~\ref{ModelDef}). The term $P_{{\rm bl},i}(p_{i,1})$ is the distribution for the absolute treatment effect of the trial-specific baseline. It is here common to use a non-informative distribution, such as a normal distribution for $\mathrm{logit}(p_{i,1})$ with large variance \cite{TSD2, Greco:2016}.

\subsubsection{Distribution of data for given nuisance parameters.}
Finally, the data is given by the number of `events' in each arm of each of the trials. The nuisance parameters $p_{i,1}$ and $\Delta_{i,1\ell}$ ($\ell=2,\dots,m_i$) for trial $i$ translate into event probabilities for the treatments in this trial via Eq.~(\ref{eq:p_i_ell}) which we re-state here,
\be
p_{i,\ell} = p_{i,\ell}(\bnu)= \frac{p_{i,1}e^{\Delta_{i,1\ell}}}{1+p_{i,1}(e^{\Delta_{i,1\ell}}-1)}.
\ee

For binary outcomes the distribution of the data for given nuisance parameters is the binomial for each arm, with parameters $n_{i,\ell}$ and $p_{i,\ell}$. Combining this for all arms of all trials in the network we arrive at
 \begin{equation}
 \label{eq:P-}
    P_{\rm out}(\bD|\bnu)=\prod_{i=1}^M \prod_{\ell=1}^{m_i}\left(\begin{array}{c}n_{i,\ell}\\
    r_{i,\ell}\end{array}\right)p_{i,\ell}(\bnu)^{r_{i,\ell}}[1-p_{i,\ell}(\bnu)]^{n_{i,\ell}-r_{i,\ell}}.
\end{equation}

\subsection{Computational techniques}\label{sec:computational_techniques}
The posterior distribution for the model parameters is obtained from
 \be
P(\btheta|\bD)=\mbox{const}\times\int d\bnu ~U(\bD,\bnu,\btheta),
\ee
where $U$ is defined in Eq.~(\ref{eq:q}). Even though $U$ is known in closed form, the integral over $\bnu$ cannot be performed analytically. Due to the high-dimensionality of the integral, direct numerical integration is not always viable either.

One therefore resorts to computational methods to sample combined values for $\bnu$ and $\btheta$, and then considers the resulting marginal distribution for $\btheta$. The most common  techniques to do this in meta-analysis are Markov Chain Monte Carlo (MCMC) methods. Popular MCMC software include WinBUGS \cite{Spiegelhalter:2003}, JAGS \cite{JAGS} and Stan \cite{STAN}. \\

MCMC methods are a class of algorithm based on the construction of a Markov chain with a stationary distribution given by the target distribution. By observing the chain after a large number of steps using Monte Carlo simulations, one eventually produces samples from the target distribution.

The MCMC implemented in the WinBUGS software combines the celebrated Metropolis-Hastings  algorithm with so-called Gibbs Sampling, resulting in what is called `Metropolis-in-Gibbs sampling'. We here briefly outline the main principles.

\subsubsection{Metropolis-Hastings Algorithm.}
We discuss this topic in a more general sense (independent of NMA) and write the distribution from which we would like to sample as $p(\boldsymbol{x})$. The Metropolis-Hastings algorithm \cite{Hastings:1970} is based on a Markov chain producing a trajectory $\boldsymbol{x}^t$, $t=0,1,2,\dots$. Each step consists of proposing a value for $\boldsymbol{x}^{t+1}$, followed by a decision whether to accept or reject this proposed value. The distribution of proposed values and the acceptance criterion only depend on $\boldsymbol{x}^t$, but not on states visited earlier in the sequence. They are constructed such that the resulting process has stationary distribution $p(\boldsymbol{x})$ \cite{Geyer:2011}. Provided one allows for a sufficiently long equilibration time $t_{\rm eq}$, the set  $\{\boldsymbol{x}^t:t>t_{\rm eq}\}$ represents a statistically faithful sample of the distribution $p(\boldsymbol{x})$.
 
 \medskip
 
\begin{shaded}
\noindent\textbf{The Metropolis-Hastings Algorithm:} 
\begin{enumerate}
    \item[1.] Initialise $t=0$, and $\boldsymbol{x}^t=\boldsymbol{x}^0$ for some starting value $\boldsymbol{x}^0$. 
    \item[2.] Generate a proposed value $\boldsymbol{x}'$ from the proposal distribution $q(\boldsymbol{x}'|\boldsymbol{x}^t)$.
    \item[3.] Calculate the acceptance probability 
   \be
        p_a(\boldsymbol{x}'|\boldsymbol{x}^t) = \min \left(1, \frac{p(\boldsymbol{x}')}{p(\boldsymbol{x}^t)}\frac{q(\boldsymbol{x}^t|\boldsymbol{x}')}{q(\boldsymbol{x}'|\boldsymbol{x}^t)}\right).
    \ee
   Then accept the proposal with probability $p_a(\boldsymbol{x}'|\boldsymbol{x})$, i.e., set $\boldsymbol{x}^{t+1}=\boldsymbol{x}'$. Alternatively, with probability $1-p_a(\boldsymbol{x}'|\boldsymbol{x})$ reject the proposed update, and set $\boldsymbol{x}^{t+1}=\boldsymbol{x}^t$.
    \item[4.] Increment time by one, and go to 2.
\end{enumerate}
\end{shaded}
The algorithm results in an overall probability $A(\boldsymbol{x}|\boldsymbol{y})=q(\boldsymbol{x}|\boldsymbol{y})p_a(\boldsymbol{x}|\boldsymbol{y})$ to transition to $\boldsymbol{x}$ if the chain is currently at $\boldsymbol{y}$. Using the fact that exactly one of $p_a(\boldsymbol{x}|\boldsymbol{y})$ and $p_a(\boldsymbol{y}|\boldsymbol{x})$ is equal to one for each pair of states $\boldsymbol{x}$ and $\boldsymbol{y}$, one has $p(\boldsymbol{x}) A(\boldsymbol{y}|\boldsymbol{x})=p(\boldsymbol{y})A(\boldsymbol{x}|\boldsymbol{y})$ for all $\boldsymbol{x}$ and $\boldsymbol{y}$, i.e., the detailed balance condition holds.  This is sufficient to demonstrate that $p(\boldsymbol{x})$ is indeed a stationary distribution of the process. We also need to choose the hopping kernel $q$ such that the stationary distribution is unique (i.e., the Markov chain must be irreducible and aperiodic). A sufficient condition to ensure this, is that $q$ is positive everywhere \cite{Robert:2015}. For further details see also \cite{Robert:2004,Toral:2014}.

\begin{figure}
   \centering
    \includegraphics[width=1\linewidth]{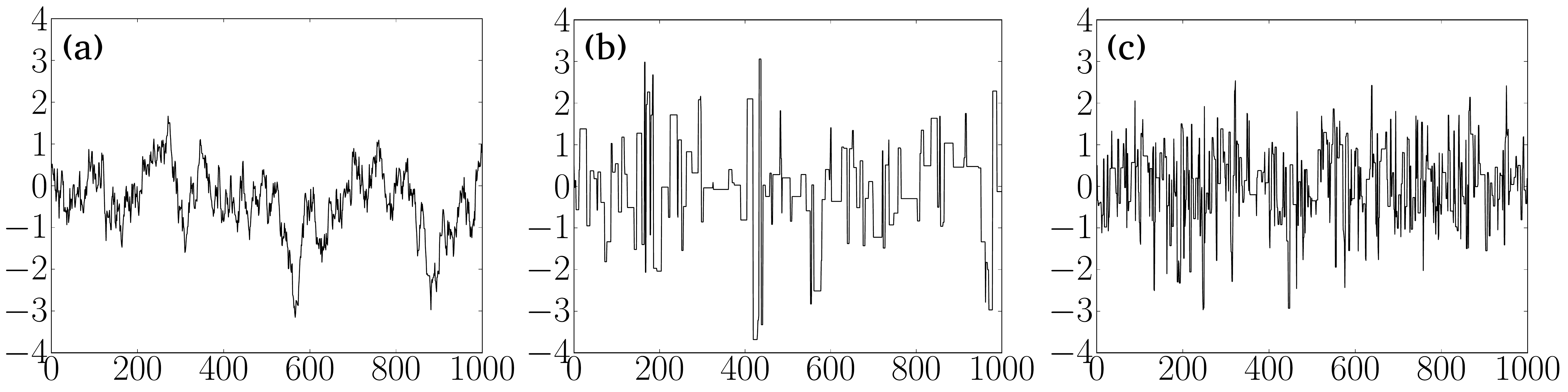}
    \caption{ (a) Sample path of a Markov chain  with small proposal variance and high acceptance rate. (b) Sample path with high proposal variance and low acceptance rate. (c) An efficient Markov chain with proposal variance tuned to obtain a `reasonable' acceptance rate. The example is for a standard normal distribution $p(x)$. In panel (a) the standard deviation of the hopping kernel is $0.25$, resulting in an acceptance rate of $0.925$, panel (b) is for a standard deviation of $10$ (acceptance rate $0.129$), and panel (c) for a standard deviation of $2.5$ (acceptance rate $0.43$). The optimal acceptance rate for a model in one dimension is approximately $0.44$ \cite{Gelman:1996}.  
    } \label{fig:AccRate} 
   
\end{figure}

The process simplifies if the proposal function (the hopping kernel) is symmetric, $q(\boldsymbol{x}'|\boldsymbol{x}^t) = q(\boldsymbol{x}^t|\boldsymbol{x}')$. In this case the acceptance probability in step 3 of the algorithm becomes
\begin{equation}
  p_a(\boldsymbol{x}'|\boldsymbol{x}^t) = \min \left(1, \frac{p(\boldsymbol{x}')}{p(\boldsymbol{x}^t)}\right).
\end{equation}
 An example of a symmetric proposal function $q(x'|x^t)$ in a univariate system ($x^t\in{\rm I\! R}$) is a normal distribution centred on the current sample value $x^t$ with a fixed variance \cite{Lynch:2007}. The choice of variance is not straightforward. Choosing a value that is too high means that most proposed values are rejected and the sequence of $\{x^t\}$ remains constant for long periods of time. This scenario is illustrated in Fig.~\ref{fig:AccRate} (b). On the other hand, if the variance is too small the chain does not explore the state space and convergence to the stationary state is slow (see Fig.~\ref{fig:AccRate} (a)) \cite{Roberts:2001}. Both of these scenarios make the MCMC less efficient and mean that more iterations are required for the chain to reach the stationary state. If the proposal variance is tuned so that the chain has a `reasonable' acceptance rate then the state space is explored efficiently. A chain with this characteristic is shown in Fig.~\ref{fig:AccRate} (c). 

For an $n$-dimensional target distribution the optimal acceptance rate has been found to be approximately $0.44$ for $n=1$ and declines to $0.23$ as $n \rightarrow \infty$ \cite{Gelman:1996, Roberts:2001}. In the WinBUGS software, the acceptance rate of proposed values is tuned to between $0.2$ and $0.4$ \cite{Spiegelhalter:2003}.

\subsubsection{Gibbs Sampling.}
Gibbs samplers are used to sample from multivariate distributions. They update one variable at a time, and are used when sampling from conditional probabilities of individual variables is easier than direct sampling from the multivariate distribution. The algorithm cycles through the individual variables, and samples from the conditional distribution of one variable in the target distribution given the current values of all other variables \cite{Lynch:2007, Gelman:2013}. This can be shown to generate a sequence of multivariate samples faithfully representing the joint distribution  \cite{Gelman:2013}.

Here we describe details of the Gibbs sampling procedure for a target distribution $p(\boldsymbol{x})$ of $n$ variables, where $\boldsymbol{x} = (x_1,x_2,...., x_n)$. The conditional probability distribution for the variable $x_i$ given all other variables is given by
\be
p_i(x_i| \boldsymbol{x}_{-i})=\frac{p(x_1,\dots, x_n)}{p_{-i}(\boldsymbol{x}_{-i})}.
\ee
We have written $\boldsymbol{x}_{-i}=(x_1,\dots, x_{i-1},x_{i+1},\dots,x_n)$ (i.e, $\boldsymbol{x}_{-i}$ is obtained from $\boldsymbol{x}$ by removing the $i$-th entry). The expression $p_{-i}(\boldsymbol{x}_{-i})=\int dx_i\, p(x_1, \dots, x_n)$ in the denominator is the marginal distribution for $\boldsymbol{x}_{-i}$.

\begin{shaded}
\noindent\textbf{The Gibbs Sampling Algorithm:}
\begin{enumerate}
\item[1.] Initialise time at $t=0$, and set $\boldsymbol{x}^{t=0}$ to a starting value.
\item[2.] Update entries of $\boldsymbol{x}$ in turn:

\smallskip

First, sample $x_{1}^{t+1}$ from the conditional distribution $p_1(\cdot|x_{2}^t,\dots,x_{n}^t)$.

\smallskip

Then sample $x_{2}^{t+1}$ from $p_2(\cdot|x_{1}^{t+1},x_{3}^{t},\dots,x_{n}^{t})$.

\smallskip
 
Then sample $x_{3}^{t+1}$ from $p_3(\cdot|x_{1}^{t+1},x_{2}^{t+1},x_{4}^{t}\dots,x_{n}^{t})$.

 \dots
 
 \smallskip
 
 Then sample $x_{n-1}^{t+1}$ from $p_{n-1}(\cdot|x_{1}^{t+1},\dots, x_{n-2}^{t+1},x_{n}^{t})$.

\smallskip
 
 Finally, sample $x_{n}^{t+1}$ from $p_{n}(\cdot|x_{1}^{t+1},\dots, x_{n-1}^{t+1})$.
 
 \smallskip
 
At the end of this process $\boldsymbol{x}^{t+1}$ is available. 
 \item[3.] Increment time by one, and go to 2
 \end{enumerate}
\end{shaded}

\subsubsection{Metropolis-in-Gibbs.} 
In NMA we are interested in samples from the distribution $U(\bD,\bnu,\btheta)=P_{\rm out}(\bD|\bnu)P_{\rm in}(\bnu|\btheta)P(\btheta)$ in Eq.~(\ref{eq:q}), for a fixed $\bD$. To generate these samples we use the `Metropolis-in-Gibbs' algorithm. This is a Gibbs sampling algorithm with a Metropolis-Hastings accept/reject step used to sample from the conditional distributions for the individual variables, i.e., we use the Metropolis-Hastings algorithm at each single-variable stage of the Gibbs Sampling algorithm. 

\begin{shaded}
\noindent \textbf{Metropolis-in-Gibbs for a random effects NMA:}
\begin{enumerate}
\item[1.] Initialise time at $t=0$ and initialise the parameters:  $\Delta_{i,\ell}^{0}$, $p_{i,1}^{0}$, $\tau^0$ and $d_{T_1a}^0$  for $i=1,\dots, M$, $\ell=1,\dots, m_i$ and $a=T_2,\dots, T_N$
\item[2.] Update the trial-specific treatment effects $\Delta_{i,\ell}$. For each study $i=1,2,...,M$ and
each $\ell=2,...,m_i$:
\begin{enumerate} 
\item Draw $\Delta_{i,\ell}'$ from the normal distribution $\mathcal{N}(\Delta_{i,\ell}^t,v_{\Delta})$.
\item Set
\begin{equation}
    \Delta_{i,\ell}^{t+1} = \left\{\begin{array}{ll}
    \Delta_{i,\ell}' & \mbox{with probability } p_\Delta \\
    \Delta_{i,\ell}^t & \mbox{with probability } 1-p_\Delta,
    \end{array}\right.
\end{equation}
where 
\end{enumerate}
\begin{equation}
\hspace{-3em}p_{\Delta} = \min\left(1,\frac{U(\Delta_{i,\ell}'|\Delta_{i,2}^{t+1},...,\Delta_{i,\ell-1}^{t+1},\Delta_{i,\ell+1}^{t},...,\Delta_{i,m_i}^t, p_{i,1}^t,\boldsymbol{d}^{t},\tau^{t},\mathbf{r})}{U(\Delta_{i,\ell}^t|\Delta_{i,2}^{t+1},...,\Delta_{i,\ell-1}^{t+1},\Delta_{i,\ell+1}^{t},...,\Delta_{i,m_i}^t, p_{i,1}^t,\boldsymbol{d}^{t},\tau^{t},\mathbf{r})}\right).
\end{equation}
NB: The parameters $\Delta_{i,\ell}$, specific to trial $i$, are independent (under the distribution $U$) of parameters that are specific to other trials $j \neq i$. 
\item[3.] Update the trial-specific baselines. For each trial $i=1,...,M$
\begin{enumerate}
\item Draw $\mbox{logit}\,p_{i,1}'$ from  $\mathcal{N}(\mbox{logit}\,p_{i,1}^t,v_b)$. From this obtain $p_{i,1}'$.
\item Set 
\begin{equation}
    p_{i,1}^{t+1}=\left\{\begin{array}{ll}
    p_{i,1}' & \mbox{with probability } p_b \\
    p_{i,1}^t & \mbox{with probability } 1-p_b,
    \end{array}\right.
\end{equation}
where 
\begin{equation}
    p_b= \min\left(1, \frac{U(p_{i,1}'|\bDelta_i^{t+1}, \boldsymbol{d}^{t},\tau^{t}, \boldsymbol{r})}{U(p_{i,1}^t|\bDelta_i^{t+1}, \boldsymbol{d}^{t},\tau^{t}, \boldsymbol{r})}\right).
\end{equation}
\end{enumerate}
We note that $p_{i,1}$ is independent of all $p_{j,1}$, $j\neq i$ under the distribution $U$.

\item[4.] Update the heterogeneity parameter: 
\begin{enumerate}
    \item Draw $\tau'$ from $\mathcal{N}(\tau^{t},v_{\tau})$.
    \item Set
    \begin{equation}
        \tau^{t+1} = \left\{\begin{array}{ll}
        \tau' & \mbox{with probability } p_{\tau} \\
        \tau^{t} & \mbox{with probability } 1-p_{\tau},
        \end{array}\right.
    \end{equation}
    where
    \begin{equation}
        p_{\tau} = \min\left(1,\frac{U(\tau'|\bDelta^{t+1}, \boldsymbol{p}_1^{t+1}, \boldsymbol{d}^{t}, \boldsymbol{r})}{U(\tau^t|\bDelta^{t+1}, \boldsymbol{p}_1^{t+1}, \boldsymbol{d}^{t}, \boldsymbol{r})}\right).
    \end{equation}
\end{enumerate}
We note that the acceptance probability $p_\tau$ in step (b) is zero by construction if $\tau'<0$ [see Eq.~(\ref{eq:tau_d_priors})].
\item[5.] Update the basic parameters. For each treatment $a= T_2,...,T_N$
\begin{enumerate}
    \item Draw $d_{T_1 a}'$ from $\mathcal{N}(d_{T_1a}^{t},v_{d})$.
    \item Set
    \begin{equation}
        d_{T_1a}^{(t+1)} = \left\{\begin{array}{ll}
        d_{T_1a}' & \mbox{with probability } p_d \\
        d_{T_1a}^{t} & \mbox{with probability } 1-p_d, \\ 
        \end{array}\right. 
    \end{equation}
    where
\begin{equation}
  \hspace{-7em}      p_d = \min\left(1, \frac{U(d_{T_1a}'|\bDelta^{t+1}, \boldsymbol{p}_1^{t+1}, \tau^{t+1}, d_{T_1T_2}^{t+1},...,d_{T_1T_{\alpha-1}}^{t+1},d_{T_1T_{\alpha+1}}^{t},...,d_{T_1T_N}^{t},\boldsymbol{r})}{U(d_{T_1a}^t|\bDelta^{t+1}, \boldsymbol{p}_1^{t+1}, \tau^{t+1}, d_{T_1T_2}^{t+1},...,d_{T_1T_{\alpha-1}}^{t+1},d_{T_1 T_{\alpha+1}}^{t},...,d_{T_1T_N}^{t},\boldsymbol{r})}\right),
\end{equation}
with $a=T_{\alpha}$ and $\alpha=2,\dots,N$.
\end{enumerate}
\item[6.] Increment time from $t$ to $t+1$. Go to 2
\end{enumerate}
\end{shaded}
We note that the acceptance probabilities $p$ in each step of the algorithm are of the form
\be\label{eq:prop_param}
p = \min\left(1,\frac{U(\mbox{parameter}'|\mbox{other parameters, data})}{U(\mbox{parameter}^t|\mbox{other parameters, data})}\right),
\ee
where $\mbox{parameter}'$ is the proposed value for the model or nuisance parameter that is being updated, and $\mbox{paramater}^t$ its value in the previous iteration. Crucially, the `other parameters' and the data in the numerator and denominator in Eq.~(\ref{eq:prop_param}) are the same. Using the definition $P(A|B)=P(A,B)/P(B)$ of conditional probabilities we can then write
\be\label{eq:aux3}
p = \min\left(1,\frac{U(\mbox{parameter}',\mbox{other parameters, data})}{U(\mbox{parameter}^t,\mbox{other parameters, data})}\right).
\ee
This means that we can use joint probabilities (or probability densities) instead of conditional probabilities. 

Using the product form of the distribution $U$ in Eq.~(\ref{eq:q}) and the fact that $P_{\rm in}$, $P_{\rm out}$ and the prior further factorise, the ratios in Eq.~(\ref{eq:aux3}) can be simplified even more by cancelling factors that do not depend on the parameter that is being updated.

\subsubsection{Assessing Convergence.}
The MCMC dynamics define a stochastic process with a stationary probability distribution for the model parameters. The process is constructed such that this stationary distribution is the target distribution we set out to sample from. Formally, the stationary distribution is reached only at infinite time, but in practice samples are effectively drawn from the target distribution after sufficiently many iterations. In simulations, we therefore discard the samples of the first $n_c$ iterations (this is referred to as the `burn in' in the statistics community, physicists know this as equilibration time or transient). We then make inferences about our parameters based on samples taken in the subsequent iterations.

To obtain accurate inferences on the parameter values we must, therefore, assess the number of iterations required to reach stationarity. A common method to assess this was developed by Gelman and Rubin  \cite{Gelman:Rubin:1992} and later modified by Brooks and Gelman  \cite{Brooks:Gelman:1998}. The latter article gives a detailed description of the approach, we summarise the main ideas here.

A Markov chain has converged (reached stationarity) when the statistics of the samples taken do not depend on distribution of initial conditions for the process. The Brooks-Gelman-Rubin approach is therefore based on assessing the similarity of samples (more precisely, distributions of samples) obtained from multiple independent chains (realisations of the process) with different starting points. 

Assume a target distribution for a scalar parameter with mean $\mu$ and variance $\sigma^2$. We now consider $m$ realisations of the process, $i=1,\dots,m$, with a set of over-dispersed (i.e, widely spread compared to the expected scale of the parameter) starting values. Each realisation is run for a burn-in of $n$ iterations, followed by another $n$ iterations during which samples are taken. This generates $m$ sets of samples of the parameter. The sample mean of each realisation $i$ provides an estimate $\hat\mu_i$ for the mean $\mu$.  We then have $m$ inferences about the parameter $\mu$ from the $m$ chains. The variance between samples {\em within} realisation $i$ is labelled $v_i$ (the `within-chain variance').

\begin{figure}
   \centering
    \includegraphics[width=0.8\linewidth]{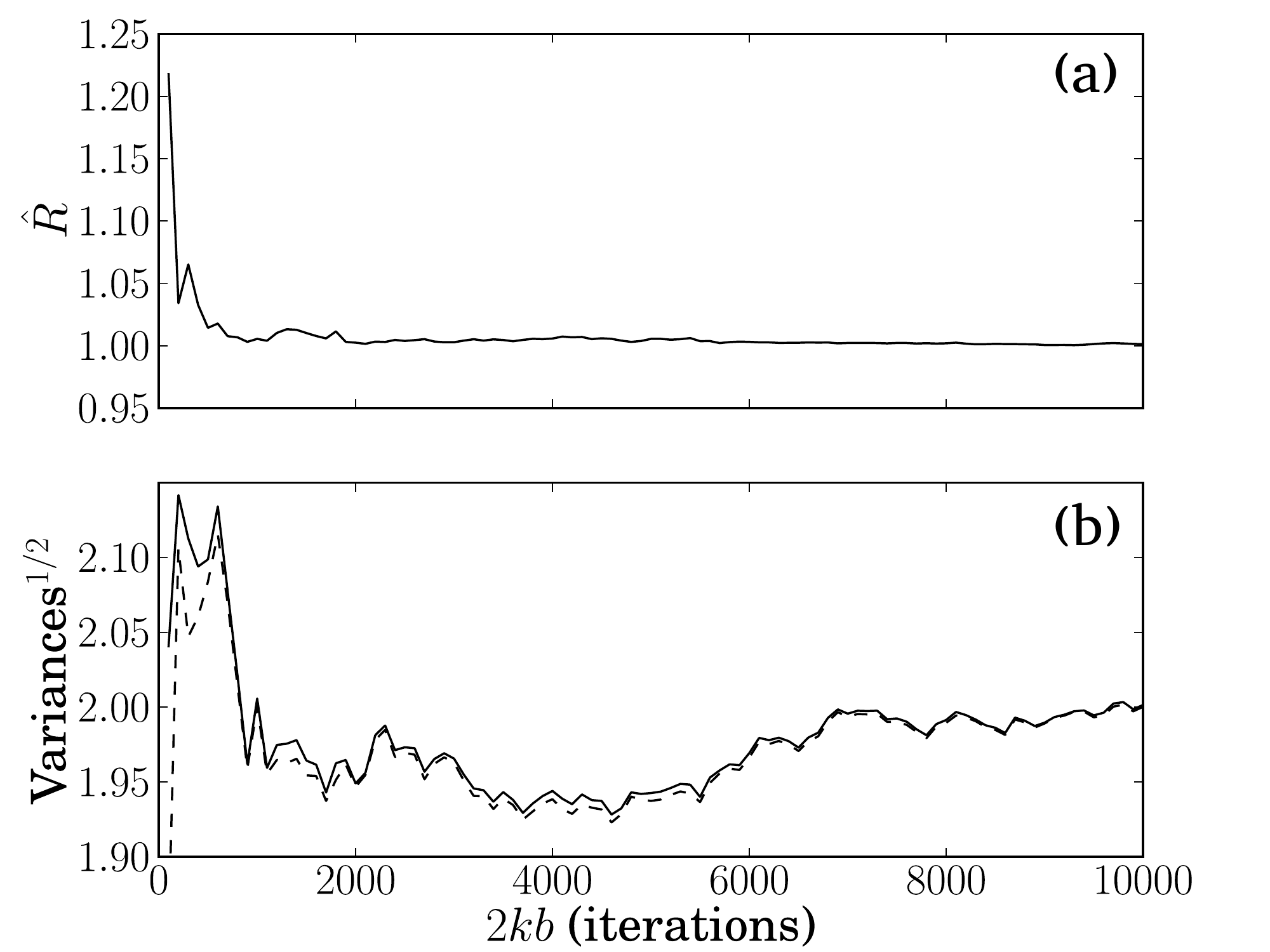}
    \caption{ Examples of Brooks-Gelman-Rubin convergence plots with $m=5$ chains and batch lengths of $b=50$. (a) The ratio $\hat{R}$ of the pooled variance and the average within-chain variance against the number of iterations. (b) The solid line shows (the square root of) the pooled variance $\hat{V}$ as a function of the number of iterations. The dotted line is (the square root of) the average within-chain variance. For this example convergence is reached after approximately $2kb=7000$ iterations (or a burn in of 3500).  \label{fig:BRG}}
\end{figure}

We can also obtain an inference about $\mu$ from the combined set of $mn$ samples from all realisations. The overall mean of this sample $\hat{\mu}$ is  the mean of the $\hat{\mu}_i$. One can then construct a measure of the so-called `pooled variance' $\hat{V}$ that accounts for the average within-chain variance, the variance in the value of $\hat{\mu}_i$ between realisations, and the sampling variability. For details of this procedure see \cite{Brooks:Gelman:1998}.

As the number of iterations $n$ (before and after burn-in) increases, we expect the value of the pooled variance and the average within-chain variance to stabilise and for these variances to converge to the same value \cite{Gelman:Rubin:1992}. To assess the number of iterations required for convergence we split each chain (total length $2n$) into $n/b$ batches of length $2b$.

For a given integer $k=1,\dots,n/b$ we then calculate the pooled variance, the average within-chain variance and their ratio $\hat{R}$ based on the samples in the first $k$ batches, where the first half of this data is discarded as burn-in. I.e., for $k=1$, we use the first batch (length $2b$), discard the first $b$ iterations of it, and compute the variances and their ratio based on iterations $b+1,...,2b$. For $k=2$, we use the first $4b$ iterations (two batches), but again discard the first half of this, and compute the variances from iterations $2b+1,\dots, 4b$. For higher values of $k$ we proceed analogously.

Plotting the variances and their ratio against $k$ (or $2kb$) as shown in Fig.~\ref{fig:BRG}, we can assess the approximate number of iterations required for the variances to stabilise and for $\hat{R}$ to be sufficiently close to unity so that the chain can be assumed to have converged \cite{Brooks:Gelman:1998}. More recent refinements of this method are discussed in \cite{Gelman:2013}. 

Once the Markov chain has reached stationarity, we make inferences about the model parameters from the samples taken in the simulations. Usually this means calculating a central value of the distribution of samples (such as the mean or median) and some measure of spread (such as standard deviation). We discuss how exactly the results of the NMA are reported in Sec.~\ref{report}.

\section{Frequentist Network Meta-Analysis}\label{sec:freq}
We now move on to the frequentist approach to network meta-analysis. In this section we use the contrast-based variant of NMA and treat the relative treatment effects measured in each trial as the raw data in the model. For binomial outcomes in each trial arm, these are the log odds ratios $y_{i,1\ell}$.

This section is structured as follows: In Sec.~\ref{sec:freq_intro} we introduce the contrast-based NMA model and write it as a linear regression problem. Then, in Sec.~\ref{sec:freq_gen} we discuss the frequentist approach in more general terms and describe how we estimate regression coefficients in a linear regression model. We start by explaining the ordinary least squares (OLS) method and then use this to derive the generalised least squares (GLS) problem. We then explain the maximum likelihood (ML) approach and show that this leads to the same condition as GLS. We solve the GLS/ML problem to obtain the so-called `Aitken estimator' of the regression coefficients. In Sec.~\ref{freq_NMA} we go back to the NMA model. First,  we use the Aitken estimator to estimate the relative treatment effects $\boldsymbol{d}$ (Sec.~\ref{Freq_d_est}). Finally, we explain two common methods for estimating the heterogeneity variance $\tau^2$ (Sec.~\ref{freq_tau}). 

\subsection{Introduction and notation}\label{sec:freq_intro}

We begin by outlining the assumptions of the contrast-based NMA model and writing the inference task as a linear regression problem.

As in previous sections, the vector of relative treatment effects in each trial is assumed to follow a normal distribution. We recall Eq.~(\ref{eq:multi_norm}), which we can write as
\be\label{eq:lin_trial}
\Delta_{i,1\ell}=d_{t_{i,1},t_{i,\ell}}+\eta_{i,1\ell},
\ee
where $i=1,\dots, M$, $\ell=1,\dots, m_i$ and $\eta_{i,1\ell}$ is a Gaussian random variable of mean zero. Using transitivity [Eq.~(\ref{eq:transitivity})] the mean $d_{t_{i,1},t_{i,\ell}}$ can be constructed via a linear relation from the vector of basic parameters $\boldsymbol{d}=(d_{T_1,T_2},\dots,d_{T_1,T_N})^\top$ [cf. Eq.~(\ref{eq:linear_X})]. The variance of $\eta_{i,1\ell}$ is given by the heterogeneity $\tau^2$, and we note that for a fixed $i$ the different $\eta_{i,1\ell}$, $\ell=1,\dots,m_i$ will in general be correlated [see Eq.~(\ref{eq:multi_norm})]. We can collect the relations in Eq.~(\ref{eq:lin_trial}) for all trials $i$ and all basic comparisons within each trial, and write more compactly
\be\label{eq:linear}
\boldsymbol{\Delta}=\mathbf{X}\boldsymbol{d}+\boldsymbol{\eta}.
\ee
Here, $\mathbf{X}$ is the design matrix of the \textit{network} which can be constructed from the trial-specific design matrices  described in Sec.~\ref{ModelDef}, $\mathbf{X}=(\mathbf{X}_1, \dots, \mathbf{X}_M)^\top$. 

The matrix $\mathbf{X}$ has $N-1$ columns and $\sum_{i=1}^M (m_i-1)$ rows, and each entry is either $-1$, $0$ or $1$. Each column of $\mathbf{X}$ represents one of the treatments $T_2,...,T_{N}$ (treatment $T_1$ is the overall baseline). The rows represent comparisons to the trial-specific baseline in each study.

As before, we write $y_{i,1\ell}$ for the \textit{observed} relative effects in each trial [e.g. the log odds ratios in Eq.~(\ref{eq:LOR})]. These are assumed to follow a normal distribution centred on the mean value $\Delta_{i,1\ell}$ with some random sampling error, $\epsilon_{i,1\ell}$. That is,
\be
y_{i,1\ell} = \Delta_{i,1\ell}+\epsilon_{i,1\ell},
\ee
where the sampling errors $\epsilon_{i,1\ell}$ within a trial are correlated. Trial $i\in\{1,\dots,M\}$  compares $m_i$ treatments and therefore contributes $m_i-1$ relative treatment effects (comparisons of treatments $\ell=2,\dots,m_i$ to the trial-specific baseline treatment).

Collecting the $\sum_{i=1}^{M} (m_i-1)$ observations $y_{i,1\ell}$ in the vector $\boldsymbol{y}$, we can write the linear model as
\begin{equation}
\label{eq:RE-regress}
    \boldsymbol{y} = \boldsymbol{\Delta} + \boldsymbol{\epsilon} = \mathbf{X}\boldsymbol{d} + \boldsymbol{\eta} + \boldsymbol{\epsilon}.
\end{equation}
The vectors $\boldsymbol{\eta}$  and $\boldsymbol{\epsilon}$ represent the two levels of stochasticity in the RE model described in Sec.~\ref{sec:FEM_REM}; $\boldsymbol{\eta} \sim \mathcal{N}(\boldsymbol{0}, \boldsymbol{\Sigma})$ models the trial-to-trial variation of relative treatment effects, and $\boldsymbol{\epsilon} \sim \mathcal{N}(\boldsymbol{0},\mathbf{V})$ models the noise on the observed relative treatment effects resulting from the sampling in the trial arms (we recall that we are working within a contrast-based model, where the sampling noise in the trial arms is assumed to be Gaussian). The covariance matrices for the two types of stochasticity are $\boldsymbol{\Sigma}$ and $\mathbf{V}$ respectively, we will discuss their mathematical form below. The  two types of noise are independent of each other, and the overall covariance matrix is then $\mathbf{C}=\boldsymbol{\Sigma}+\mathbf{V}$, such that the model in Eq.~(\ref{eq:RE-regress}) can be written as
\begin{eqnarray}
    \boldsymbol{y} \sim \mathcal{N}(\mathbf{X}\boldsymbol{d}, \boldsymbol{\Sigma}+\mathbf{V}).
\end{eqnarray}

The covariance matrix associated with the sampling errors in the trials is of block diagonal form, $\mathbf{V} = \mathrm{diag}(\mathbf{V}_i)$, where each trial $i$ contributes an $(m_i-1)\times(m_i-1)$ matrix,
\begin{equation}
\label{eq:Vi}
    \mathbf{V}_i = \left(\matrix{
    \sigma_{i,12}^2 & \mathrm{Cov}(y_{i,12},y_{i,13}) & \dots & \mathrm{Cov}(y_{i,12},y_{i,1m_i})\cr
    \mathrm{Cov}(y_{i,13},y_{i,12}) & \sigma_{i,13}^2 & \dots & \vdots\cr
    \vdots & \vdots & \ddots &\vdots \cr
     \mathrm{Cov}(y_{i,1m_i},y_{i,12}) & \dots &\dots & \sigma_{i,1m_i}^2
    }\right).
\end{equation}
We stress that this describes sampling errors only, i.e., the matrix entries are the variances of the components $\epsilon_{i,1\ell}$, $\ell=2,\dots,m_i$, and the correlations between these variables. The measurements of relative treatment effects within a multi-arm trial are correlated because they involve a common treatment arm (the trial-specific baseline treatment). The values that make up the matrices $\mathbf{V}_i$ are assumed to be known (i.e. they are reported in the study, or can be directly calculated from the data - see Appendix~\ref{App:SigCovHat} for details). Further details can also be found in  \cite{DerSimonian:1986, Chang:2001, TSD2, Higgins:2009, Hamza:2008, Riley:2012}.

The covariance matrix associated with the random effects $\boldsymbol{\Sigma}$ represents the heterogeneity \textit{between} trials. Similarly to $\mathbf{V}$, it has block diagonal form, $\boldsymbol{\Sigma} = \mathrm{diag}(\boldsymbol{\Sigma}_i)$, where the blocks are the $(m_i-1)\times (m_i-1)$ matrices $\boldsymbol{\Sigma}_i$ defined in Sec.~\ref{ModelDef}. The diagonal elements of $\boldsymbol{\Sigma}_i$ are the variances associated with the random effects and the off diagonal elements relate to the correlations between the random effects within a multi-arm trial. We assume that these are determined by the unknown heterogeneity variance, $\tau^2$, as described in Sec.~\ref{ModelDef}. Determining $\tau^2$ is therefore part of the inference problem.

\begin{figure}
   \centering
    \includegraphics[width=0.7\linewidth]{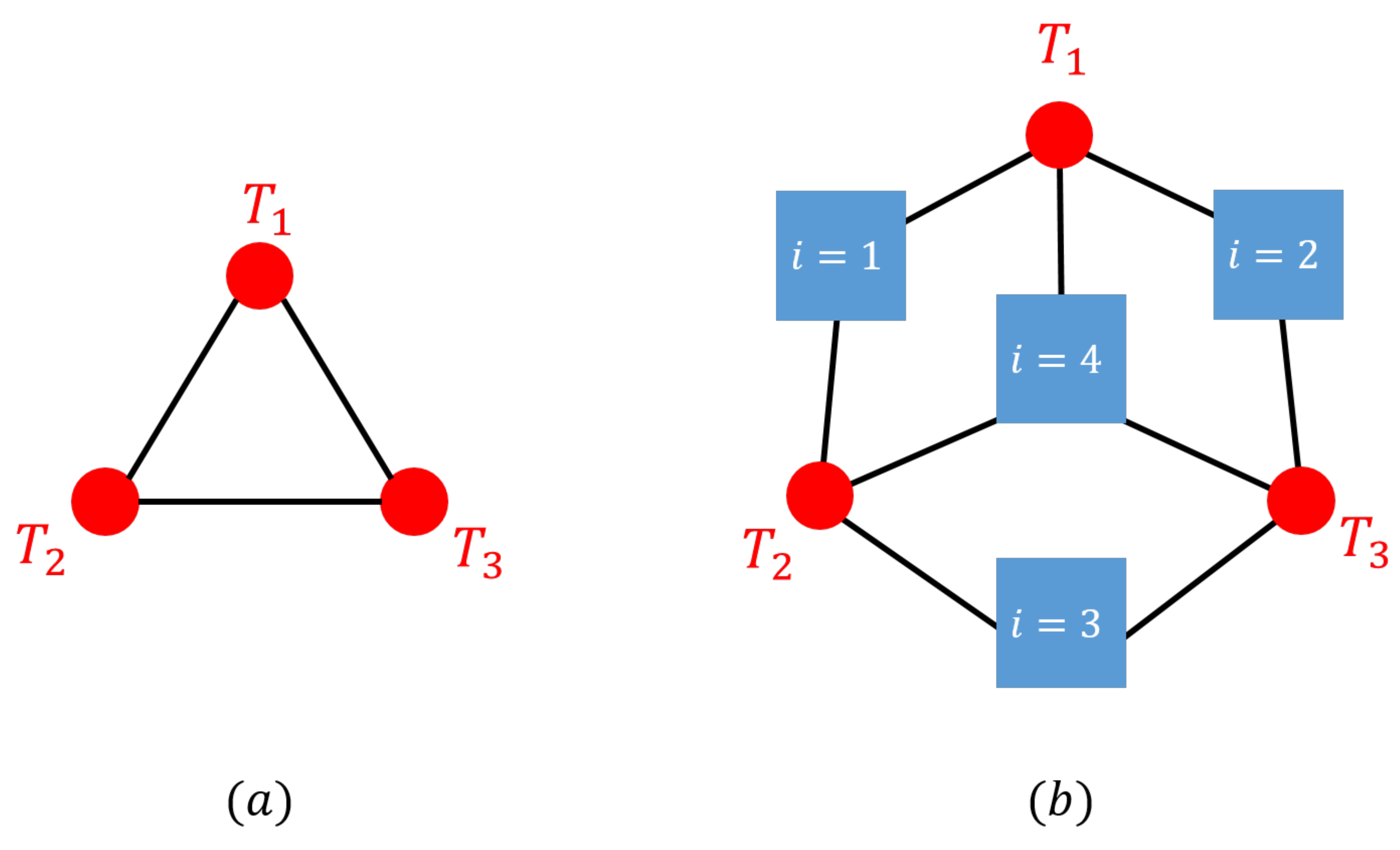}
    \caption{A fictional example of a network meta-analysis of $N=3$ treatments, $\{T_1,T_2,T_3\}$, and $M=4$ trials. (a) Standard network representation, all treatments are included in three trials (each pair of treatments appears in two trials). (b) Representation as a bipartite graph indicating which treatments are compared in each trial $i=1,\dots,4$. \label{fig:FictionEG}}
\end{figure}

\begin{shaded}
\noindent\textbf{Example - NMA as a linear regression model:}
 
\noindent Consider a network of $M=4$ trials comparing $N=3$ treatments, $T_1,T_2,T_3$. Trials $i=1,2,3$ are two arm trials comparing $(T_1,T_2)$, $(T_1,T_3)$ and $(T_2,T_3)$ respectively. Trial $i=4$ is a three-arm trial comparing all three treatments. This network is shown in Fig.~\ref{fig:FictionEG}. The regression model is then
\begin{equation*}
    \left(\matrix{
    y_{1,T_1 T_2}\cr
    y_{2,T_1 T_3}\cr
    y_{3,T_2 T_3}\cr
    y_{4,T_1 T_2}\cr
    y_{4,T_1 T_3}
    }\right)=\left(\matrix{
    1 & 0\cr
    0 & 1\cr
    -1 &1\cr
    1 & 0 \cr
    0 & 1
    }\right)\left(\matrix{
    d_{T_1 T_2}\cr
    d_{T_1 T_3}
    }\right)+\left(\matrix{
    \eta_{1,T_1 T_2}\cr
    \eta_{2,T_1 T_3}\cr
    \eta_{3,T_2 T_3}\cr
    \eta_{4,T_1 T_2}\cr
    \eta_{4,T_1 T_3}
    }\right)+\left(\matrix{
    \epsilon_{1,T_1 T_2}\cr
    \epsilon_{2,T_1 T_3}\cr
    \epsilon_{3,T_2 T_3}\cr
    \epsilon_{4,T_1 T_2}\cr
    \epsilon_{4,T_1 T_3}
    }\right)
\end{equation*}
where
\begin{equation*}
    \hspace{-80pt}\left(\matrix{
    \epsilon_{1,T_1 T_2}\cr
    \epsilon_{2,T_1 T_3}\cr
    \epsilon_{3,T_2 T_3}\cr
    \epsilon_{4,T_1 T_2}\cr
    \epsilon_{4,T_1 T_3}
    }\right)\!\sim\!\mathcal{N}\left( \hspace{-3pt}\left(\matrix{0\cr
    0\cr
    0\cr
    0\cr
    0
    }\right),\left(\matrix{
    \sigma_{1,T_1 T_2}^2 \hspace{-5pt}&\hspace{-5pt} 0 \hspace{-5pt}&\hspace{-5pt} 0 \hspace{-5pt}&\hspace{-5pt} 0 \hspace{-5pt}&\hspace{-5pt} 0\cr
    0 \hspace{-5pt}&\hspace{-5pt} \sigma_{2,T_1 T_3}^2 \hspace{-5pt}&\hspace{-5pt} 0 \hspace{-5pt}&\hspace{-5pt} 0 \hspace{-5pt}&\hspace{-5pt} 0\cr
    0 \hspace{-5pt}&\hspace{-5pt} 0 \hspace{-5pt}&\hspace{-5pt} \sigma_{3,T_2 T_3}^2 \hspace{-5pt}&\hspace{-5pt} 0 \hspace{-5pt}&\hspace{-5pt} 0\cr
    0 \hspace{-5pt}&\hspace{-5pt} 0 \hspace{-5pt}&\hspace{-5pt} 0 \hspace{-5pt}&\hspace{-5pt} \sigma_{4,T_1 T_2}^2 \hspace{-5pt}&\hspace{-5pt} \mathrm{Cov}(y_{4,T_1 T_2}, y_{4,T_1 T_3}) \cr
    0 \hspace{-5pt}&\hspace{-5pt} 0 \hspace{-5pt}&\hspace{-5pt} 0 \hspace{-5pt}&\hspace{-5pt} \mathrm{Cov}(y_{4,T_1 T_3}, y_{4,T_1 T_2})\hspace{-5pt}&\hspace{-5pt}\sigma_{4,T_1 T_3}^2 
    }\right)\hspace{-3pt} \right)\hspace{-3pt} ,
\end{equation*}
where the matrix entries are the variances and covariances of the  $\epsilon_{i,1\ell}$. We also have
\begin{equation*}
    \left(\matrix{
    \eta_{1,T_1 T_2}\cr
    \eta_{2,T_1 T_3}\cr
    \eta_{3,T_2 T_3}\cr
    \eta_{4,T_1 T_2}\cr
    \eta_{4,T_1 T_3}
    }\right)\sim\mathcal{N}\left(\left(\matrix{0\cr
    0\cr
    0\cr
    0\cr
    0
    }\right),\left(\matrix{
    \tau^2 & 0 & 0 & 0 & 0 \cr
    0 & \tau^2 & 0 & 0 & 0 \cr
    0 & 0 & \tau^2 & 0 & 0\cr
    0 & 0 & 0 & \tau^2 & \tau^2/2 \cr
    0 & 0 & 0 & \tau^2/2 & \tau^2
    }\right) \right).
\end{equation*}
\end{shaded}

The aim of NMA is to estimate the unknown parameters $\boldsymbol{d}$ and $\tau^2$. In the language of regression models, $\boldsymbol{d}$ are the `regression coefficients' of the linear model and $\tau^2$ is the `variance parameter'. In frequentist NMA, the variance parameter is estimated first, then this estimate is used in the estimate of the regression coefficients. In the following we describe two frequentist approaches for estimating the regression coefficients assuming knowledge of the variance parameter. We then relate this to the contrast-based NMA model. Finally, we describe some common methods for estimating the variance parameter.

\subsection{General frequentist approach} \label{sec:freq_gen}

Two frequentist approaches to inferring the regression coefficients of a model are based on `maximum likelihood' and `least squares'. 

In the maximum likelihood (ML) approach one finds the values of the parameters that maximise the likelihood or, equivalently, minimise the negative log likelihood. In least squares regression, we start from Eq.~(\ref{eq:RE-regress}). The vector $\boldsymbol{y}$ is observed from the trials, and the matrix $\mathbf{X}$ is known from the design of the trials (what treatments are tested in each trial). We therefore wish to find the vector $\boldsymbol{d}$ that best fits the observed data via Eq.~(\ref{eq:RE-regress}). To do this a `residual' is defined as the difference between the observed value of the response variable (here $\boldsymbol{y}$) and the mean value $\mathbf{X}\boldsymbol{d}$ predicted by the regression model. The model parameters (here $\boldsymbol{d}$) are then estimated by minimising the sum of the squared residuals, i.e. we find the values of the model parameters that `best fit' the data. When measurements are associated with correlated random errors we must use so-called `generalised least squares' (GLS) regression \cite{Amemiya:1985}. This will be explained in more detail later. For a linear regression model under the assumption of normally distributed errors, the ML estimates and GLS estimates are equivalent \cite{Charnes:1976, Dolby:1972} and can be found analytically. 

We now derive these estimates using the GLS procedure and show that this is equivalent to obtaining the maximum likelihood estimates.

\subsubsection{Ordinary least squares problem.} \label{OLS}
We first describe this in general terms, and discuss the application to NMA further below.  Assume we observe data $\{y_i, x_{i,j}\}$ on $n$ statistical units such that $i=1,\dots,n$ and $j=1,\dots,p$. In the context of an NMA these would be the relative treatment effects and the design matrix elements. The latter are treated as part of the `data' for the discussion in this section, as they are specific to individual instances of a real-world NMA. More generally, $\mathbf{X}$ may contain a set of observed covariates. 

The values of the response variable are collected in the vector $\boldsymbol{y}=(y_1, y_2, \dots, y_n)^\top$. The predictor variables are placed in the $n \times p$ design matrix $\mathbf{X}=(\boldsymbol{x}_1^\top, \boldsymbol{x}_2^\top, \dots, \boldsymbol{x}_n^\top)^\top$, where $\boldsymbol{x}_{i}=(x_{i,1}, x_{i,2}, \dots, x_{i,p})^\top$ and $\mathbf{X}$ is assumed to have full rank (in NMA this is true by construction). We consider the linear regression model
\begin{eqnarray}
\label{eq:OLS}
    \boldsymbol{y} = \mathbf{X} \boldsymbol{\beta} + \boldsymbol{\epsilon},
\end{eqnarray}
where $\boldsymbol{\beta}$ is a column vector of length $p$ containing the parameters that we wish to estimate. The error term is assumed to be normally distributed, $\boldsymbol{\epsilon} \sim \mathcal{N}(0,\mathbf{C}(\boldsymbol{\phi}))$, with an $n \times n$ covariance matrix $\mathbf{C}(\boldsymbol{\phi})$ that depends on a set of parameters $\boldsymbol{\phi}$. For convenience we will write $\mathbf{C}$ for $\mathbf{C}(\boldsymbol{\phi})$. The parameters $\boldsymbol{\beta}$ are the regression coefficients of the model while $\boldsymbol{\phi}$ represents the variance parameters. In the simplest case we assume uncorrelated errors and equal variances such that $\mathbf{C}$ is a multiple of the identity matrix. This assumption is known as the ordinary least squares (OLS) condition. 

The vector of residuals, $\boldsymbol{y}-\mathbf{X}\boldsymbol{\beta}$, represents the difference between the observed outputs, $\boldsymbol{y}$, and the mean values predicted by the model in Eq.~(\ref{eq:OLS}). The ordinary least squared estimates of the parameters are then obtained by minimising the sum of the squared residuals,
\begin{eqnarray}
    \hat{\boldsymbol{\beta}}^{\mathrm{OLS}}&=&\underset{\boldsymbol{\beta}}{\mathrm{argmin}} \sum_{i=1}^n [(\boldsymbol{y}-\mathbf{X}\boldsymbol{\beta})_i]^2  \nonumber \\
    &=&\underset{\boldsymbol{\beta}}{\mathrm{argmin}}  \left[(\boldsymbol{y}-\mathbf{X}\boldsymbol{\beta})^\top (\boldsymbol{y}-\mathbf{X}\boldsymbol{\beta})\right].
\end{eqnarray}

\subsubsection{Generalised least squares problem.} \label{GLS}
In generalised least squares regression we relax the ordinary least squares assumption. That is, we make no assumptions on the form of the covariance matrix $\mathbf{C}$. If errors are uncorrelated but do not necessarily have equal variances then $\mathbf{C}$ is a diagonal matrix. Regression under these conditions is a special case of GLS known as `weighted least squares'. If errors are correlated then $\mathbf{C}$ has non-zero off diagonal elements representing the covariance between error terms.

To find the generalised least squares estimator we carry out a transformation of the GLS model so that it fulfills the ordinary least squares condition. To this end we note that given that $\mathbf{C}$ is a covariance matrix, it must be symmetric and positive-definite. Therefore we can write $\mathbf{C}=\mathbf{K}^\top \mathbf{K}=\mathbf{K}\mathbf{K}$ where $\mathbf{K}$ is the (symmetric) square root of $\mathbf{C}$ \cite{Amemiya:1985}. We now multiply both sides of Eq.~(\ref{eq:OLS}) with $\mathbf{K}^{-1}$ from the left,
\begin{eqnarray}
    \mathbf{K}^{-1}\boldsymbol{y} = \mathbf{K}^{-1}\mathbf{X} \boldsymbol{\beta} + \mathbf{K}^{-1}\boldsymbol{\epsilon}.
\end{eqnarray}
Defining the variables
\begin{eqnarray}
    \tilde{\boldsymbol{y}}=\mathbf{K}^{-1}\boldsymbol{y}, 
    \hspace{20pt} \tilde{\mathbf{X}}=\mathbf{K}^{-1}\mathbf{X}, 
    \hspace{20pt} \tilde{\boldsymbol{\epsilon}}&=\mathbf{K}^{-1}\boldsymbol{\epsilon} \label{eq:tilde}
\end{eqnarray}
we obtain the model
\begin{eqnarray}
\label{eq:GLS-tilde}
    \tilde{\boldsymbol{y}} = \tilde{\mathbf{X}} \boldsymbol{\beta} + \tilde{\boldsymbol{\epsilon}}.
\end{eqnarray}
Now let us inspect the error term $\tilde{\boldsymbol{\epsilon}}$ of this model to see if it fulfills the OLS condition. The expected value of the error is $\mathbb{E}(\tilde{\boldsymbol{\epsilon}})=\mathbb{E}(\mathbf{K}^{-1}\boldsymbol{\epsilon})=\mathbf{K}^{-1}\mathbb{E}(\boldsymbol{\epsilon})=0$, as required. To obtain the covariance matrix, we use the relation
$\mathrm{Cov}(\mathbf{A}\boldsymbol{z}) = \mathbf{A}\mathrm{Cov}(\boldsymbol{z}) \mathbf{A}^\top$ which is valid for any random vector $\boldsymbol{z}$ and fixed matrix $\mathbf{A}$. Therefore, 
\begin{eqnarray}
    \mathrm{Cov}(\tilde{\boldsymbol{\epsilon}})&=\mathrm{Cov}(\mathbf{K}^{-1}\boldsymbol{\epsilon})=\mathbf{K}^{-1}\mathrm{Cov}(\boldsymbol{\epsilon})(\mathbf{K}^{-1})^\top \nonumber \\
    &=\mathbf{K}^{-1}\mathbf{C}\mathbf{K}^{-1} = \mathbf{I},
\end{eqnarray}
where we have used $\mathrm{Cov}(\boldsymbol{\epsilon})=\mathbf{C}$, $\mathbf{K}^{-1}$ is symmetric, and $\mathbf{C}=\mathbf{K}\mathbf{K}$. The errors $\tilde{\boldsymbol{\epsilon}}$ therefore fulfil the OLS condition, and  Eq.~(\ref{eq:GLS-tilde}) hence defines an ordinary least squares problem. 

We now obtain the generalised least squares estimator by using the OLS estimator with Eq.~(\ref{eq:GLS-tilde}), that is,
\begin{eqnarray}
    \hat{\boldsymbol{\beta}}^{\mathrm{GLS}} = \underset{\boldsymbol{\beta}}{\mathrm{argmin}} \left[(\tilde{\boldsymbol{y}}-\tilde{\mathbf{X}}\boldsymbol{\beta})^\top (\tilde{\boldsymbol{y}}-\tilde{\mathbf{X}}\boldsymbol{\beta})\right].
\end{eqnarray}
Using the definitions in Eq.~(\ref{eq:tilde}) we find
\begin{eqnarray}
    (\tilde{\boldsymbol{y}}-\tilde{\mathbf{X}}\boldsymbol{\beta})^\top (\tilde{\boldsymbol{y}}-\tilde{\mathbf{X}}\boldsymbol{\beta}) &= (\mathbf{K}^{-1}(\boldsymbol{y}-\mathbf{X}\boldsymbol{\beta}))^\top (\mathbf{K}^{-1}(\boldsymbol{y}-\mathbf{X}\boldsymbol{\beta})) \nonumber\\
    &=(\boldsymbol{y}-\mathbf{X}\boldsymbol{\beta})^\top (\mathbf{K}^{-1})^\top \mathbf{K}^{-1}(\boldsymbol{y}-\mathbf{X}\boldsymbol{\beta})\nonumber \\
    &=(\boldsymbol{y}-\mathbf{X}\boldsymbol{\beta})^\top \mathbf{C}^{-1}(\boldsymbol{y}-\mathbf{X}\boldsymbol{\beta}).
\end{eqnarray}
The GLS estimator is therefore
\begin{eqnarray}
\label{eq:GLS-eq}
    \hat{\boldsymbol{\beta}}^{\mathrm{GLS}} &=\underset{\boldsymbol{\beta}}{\mathrm{argmin}} \left[(\boldsymbol{y}-\mathbf{X}\boldsymbol{\beta})^\top \mathbf{C}^{-1}(\boldsymbol{y}-\mathbf{X}\boldsymbol{\beta})\right].
\end{eqnarray}
This can be solved analytically as we will see in Sec.~\ref{GLS_ML_Solution}. Before we return to this we derive the maximum likelihood estimator and show that this leads to the same condition as in Eq.~(\ref{eq:GLS-eq}).

\subsubsection{Maximum likelihood approach.}
\label{ML}
The linear model in Eq.~(\ref{eq:OLS}) with normally distributed errors $\boldsymbol{\epsilon}\sim \mathcal{N}(0,\mathbf{C})$ can be written equivalently as
\begin{eqnarray}
    \boldsymbol{y}\sim \mathcal{N}(\mathbf{X}\boldsymbol{\beta}, \mathbf{C}),
\end{eqnarray}
where we place no assumptions on the covariance matrix $\mathbf{C}$ except that it depends on a set of variance parameters $\boldsymbol{\phi}$. The likelihood of this model is then simply the multivariate normal distribution with mean vector $\mathbf{X}\boldsymbol{\beta}$ and covariance matrix $\mathbf{C}$, 
\begin{eqnarray} \label{eq:Like_y}
    \hspace{-4em}L(\boldsymbol{\beta}, \boldsymbol{\phi}|\boldsymbol{y},\mathbf{X}) = \frac{1}{(2\pi)^{n/2}(\det \mathbf{C})^{1/2}}\exp \left(-\frac{1}{2} (\boldsymbol{y}-\mathbf{X}\boldsymbol{\beta})^\top \mathbf{C}^{-1}(\boldsymbol{y}-\mathbf{X}\boldsymbol{\beta})\right),
\end{eqnarray}
and the log likelihood is
\begin{eqnarray}
\label{eq:LogL}
    \hspace{-20pt}\ln L(\boldsymbol{\beta}, \boldsymbol{\phi}|\boldsymbol{y},\mathbf{X}) = -\frac{1}{2}\ln \det \mathbf{C} -\frac{1}{2} (\boldsymbol{y}-\mathbf{X}\boldsymbol{\beta})^\top \mathbf{C}^{-1}(\boldsymbol{y}-\mathbf{X}\boldsymbol{\beta}) + \mathrm{const}.
\end{eqnarray}
Treating the variance parameters as known, we infer the regression coefficients $\boldsymbol{\beta}$ by maximising the (log) likelihood with respect to $\boldsymbol{\beta}$. This is equivalent to minimising the term $(\boldsymbol{y}-\mathbf{X}\boldsymbol{\beta})^\top \mathbf{C}^{-1}(\boldsymbol{y}-\mathbf{X}\boldsymbol{\beta})$, 
\begin{eqnarray}
\label{eq:ML-eq}
    \hat{\boldsymbol{\beta}}^{\mathrm{ML}} = \underset{\boldsymbol{\beta}}{\mathrm{argmin}} \left[ (\boldsymbol{y}-\mathbf{X}\boldsymbol{\beta})^\top \mathbf{C}^{-1}(\boldsymbol{y}-\mathbf{X}\boldsymbol{\beta}) \right],
\end{eqnarray}
which is identical to the generalised least squares problem in Eq.~(\ref{eq:GLS-eq}).

\subsubsection{Solution to the GLS and ML problem (The Aitken estimator).}
\label{GLS_ML_Solution}
Now we proceed solve Eq. (\ref{eq:GLS-eq}) [= Eq. (\ref{eq:ML-eq})] to find the estimator $\hat{\boldsymbol{\beta}} = \hat{\boldsymbol{\beta}}^{\mathrm{GLS}} = \hat{\boldsymbol{\beta}}^{\mathrm{ML}}$. We first multiply out the product $(\boldsymbol{y}-\mathbf{X}\boldsymbol{\beta})^\top \mathbf{C}^{-1}(\boldsymbol{y}-\mathbf{X}\boldsymbol{\beta})$, and then set the partial derivative with respect to $\boldsymbol{\beta}$ equal to zero. This leads to 
\begin{eqnarray}
\label{eq:diff-beta}
    \frac{\partial}{\partial \boldsymbol{\beta}}\left( \boldsymbol{y}^\top\mathbf{C}^{-1}\boldsymbol{y}-\boldsymbol{y}^\top\mathbf{C}^{-1}\mathbf{X}\boldsymbol{\beta} - \boldsymbol{\beta}^\top\mathbf{X}^\top\mathbf{C}^{-1}\boldsymbol{y}+\boldsymbol{\beta}^\top\mathbf{X}^\top\mathbf{C}^{-1}\mathbf{X}\boldsymbol{\beta}\right) = 0.
\end{eqnarray}
We address this term by term. The first term $\boldsymbol{y}^\top\mathbf{C}^{-1}\boldsymbol{y}$ is independent of $\boldsymbol{\beta}$ and therefore differentiates to zero. The second and third terms take the forms $\boldsymbol{a}^\top \boldsymbol{\beta}$ and $\boldsymbol{\beta}^\top\boldsymbol{a}$ respectively, where $\boldsymbol{a}=-\mathbf{X}^\top\mathbf{C}^{-1}\boldsymbol{y}$ is a column vector of length $p$. We have  $\boldsymbol{a}^\top \boldsymbol{\beta}=\boldsymbol{\beta}^\top\boldsymbol{a}$, and  the second and third terms of Eq.~(\ref{eq:diff-beta}) each evaluate to $-\mathbf{X}^\top\mathbf{C}^{-1}\boldsymbol{y}$. The last term on the right-hand side of  Eq. (\ref{eq:diff-beta}) is quadratic in $\boldsymbol{\beta}$. We also note that the matrix $\mathbf{X}^\top \mathbf{C}^{-1} \mathbf{X}$ is symmetric. Therefore the final term in Eq.~(\ref{eq:diff-beta}) evaluates to $2 \mathbf{X}^\top \mathbf{C}^{-1} \mathbf{X} \boldsymbol{\beta}$.

Combining these results, Eq.~(\ref{eq:diff-beta}) reduces to
\begin{eqnarray}
\label{eq:diff-beta-result}
    -2 \mathbf{X}^\top\mathbf{C}^{-1}\boldsymbol{y} + 2\mathbf{X}^\top \mathbf{C}^{-1} \mathbf{X} \boldsymbol{\beta} = 0.
\end{eqnarray}
Solving for $\boldsymbol{\beta}$ yields the GLS and ML estimator of the vector of regression coefficients,
\begin{eqnarray}
\label{eq:Aitken}
    \hat{\boldsymbol{\beta}} = \hat{\boldsymbol{\beta}}^{\mathrm{GLS}} = \hat{\boldsymbol{\beta}}^{\mathrm{ML}} = \left(\mathbf{X}^\top \mathbf{C}^{-1} \mathbf{X} \right)^{-1} \mathbf{X}^\top\mathbf{C}^{-1}\boldsymbol{y},
\end{eqnarray}
also known as the `Aitken estimator' \cite{Aitken:1936}. 

Recalling that $\mathbb{E}(\boldsymbol{y})=\mathbf{X}\boldsymbol{\beta}$ [see Eq.~({\ref{eq:OLS})] the expectation of this estimate is 
\begin{eqnarray}
\label{eq:E_beta_hat}
    \mathbb{E}( \hat{\boldsymbol{\beta}})
    &=\left(\mathbf{X}^\top \mathbf{C}^{-1} \mathbf{X} \right)^{-1} \mathbf{X}^\top\mathbf{C}^{-1}\mathbf{X}\boldsymbol{\beta} = \boldsymbol{\beta},
\end{eqnarray}
indicating that the Aitken estimator is an unbiased estimate of $\boldsymbol{\beta}$.

To find the ($p\times p$) covariance matrix of this estimate we once again make use of the result $\mathrm{Cov}(\mathbf{A}\boldsymbol{z}) = \mathbf{A}\mathrm{Cov}(\boldsymbol{z}) \mathbf{A}^\top$ and find 
\begin{eqnarray}\label{eq:varbeta}
    \hspace{-20pt}\mathrm{Cov}(\hat{\boldsymbol{\beta}})  &= \left[\left(\mathbf{X}^\top \mathbf{C}^{-1} \mathbf{X} \right)^{-1} \mathbf{X}^\top\mathbf{C}^{-1}\right] \mathrm{Cov}(\boldsymbol{y}) \left[\left(\mathbf{X}^\top \mathbf{C}^{-1} \mathbf{X} \right)^{-1} \mathbf{X}^\top\mathbf{C}^{-1}\right]^\top \nonumber\\
    &=\left(\mathbf{X}^\top \mathbf{C}^{-1} \mathbf{X} \right)^{-1},
\end{eqnarray}
where we have used $\mathrm{Cov}(\boldsymbol{y})=\mathbf{C}$ and the fact that the matrices $\mathbf{C}$ and $\mathbf{X}^\top \mathbf{C}^{-1} \mathbf{X}$ are symmetric. 

\subsection{Frequentist inference for NMA}
\label{freq_NMA}
In Sec.~\ref{sec:freq_gen} we derived the GLS/ML estimator of the regression coefficients for a linear regression model [Eq.~(\ref{eq:Aitken})]. We now use this result to estimate the relative treatment effects $\boldsymbol{d}$ in our NMA model from Sec.~\ref{sec:freq_intro} [Eq.~(\ref{eq:RE-regress})]. Following this, we discuss frequentist methods of estimating the heterogeneity variance $\tau^2$.  }

\subsubsection{Estimating the mean relative treatment effects.} \label{Freq_d_est} We start from the linear regression model for a RE network meta-analysis,
\begin{equation}
    \boldsymbol{y} = \mathbf{X}\boldsymbol{d} + \boldsymbol{\eta} + \boldsymbol{\epsilon}, \hspace{10pt} \boldsymbol{\eta} \sim \mathcal{N}(0,\boldsymbol{\Sigma}), \hspace{10pt} \boldsymbol{\epsilon} \sim \mathcal{N}(0,\mathbf{V}).
\end{equation}
The within-study covariance matrix $\mathbf{V}$ (describing the statistics of sampling noise) is assumed to be known, whereas the between-study covariance matrix  $\boldsymbol{\Sigma}$ depends on the unknown heterogeneity variance $\tau^2$.

Assuming an estimate of the heterogeneity variance $\hat{\tau}^2$ (and therefore the covariance matrix $\hat{\boldsymbol{\Sigma}}$), we  find the mean relative treatment effects $\boldsymbol{d}$ via the Aitken estimator in Eq.~(\ref{eq:Aitken}),
\begin{equation}
    \boldsymbol{\hat{d}}^{\mathrm{RE}} = (\mathbf{X}^{\top} (\mathbf{V}+\boldsymbol{\hat\Sigma})^{-1} \mathbf{X})^{-1} \mathbf{X}^{\top} (\mathbf{V}+\boldsymbol{\hat\Sigma})^{-1} \boldsymbol{y},
\end{equation}
where we have labelled this explicitly as a random effects (RE) estimate, since the estimator depends on the heterogeneity $\tau^2$ (or an estimate $\hat\tau^2$). It is useful to define the inverse-variance weight matrix $\mathbf{W} = (\mathbf{V}+\boldsymbol{\hat{\Sigma}})^{-1}$. We can then write
\begin{equation}
\label{eq:d-freq}
    \boldsymbol{\hat{d}}^{\mathrm{RE}} = (\mathbf{X}^{\top} \mathbf{W} \mathbf{X})^{-1} \mathbf{X}^{\top} \mathbf{W} \boldsymbol{y}.
\end{equation}
Using Eq.~(\ref{eq:varbeta}) the covariance matrix associated with this estimator is given by 
\begin{equation}\label{eq:freq_var}
    \mathrm{Cov}(\boldsymbol{\hat{d}}^{\mathrm{RE}}) = (\mathbf{X}^{\top} \mathbf{W} \mathbf{X})^{-1}.
\end{equation}
To obtain the estimator for a fixed effects (FE) model, $\boldsymbol{\hat{d}}^{\mathrm{FE}}$, we simply set $\tau^2=0$. We then have  $\boldsymbol{\Sigma}=0$ and hence $\mathbf{W}=\mathbf{V}^{-1}$. 
\medskip

{\em Special case: Pairwise meta-analysis.} In a pairwise meta-analysis of $N=2$ treatments, each trial provides an estimate $y_i$ of the same relative treatment effect (we write $d$ for its true value). The design matrix $\mathbf{X}$ is then an $M\times 1$ matrix, and all entries are equal to one. The covariance matrix $\mathbf{C}$ is an $M \times M$ diagonal matrix with elements equal to $\sigma_i^2+\tau^2$ and the within-study variances $\sigma_i^2$ are assumed known. The RE model is then $y_i \sim \mathcal{N}(d, \sigma_i^2 + \tau^2)$ for $i=1,\dots,M$. The weight matrix is also diagonal and, for a given estimate of heterogeneity $\hat{\tau}^2$, its elements are equal to $w_i = (\sigma_i^2+\hat{\tau}^2)^{-1}$. We then find that the Aitken estimator of the relative treatment effect reduces to
\begin{eqnarray}
\label{eq:dhat_pair}
    \hat{d}^{\mathrm{RE}} = \frac{\sum_{i=1}^{M} w_i y_i}{\sum_{i=1}^{M} w_i} = \frac{\sum_{i=1}^{M} (\sigma_i^2+\hat{\tau}^2)^{-1} y_i}{\sum_{i=1}^{M} (\sigma_i^2+\hat{\tau}^2)^{-1}}.
\end{eqnarray}
The variance of this estimate is
\begin{eqnarray}
\label{eq:var_dhat_pair}
    \mathrm{Var}(\hat{d}^{\mathrm{RE}}) =\frac{1}{\sum_{i=1}^{M} w_i} = \frac{1}{\sum_{i=1}^{M} (\sigma_i^2+\hat{\tau}^2)^{-1}}.
\end{eqnarray}
Therefore, for pairwise meta-analysis, the GLS and ML approaches recover the results for a simple weighted mean of the sample. Again, for a fixed effect model, $\hat{d}^{\mathrm{FE}}$ is obtained by setting $\tau^2=0$, that is, $w_i=\sigma_i^{-2}$.

\subsubsection{Estimating the heterogeneity variance.}
\label{freq_tau}
So far, we have assumed knowledge of the heterogeneity variance $\tau^2$. We now discuss how this is estimated. There are numerous methods for obtaining a frequentist estimate of $\tau^2$ in pairwise meta-analysis \cite{DerSimonian:1986, DerSimonian:2007, Hartung:2003, Sidik:2005, Rukhin:2013, Paule:1982}, and there is much debate over which method is most appropriate \cite{Langan:2019, Veroniki:2016b, Petropoulou:2017b}. Some of these methods have also been extended to network meta-analysis \cite{Jackson:2017,Jackson:2010, Law:2016, Jackson:2016, Jackson:2012}. 

The most widely used methods fall into two categories: (i) the so-called method of moments \cite{Kacker:2004}, and (ii) what is known as restricted maximum likelihood (REML) approaches \cite{Patterson:1971}. The former involves defining a measure of heterogeneity based on the sum of squared residuals. The latter involves modifying the likelihood function of the random effects model to remove dependence on the relative treatment effects and then maximising this modified likelihood with respect to $\tau^2$. We describe both approaches in turn.

\medskip

\textit{Method of moments.} Multiple heterogeneity estimators in pairwise meta-analysis are based on the so-called `method-of-moments' \cite{Kacker:2004, DerSimonian:2007}. Inference of the regression coefficients in pairwise meta-analysis involves a weighted mean of the sample [Eq.~ (\ref{eq:dhat_pair})]. For general weights $a_i$ ($i=1,\dots,M$) associated with observations $y_i$ we define the weighted mean,
\begin{eqnarray}
\label{eq:yhat}
    \hat{y} = \frac{\sum_{i=1}^{M} a_i y_i }{\sum_{i=1}^{M} a_i }.
\end{eqnarray}
If we set $a_i = (\sigma_i^2+\tau^2)^{-1}$ we recover the random effects estimate $\hat{d}^{\mathrm{RE}}$ in Eq.~(\ref{eq:dhat_pair}). Other choices of the weights will be discussed below.

We can then define a generalised version of the so-called `Q statistic' \cite{DerSimonian:1986, DerSimonian:2007, Langan:2019} as the weighted sum of squared residuals,
\begin{eqnarray}
\label{eq:Q}
    Q = \sum_{i=1}^{M} a_i (y_i-\hat{y})^2.
\end{eqnarray}
The estimate of $\tau^2$ is obtained by assuming that the empirical value of $Q$ obtained via Eq.~(\ref{eq:Q}) from the observed data is equal to its expectation under the random effects model \cite{Kacker:2004}. That is, 
\begin{eqnarray}
\label{eq:EQ}
    \sum_{i=1}^{M} a_i (y_i-\hat{y})^2=\mathbb{E}_{\mathrm{RE}}(Q) \equiv \mathbb{E}_{\mathrm{RE}}\left(\sum_{i=1}^{M} a_i (y_i-\hat{y})^2\right). 
\end{eqnarray}
The expectation on the right is calculated assuming that observations follow the RE model, $y_i \sim \mathcal{N}(d, \sigma_i^2 + \tau^2)$. In Appendix~\ref{App:EQ-Pair} we show that
\begin{eqnarray} \label{eq:EQ_pair}
    \mathbb{E}_{\mathrm{RE}}(Q) = \tau^2 \left(\sum_{i=1}^{M} a_i -  \frac{\sum_{i=1}^{M} a_i^2 }{\sum_{i=1}^{M} a_i }\right) + \left(\sum_{i=1}^{M} a_i \sigma_i^2  - \frac{\sum_{i=1}^{M} a_i^2  \sigma_i^2}{\sum_{i=1}^{M} a_i }\right).
\end{eqnarray}
Using this in Eq.~(\ref{eq:EQ}) and re-arranging for $\tau^2$ we find
\begin{eqnarray}
\label{eq:MoM}
    \hat\tau^2 = \frac{\sum_{i=1}^M a_i\left(y_i - \frac{\sum_j a_j y_j}{\sum_j a_j} \right)^2 - \left(\sum_{i=1}^{M} a_i \sigma_i^2  - \frac{\sum_{i=1}^{M} a_i^2  \sigma_i^2}{\sum_{i=1}^{M} a_i }\right) }{\sum_{i=1}^{M} a_i -  \frac{\sum_{i=1}^{M} a_i^2 }{\sum_{i=1}^{M} a_i }},
\end{eqnarray}
where we have used the definition of $\hat{y}$ in Eq.~(\ref{eq:yhat}). This is the general method-of-moments estimator for $\tau^2$ in pairwise meta-analysis. In practice, the expression on the right-hand side of Eq.~(\ref{eq:MoM}) can come out negative. In this case one sets $\hat \tau^2=0$.  

Different choices can be made for the weights $a_i$ in Eqs.~(\ref{eq:yhat}) and (\ref{eq:Q}). For example, the widely used DerSimonian and Laird (DL) estimator \cite{DerSimonian:1986} uses the fixed effect weights, $a_i=\sigma_i^{-2}$,  so that $\hat{y}=\hat{d}^{\mathrm{FE}}$. Cochran's ANOVA (CA) estimator \cite{Cochran:1954} uses equal weights $a_i=1/M$ while the Paule Mandel (PM) estimator \cite{Paule:1982} uses the random effects weights $a_i=(\sigma_i^2+\tau^2)^{-1}$. The DL and CA estimators lead to a closed form solution for the estimate of $\tau^2$ [in the sense that the right-hand side of Eq.~(\ref{eq:MoM}) becomes independent of $\tau^2$]. This is not the case for the PM estimator since these weights depend on $\tau^2$. Eq.~(\ref{eq:MoM}) must then be solved numerically.

\medskip

Extending the DL estimator to the case of network meta-analysis is straightforward \cite{Jackson:2016, Jackson:2010, Jackson:2012}. We generalise Eq.~(\ref{eq:Q}) using the inverse-variance weight matrix and obtain
\begin{eqnarray}
\label{eq:Q-NMA}
    Q = (\boldsymbol{y}-\hat{\boldsymbol{y}})^\top \mathbf{V}^{-1} (\boldsymbol{y}-\hat{\boldsymbol{y}}). 
\end{eqnarray}

We recall that $\mathbf{V}^{-1}$ represents the observed within-study variances and correlations so the expression in Eq.~(\ref{eq:Q-NMA}) is analogous to using $a_i=\sigma_i^{-2}$ in the pairwise case. The vector $\hat{\boldsymbol{y}}$ is the set of network estimates of $\boldsymbol{y}$ obtained using the fixed effects weights $\mathbf{V}^{-1}$. That is
\begin{eqnarray}
\label{eq:yhat-NMA}
    \hat{\boldsymbol{y}} = \mathbf{X} \boldsymbol{\hat{d}}^{\mathrm{FE}} = \mathbf{X}(\mathbf{X}^{\top} \mathbf{V}^{-1} \mathbf{X})^{-1} \mathbf{X}^{\top} \mathbf{V}^{-1} \boldsymbol{y},
\end{eqnarray}
where $\boldsymbol{\hat{d}}^{\mathrm{FE}}$ are the estimates of the mean relative treatment effects under the fixed effect model, obtained by setting $\mathbf{W}=\mathbf{V}^{-1}$ (i.e., $\boldsymbol{\Sigma}=0$) in Eq.~(\ref{eq:d-freq}). Eqs.~(\ref{eq:Q-NMA}) and (\ref{eq:yhat-NMA}) are the NMA analogue of Eqs.~(\ref{eq:Q}) and (\ref{eq:yhat}) in pairwise MA (when $a_i=\sigma_i^{-2}$).

In Appendix~\ref{App:EQ-Net} we evaluate the expectation of $Q$ in Eq.~(\ref{eq:Q-NMA}) and find
\begin{eqnarray}\label{eq:expected-Q-NMA}
    \mathbb{E}_{\mathrm{RE}}(Q) =\sum_{i=1}^{M}(m_i-1) - (N-1)  + \tau^2 \mathrm{tr}(\mathbf{A}\mathbf{P}),
\end{eqnarray}
where, following Jackson et al (2016) \cite{Jackson:2016}, we have defined the matrix
\begin{eqnarray}
    \mathbf{A} = \mathbf{V}^{-1} - \mathbf{V}^{-1}\mathbf{X}(\mathbf{X}^{\top} \mathbf{V}^{-1} \mathbf{X})^{-1} \mathbf{X}^{\top} \mathbf{V}^{-1}.
\end{eqnarray}
We have also defined the block diagonal matrix $\mathbf{P}$ such that $\boldsymbol{\Sigma}=\tau^2 \mathbf{P}$. Each $(m_i-1) \times (m_i-1)$ block in $\mathbf{P}$ represents a trial $i$, and has diagonal elements equal to 1 and off-diagonal elements equal to $1/2$. All other elements of $\mathbf{P}$ are zero.

Similar to the pairwise case we equate the expectation in Eq.~(\ref{eq:expected-Q-NMA}) with the empirically observed value of $Q$ in Eq.~(\ref{eq:Q-NMA}). Re-arranging for $\tau^2$ we find
\begin{eqnarray}
    \hat{\tau}^2 = \frac{(\boldsymbol{y}-\hat{\boldsymbol{y}})^\top \mathbf{V}^{-1} (\boldsymbol{y}-\hat{\boldsymbol{y}}) - [\sum_{i=1}^{M}(m_i-1) - (N-1)] }{\mathrm{tr}(\mathbf{A}\mathbf{P})}.
\end{eqnarray}
This is the DerSimonian and Laird estimator of $\tau^2$ in network meta-analysis. Again, if $\hat{\tau}^2$ comes out negative we set its value equal to zero. \\

\textit{Restricted maximum likelihood. } \label{REML} We now explain the restricted maximum likelihood method for estimating the variance parameter. We start by discussing this method in a more general sense [for a linear regression problem] and then relate this to the NMA model. 

In Sec.~\ref{ML} we showed that we can obtain estimates for the regression coefficients $\boldsymbol{\beta}$ in a linear model of the form $\boldsymbol{y} = \mathbf{X} \boldsymbol{\beta} + \boldsymbol{\epsilon}$ with $ \boldsymbol{\epsilon} \sim \mathcal{N}(0,\mathbf{C})$ by maximising the log likelihood in Eq.~(\ref{eq:LogL}) with respect to these parameters. This led to the Aitken estimator in Eq.~(\ref{eq:Aitken}). 

The likelihood of this model is also a function of the variance parameters $\boldsymbol{\phi}$ characterising the covariance matrix $\mathbf{C}$. The Aitken estimator assumes that these are known, which is generally not the case. One way of obtaining the variance parameters is to maximise the log likelihood simultaneously with respect to $\boldsymbol{\beta}$ and $\boldsymbol{\phi}$.

A problem with this approach is that the resulting estimates of the variance parameters are generally biased \cite{Patterson:1971}. This is because the variance estimate fails to account for the loss in degrees of freedom that results from estimating $\boldsymbol{\beta}$ \cite{Harville:1977}. This is well known, and can be demonstrated easily for a  one dimensional normal distribution, see Appendix~\ref{App:Bias_ML_Var}.

The restricted maximum likelihood (REML) approach was proposed by Patterson and Thompson  \cite{Patterson:1971} as a method to overcome this problem. The principal idea is to carry out a linear transformation of the variables $\boldsymbol{y}$ so that the likelihood function for the transformed variables no longer depends on the parameters $\boldsymbol{\beta}$, but only on their estimates $\boldsymbol{\hat\beta}$ (which in turn depend on $\boldsymbol{\phi}$). This `restricted likelihood' is then maximised with respect to the variance parameters. 

The restricted likelihood is given by
\begin{eqnarray}
\label{eq:rl_result}
    \hspace{-70pt}RL(\boldsymbol{\phi}|\boldsymbol{y}, \mathbf{X}) &\propto  (\det \mathbf{C})^{-1/2} (\det \mathbf{X}^\top \mathbf{C}^{-1} \mathbf{X})^{-1/2}  \exp \left( -\frac{1}{2} (\boldsymbol{y}-\mathbf{X}\boldsymbol{\hat\beta})^\top \mathbf{C}^{-1}(\boldsymbol{y}-\mathbf{X}\boldsymbol{\hat\beta}) \right).
\end{eqnarray}
We can arrive at this expression in several ways. One possible method involves evaluating the marginal likelihood of the transformed variable [see \cite{Wakefield:2009} for details]. Alternatively, one can use a Bayesian interpretation \cite{Harville:1974}. Here, one assumes that nothing is known about $\boldsymbol{\beta}$, assigns an improper flat prior (a prior that is not properly normalised), and integrates out $\boldsymbol{\beta}$. This directly leads to  \begin{eqnarray}\label{eq:rl}
   \hspace{-3em} RL(\boldsymbol{\phi}|\boldsymbol{y}, \mathbf{X})
    &\propto \int_{-\infty}^{\infty} \frac{1}{(\det~\mathbf{C})^{1/2}}\exp \left(-\frac{1}{2} (\boldsymbol{y}-\mathbf{X}\boldsymbol{\beta})^\top \mathbf{C}^{-1}(\boldsymbol{y}-\mathbf{X}\boldsymbol{\beta})\right)  \d \boldsymbol{\beta}.
\end{eqnarray}
The Gaussian integral on the right-hand side of Eq.~(\ref{eq:rl}) can then be evaluated exactly to give the result in Eq.~(\ref{eq:rl_result}). In this context, we note that, for a given matrix $\mathbf{C}$, the integrand in Eq.~(\ref{eq:rl}) is maximal at
\be\label{eq:beta_hat}
    \hat{\boldsymbol{\beta}} = \left(\mathbf{X}^\top \mathbf{C}^{-1} \mathbf{X} \right)^{-1} \mathbf{X}^\top\mathbf{C}^{-1}\boldsymbol{y}
\ee
by construction [see Eq.~(\ref{eq:Aitken})].  
 
From Eq.~(\ref{eq:rl_result}), the restricted log likelihood is \cite{Harville:1974, Harville:1977}
\begin{eqnarray}
\label{eq:logRL}
    \ln~RL(\boldsymbol{\phi}|\boldsymbol{y}, \mathbf{X})&=&  -\frac{1}{2}  \ln( \det \mathbf{C}) -\frac{1}{2} (\boldsymbol{y}-\mathbf{X}\boldsymbol{\hat\beta})^\top \mathbf{C}^{-1}(\boldsymbol{y}-\mathbf{X}\boldsymbol{\hat\beta}) \nonumber \\ 
    &&- \frac{1}{2}  \ln (\det \mathbf{X}^\top \mathbf{C}^{-1} \mathbf{X}) + \mathrm{const}.
\end{eqnarray}
This is similar to the original log likelihood function in Eq.~(\ref{eq:LogL}) but with dependence on the maximum likelihood estimator $\boldsymbol{\hat\beta}$ instead of the true parameter $\boldsymbol{\beta}$ and with an additional term. 

A possible iterative procedure to obtain estimates of the regression and variance parameters is now as follows: Start with an initial choice for $\boldsymbol{\hat \phi}$. This defines $\mathbf{C}$. Use this in Eq.~(\ref{eq:beta_hat}) to obtain $\boldsymbol{\hat\beta}$. Use this value for $\boldsymbol{\hat\beta}$ in Eq.~(\ref{eq:logRL}) and then maximise $ \ln~RL(\boldsymbol{\phi}|\boldsymbol{y}, \mathbf{X})$ with respect to $\boldsymbol{\phi}$ (keeping $\boldsymbol{\hat\beta}$ constant). This delivers an updated value for $\boldsymbol{\hat \phi}$. Then repeat, and iterate until convergence.  

In this process, the maximisation of the expression in Eq.~(\ref{eq:logRL}) with respect to $\boldsymbol{\phi}$ requires numerical techniques such as the Newton-Raphson method \cite{Press:1992}, Fisher's scoring algorithm \cite{Longford:1987}, or the expectation-maximisation (EM) algorithm \cite{Dempster:1977}.

\medskip

In NMA the parameter estimates $\boldsymbol{\hat\beta}$ are the estimates of the relative treatment effects $\boldsymbol{\hat d}^{\mathrm{RE}}$ in Eq.~(\ref{eq:d-freq}). The covariance matrix $\mathbf{C}$ is given by   $\mathbf{V}+\boldsymbol{\Sigma}$ where the within-study covariance matrix $\mathbf{V}$ is assumed known, and the between-study covariance matrix depends on the unknown parameter $\tau^2$ we wish to estimate. 

In pairwise meta-analysis  $\boldsymbol{\hat\beta}$ is the  estimate of the treatment effect $\hat{d}^{\mathrm{RE}}$ in Eq.~(\ref{eq:dhat_pair}),  $\mathbf{X}$ is an $M\times1$ matrix of ones, and the covariance matrix $\mathbf{C}$ is diagonal with elements $\sigma_i^2+\tau^2$ (with $\sigma_i^2$ assumed known). The restricted log likelihood in Eq.~(\ref{eq:logRL}) then simplifies to \cite{Viechtbauer:2005, Veroniki:2016b}
\begin{eqnarray}
\label{eq:logRL_pair}
     \ln ( RL(\tau^2|\boldsymbol{y})) &=& -\frac{1}{2} \sum_{i=1}^{M} \ln(\sigma_i^2+\tau^2) - \frac{1}{2} \sum_{i=1}^M \frac{(y_i-\hat{d}^{\mathrm{RE}})^2}{\sigma_i^2 + \tau^2} \nonumber \\ 
     && - \frac{1}{2} \ln \left( \sum_{i=1}^M \frac{1}{\sigma_i^2 + \tau^2} \right) + \mathrm{const},
\end{eqnarray}
where we recall that $\hat{d}^{\mathrm{RE}}$ depends on the \textit{estimate} $\hat{\tau}^2$ [Eq.~ (\ref{eq:dhat_pair})]. Following the above procedure we now maximise $\ln(RL(\tau^2|\boldsymbol{y}))$ with respect to $\tau^2$, while keeping $\hat{d}^{\mathrm{RE}}$ fixed. This can be done analytically. Setting the partial derivative of the log likelihood with respect to $\tau^2$ equal to zero yields the REML estimator,
\begin{eqnarray}
\label{eq:tau_REML_pair}
\hat{\tau}^2_{\mathrm{REML}} &=& \max\left\{0, \frac{\sum_{i=1}^{M}  (\sigma_i^2 + \hat{\tau}^2_{\mathrm{REML}})^{-2} \left((y_i - \hat{d}^{\mathrm{RE}})^2 - \sigma_i^2 \right) }{\sum_{i=1}^{M} (\sigma_i^2 + \hat{\tau}^2_{\mathrm{REML}})^{-2}} \right. \nonumber\\ && \hspace{170pt} \left.
+ \frac{1}{\sum_{i=1}^{M} (\sigma_i^2 + \hat{\tau}^2_{\mathrm{REML}})^{-1}} \right\},
\end{eqnarray}
where the truncation at zero ensures that $\hat{\tau}^2_{\mathrm{REML}}$ remains non-negative. The joint system of Eqs.~(\ref{eq:tau_REML_pair}) and (\ref{eq:dhat_pair}) can then be solved iteratively for $\hat{\tau}^2_{\mathrm{REML}}$ and $\hat{d}^{\mathrm{RE}}$.

\section{Reporting NMA Results} \label{report}
In Secs.~\ref{sec:Bayes} and \ref{sec:freq} we have explained how to obtain estimates for the model parameters in NMA using Bayesian and frequentist methods. We now explain how these estimates are reported and summarised for use in decision making. 

\subsection{Confidence/credible intervals in frequentist and Bayesian inference}
We focus on a particular parameter $x$. What is reported at the end of the inference process is an estimate for the parameter along with a measure of precision. In frequentist inference the parameter estimate itself is the one discussed in Sec.~\ref{Freq_d_est}, and an estimate of the variance is obtained from Eq.~(\ref{eq:freq_var}). In Bayesian inference the parameter estimate is usually the mean or median of the samples from the posterior distribution, and we also record the variance of the samples for the parameter \cite{TSD2}.

In a frequentist setting uncertainty on a parameter is often expressed in terms of confidence intervals, for example a `$95\%$ confidence interval'. In Bayesian inference uncertainty is expressed in terms of `credible intervals'. We note the subtle difference between these two concepts. The Bayesian interpretation is intuitive: given the observed data, there is a $95\%$ probability that the true (unknown) parameter lies within this interval \cite{Hespanhol:2019}. In a frequentist setting this would mean that if we were to repeat the experiment and inference many times (each time constructing a $95\%$-confidence interval) then $95\%$ of these intervals would contain the true value of the parameter \cite{Hespanhol:2019}.

The $\zeta$\% confidence interval ($0\leq\zeta\leq 100$) is constructed from the parameter estimate and its variance assuming a Gaussian distribution for the parameter with mean $\hat x$ and variance $\mathrm{Var}(\hat x)$. More precisely,
\begin{eqnarray}
\label{eq:CI}
    \mathrm{CI} = \hat{x} \pm q(\zeta) \sqrt{\mathrm{Var}(\hat x)},
\end{eqnarray}
where $q(\zeta)$ is such that a total probability of $\zeta\%$ of the Gaussian distribution is in the interval of length $2q(\zeta)\sqrt{\mathrm{Var}(\hat x)}$ around the mean. (Scaling out the variance, this means $\int_{-q}^q dx~ e^{-x^2/2}/\sqrt{2\pi}=\zeta/100$.) For example, using $q=1.96$ in Eq.~(\ref{eq:CI}) indicates a $95\%$  confidence interval. In a Bayesian setting the 95\% credible interval can be obtained in a similar way from Eq.~(\ref{eq:CI}), but using the mean and variance of samples drawn from the posterior. Alternatively, one can calculate the 2.5\% and 97.5\% quantiles of the posterior samples.

\subsection{Forest Plots}
\begin{figure}
    \centering
    \includegraphics[width=0.85\linewidth]{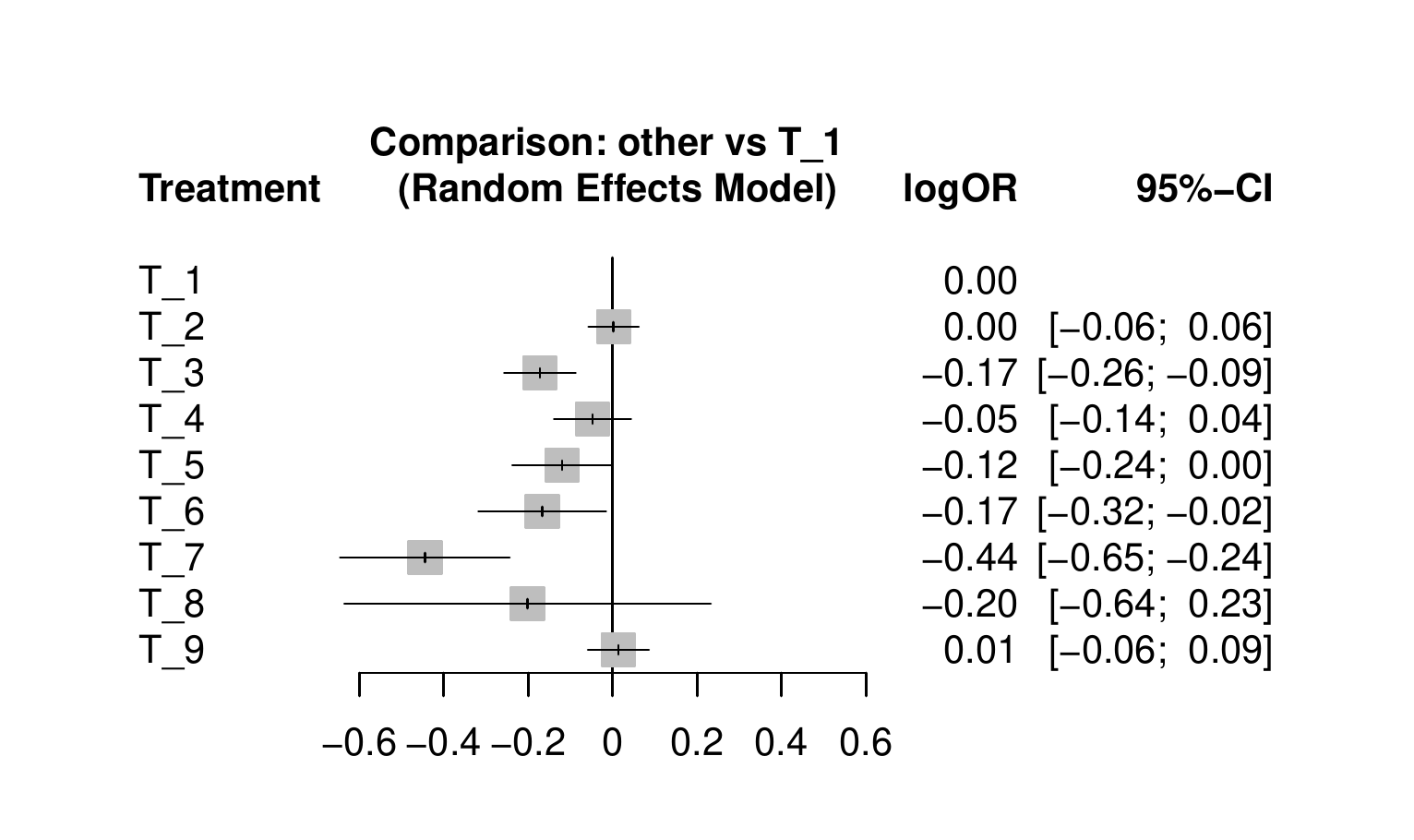}
    \caption{A forest plot showing the results of a frequentist (random effects) analysis of the Thrombolytic drug data set \cite{Boland:2003, Keeley:2003, Dias:2010}. The network graph and treatment labels are shown in Fig.~\ref{fig:RealNetEG}. The global baseline treatment is $T_1$ and has a log odds ratio of 0 by definition. The outcome of interest is the number of deaths that occur within 30 or 35 days of a heart attack. Therefore a log odds ratio $<0$ indicates that the treatment is more effective than the baseline $T_1$. The horizontal lines indicate the $95\%$-confidence intervals about the estimated LORs. Figure was created using the software \texttt{netmeta} \cite{netmeta:2021} \label{fig:Forest}. (The grey boxes highlight the central value and do not convey any additional information.) }
\end{figure}

A common way to present the relative treatment effect estimates in both frequentist and Bayesian NMA is on a forest plot \cite{netmeta:2021, White:2015, Hawkins:2009}. Each basic parameter $d_{T_1a}$ is represented by a horizontal line centered on its estimated value $\hat{d}_{T_1a}$ with length equal to its confidence (or credible) interval. For relative treatment effects measured as log odds ratios a value of zero represents no difference in treatment effect. When the outcome from the trials is the number of negative events or `failures', a log odds ratio $\hat{d}_{T_1a}<0$ indicates that treatment $a$ is more effective than the baseline $T_1$ and vice versa. A forest plot showing the results for a frequentist analysis of the Thrombolytic drug data in Sec.~\ref{NetTrials} is shown in Fig.~\ref{fig:Forest}.  A `caterpillar plot' is also sometimes used \cite{Spiegelhalter:2003}. This is essentially the same as a forest plot but the relative treatment effects are sorted in order of increasing effect size \cite{Hurley:2020}.

\subsection{Ranking}\label{sec:ranking}
The aim of a network meta-analysis is to provide clinicians with a clear statistical summary of all relevant data so that they know what the most desirable treatment options are for a particular condition. Relative treatment effect estimates and their confidence/credible intervals can be difficult to interpret and draw conclusions from, especially when many treatments have been compared \cite{Salanti:2011, Mbuagbaw:2017}.

Ranking treatments from best to worst based on their relative effect estimates is the simplest way of summarising the results of an NMA. For example, from the estimated treatment effects $\hat d_{T_1a}$ in the forest plot in Fig.~\ref{fig:Forest} the treatments would be ranked from best to worst in the order $T_7, T_8, T_3=T_6, T_5, T_4, T_1=T_2,T_9$ (where we have written $a=b$ for treatments that are observed to be equally effective within the reported number of digits for the treatment effects).

However, this summary does not account for the level of overlap between the confidence intervals or the similarity between the point estimates. For example, in the Thrombolytic drug data set treatment $T_8$ is ranked second best using this method despite the fact it has a very large confidence interval that covers almost the entire width of the other intervals. On inspection of the forest plot we cannot draw any meaningful conclusion about the effectiveness of treatment $T_8$ but this fact is not reflected in its rank.

\subsubsection{Rank Probability and Rankograms.}
A more sophisticated method of ordering the treatments is to calculate so-called rank probabilities \cite{Salanti:2011}. That is, we calculate the probability that each treatment is best, second best and so on. We use the notation $P_a(r)$ to represent the probability that treatment $a$ has rank $r$. These quantities are only meaningful in a Bayesian framework where probability describes the degree of belief in parameter values. In a Bayesian NMA, treatments are ranked at each iteration of the MCMC according to the values of relative treatment effect sampled at that iteration. The probability $P_a(r)$ is then estimated from the proportion of times treatment $a$ was ranked $r$-th. 

Although rank probabilities do not strictly make sense in a frequentist framework, so-called `re-sampling' methods have been developed to produce estimates of rank probabilities based on the results of a frequentist NMA \cite{White:2011, White:2012}. Essentially, this involves assuming that the distribution of the model parameters can be approximated by a normal distribution with mean and variance equal to the values estimated from frequentist methods (Sec.~\ref{freq_NMA}). Values of relative treatment effect are sampled multiple times from this approximate distribution and rank probabilities are estimated in the same way as before.  

\begin{figure}
    \centering
    \includegraphics[width=1\linewidth]{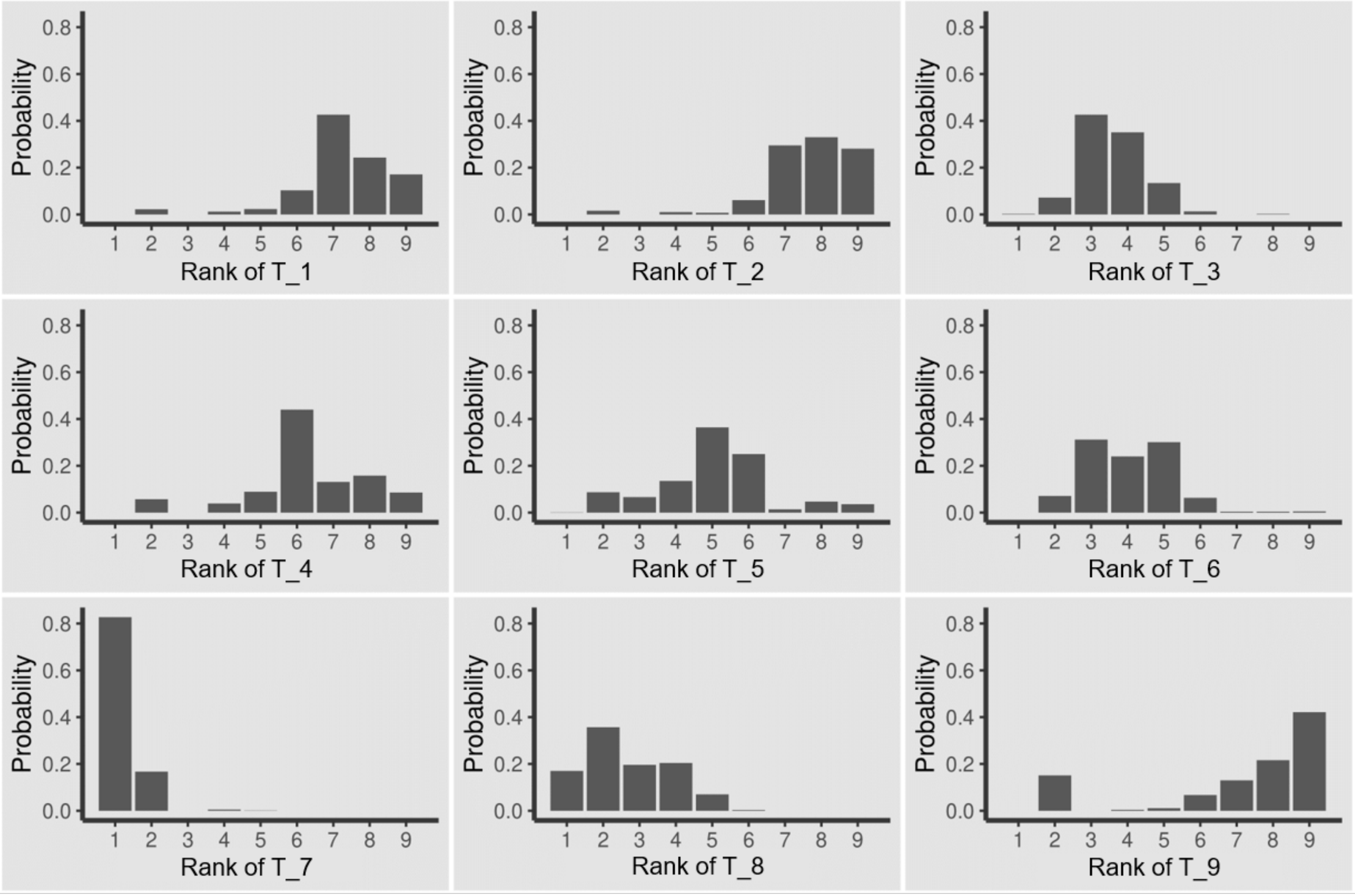}
    \caption{Rankograms for the Thrombolytic data set in Fig.~\ref{fig:RealNetEG}. Rank probabilities $P_a(r)$ are plotted against rank $r$ for each treatment in the network, $a=T_1,\dots,T_9$. Rank probabilities were obtained from frequentist re-sampling methods (based on 1000 simulations) using \texttt{netmeta} \cite{netmeta:2021}. \label{fig:Rankogram}}
\end{figure}

The rank probabilities can be displayed graphically using `rankograms' \cite{Salanti:2011}. For each treatment, the rank probabilities $P_a(r)$ are plotted against the rank $r$ either as a bar chart or a line graph. The rankograms for treatments in the Thrombolytic drug data set are shown in Fig.~\ref{fig:Rankogram} as bar charts.

Ranking using probabilities reflects not only the point estimates of relative treatment effects but also the uncertainties on these estimates and their overlapping confidence/credible intervals. Clearly, the more overlap between intervals, the flatter the rankograms will be. For example, in the Thrombolytic data set, while treatment $T_8$ has the highest probability of being second best, its rankogram is relatively flat indicating the uncertainty in its rank.  

\subsubsection{SUCRA and P values.}
The number of rank probabilities for a network increases with $N^2$. Therefore rank probabilities and rankograms become increasingly difficult to interpret as the number of treatments increases \cite{Mbuagbaw:2017}. 

Instead of rankograms, we could instead plot the cumulative probability $F_a(r)$ against rank $r$ to obtain cumulative ranking curves. Here, $F_a(r)$ is the probability that treatment $a$ has rank $r$ or better, 
\begin{eqnarray}
    F_a(r) = \sum_{s=1}^{r} P_a(s).
\end{eqnarray}
A simple summary of rank probabilities is then the area under these curves. Salanti et al (2011) \cite{Salanti:2011} termed this measurement the `surface under the cumulative ranking line' or SUCRA.  The value of SUCRA for a particular treatment $a$ is then
\begin{equation}
    \mathrm{SUCRA}_a = \frac{1}{N-1} \sum_{r=1}^{N-1} F_a (r) = \frac{1}{N-1}(N-\mathbb{E}(r)_a),
\end{equation}
where $\mathbb{E}(r)_a$ is the mean or expected rank of treatment $a$,
\begin{eqnarray}
    \mathbb{E}(r)_a = \sum_{r=1}^{N} r P_a(r). 
\end{eqnarray}
SUCRA takes values from 0 to 1, though these are often expressed as a percentage. If treatment $a$ ranks first with probability one then it will have $\mathrm{SUCRA}_a=1$ (or 100\%) whereas a treatment that ranks worst with probability one will have a  $\mathrm{SUCRA}_a=0$ (or 0\%) \cite{Salanti:2011}. $\mathrm{SUCRA}_{a}$ can be interpreted as the average proportion of treatments worse than $a$. $\mathrm{SUCRA}$ values provide a concise summary of treatment rankings that accounts for the estimated relative treatment effects, uncertainty in these estimates, and the resulting overlap in their confidence/credible intervals. 

In frequentist NMA values of SUCRA can be calculated from the rank probability estimates obtained via re-sampling methods. Alternatively, R{\"u}cker and Schwarzer \cite{Rucker:2015} proposed an analogous quantity called a `P score' that does not require re-sampling. By assuming a normal distribution for the model parameters they define
\begin{eqnarray}
    F_{ab} = \Phi\left(\frac{\hat{d}_{ab}}{\sqrt{\mathrm{Var}(\hat{d}_{ab})}}\right),
\end{eqnarray}
 where $\Phi(.)$ is the CDF of the standard normal distribution. This is interpreted as the extent of certainty that $\hat{d}_{ab}>0$ (i.e. that $a$ is more effective than $b$). The P-score for treatment $a$ is then the mean of $F_{ab}$ over treatments $b\neq a$. This is the mean extent of certainty that $a$ is more effective than any other treatment. R{\"u}cker and Schwarzer \cite{Rucker:2015} show that when the true probabilities are known, P-scores and SUCRA values are identical. In practice they give very similar results.

\section{Existing points of contact between NMA and physics} \label{sec:existing}

In the previous sections we have introduced some of the essential concepts and methods for NMA. In the remainder of the paper we now discuss how physics (in particular, statistical physics) and physicists can contribute to this area. 

For example, a number of analogies between meta-analysis and specific physical systems have been proposed in recent years. These analogies have provided insight, and they have helped to improve meta-analysis methodology and the visualisation of the problem. In this section we briefly outline these analogies.

Sec.~\ref{sec:future} then describes a few examples of more general methods used in statistical physics which have been shown to be useful in a meta-analysis context. We also present a number of more speculative ideas on how knowledge from physics might be used for NMA.

\subsection{NMA and electrical networks}
\label{NMA-Electric}
Arguably the most influential of the meta-analysis analogies was developed by R{\"u}cker (2012) \cite{Rucker:2012} who demonstrated the connection between NMA and electrical network theory. The starting point for this analogy is the observation that variance in meta-analysis combines in the same way as resistance in an electrical network.

\subsubsection{Variances in NMA combine like resistance in a network.} 
\begin{figure}
    \centering
    \includegraphics[width=0.9\linewidth]{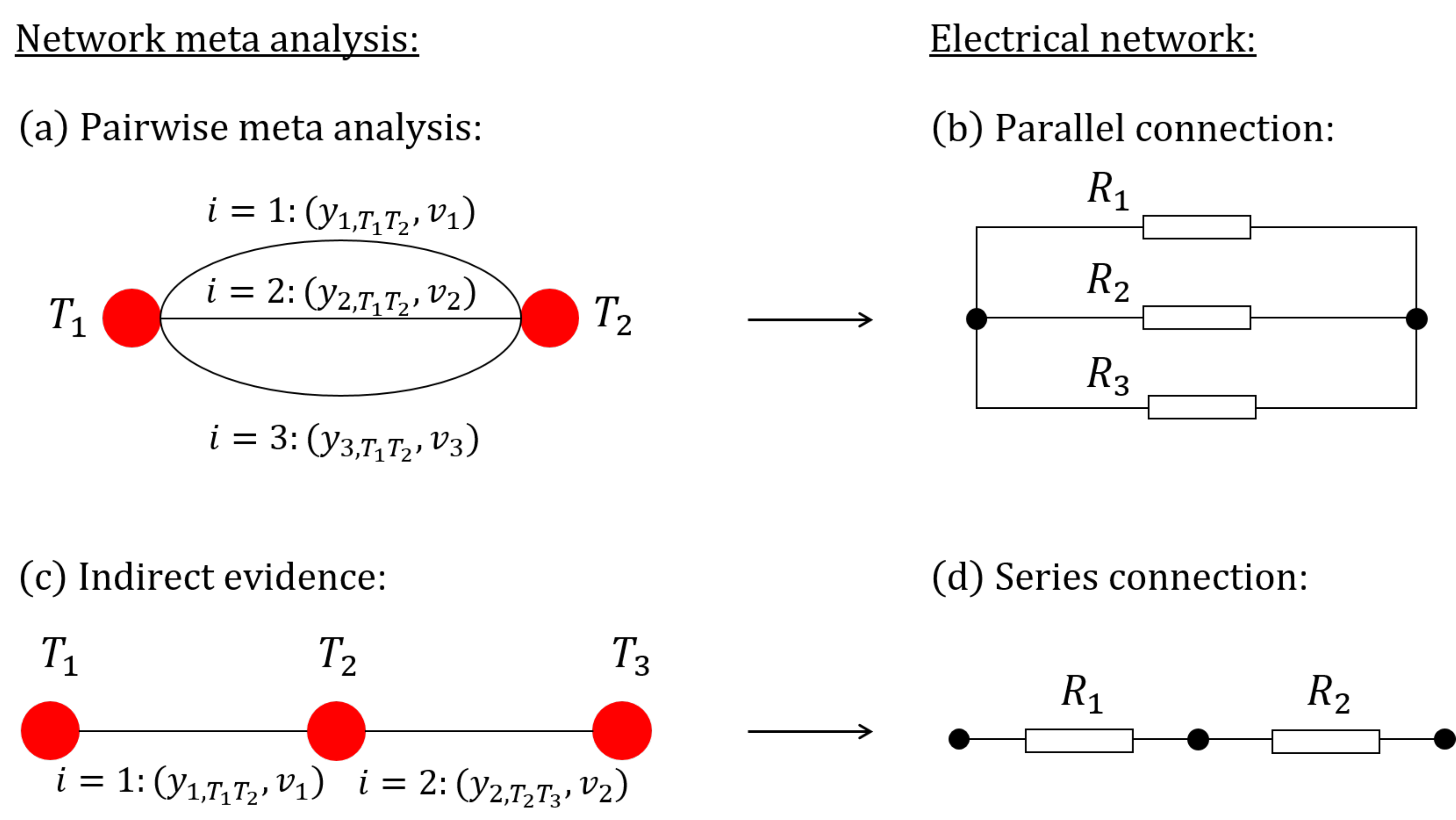}
    \caption{ An illustration of the analogy between NMA and electrical networks. (a) A pairwise meta-analysis of three trials corresponds to (b) three resistors connected in parallel. (c) A chain of two trials connecting three treatments corresponds to (d) two resistors connected in series. We label each treatment as $T_{a}$ and each resistor as $R_i$. Each trial $i$ is labelled with the measurement of relative treatment effect made in that trial, $y_{i,12}$, and the associated variance, $v_i$. 
    \label{fig:NMAelectric}}
\end{figure}

In Sec.~\ref{Freq_d_est} we saw that for a frequentist pairwise meta-analysis, the variance of the estimated treatment effect $\hat{d}$ is an expression for the variance of the weighted mean [Eq.~(\ref{eq:var_dhat_pair})] calculated in terms of the variances associated with each of the trials. Taking the reciprocal on both sides of this equation and writing $v_i$ for the variance associated with the measurement in trial $i$ we find
\begin{eqnarray}
\label{eq:V_parallel}
    \frac{1}{\mathrm{Var}(\hat{d})} = \sum_{i=1}^{M}\frac{1}{v_i}.
\end{eqnarray}
For a random effects model we have $\hat{d} = \hat{d}^{\mathrm{RE}}$ and $v_i = \sigma_i^2 + \tau^2$, whereas for a fixed effect model $\hat{d} = \hat{d}^{\mathrm{FE}}$ and $v_i = \sigma_i^2$. As illustrated in Fig.~\ref{fig:NMAelectric} (a), a pairwise meta-analysis can be represented by a graph with two nodes (representing the two treatment options) and multiple parallel connections (edges) between the nodes (representing the individual trials comparing the treatments). The same graphical representation describes an electrical network with resistors connected in parallel [Fig.~\ref{fig:NMAelectric} (b)]. The effective resistance, $R$, of a set of $M$ parallel resistors $R_i$, $i=1,\dots,M$ is 
\begin{eqnarray}
    \frac{1}{R} = \sum_{i=1}^{M}\frac{1}{R_i}.
\end{eqnarray}
Therefore, resistors in parallel combine like variances in a pairwise meta-analysis.

Now consider the network in Fig.~\ref{fig:NMAelectric} (c) comprising three treatments, $T_1, T_2, T_3$, and two trials. The first trial $i=1$ compares treatments $T_1$ and $T_2$ and measures a relative treatment effect $y_{1,T_1 T_2}$ with variance $v_1$. Trial $i=2$ compares treatments $T_2$ and $T_3$ and measures $y_{2,T_2 T_3}$ with variance $v_2$. The network estimates of the relative effect between treatments $T_1$ and $T_2$, and between $T_2$ and $T_3$, respectively, are simply the direct estimates $\hat{d}_{T_1 T_2}=y_{1,T_1 T_2}$ and $\hat{d}_{T_2 T_3}=y_{2,T_2 T_3}$. There is no direct evidence for the comparison of treatments $T_1$ and $T_3$. The network estimate of this comparison is obtained from an indirect estimate via $T_2$, 
\begin{eqnarray}
       \hat{d}_{T_1 T_3} = y_{1,T_1 T_2} + y_{2,T_2 T_3}.
\end{eqnarray}
Since the trials are independent, the variance associated with this estimate is
\begin{eqnarray}\label{eq:V_series}
       \mathrm{Var}(\hat{d}_{T_1 T_3}) = v_1 + v_2.
\end{eqnarray}
As shown in Fig.~\ref{fig:NMAelectric} (d) this set-up relates to an electrical network with resistors connected in series. The effective resistance for this network is
\begin{eqnarray}
    R = R_1+R_2.
\end{eqnarray}
Therefore resistors in series combine like variances for indirect estimates in NMA. 

The analogy with resistor circuits can be extended to networks of more than two trials, and those combining both parallel connections and connections in series. Through a more detailed analysis (which we do not describe here) and by comparing Ohm's law to the weighted mean expression in Eq.~(\ref{eq:d-freq}), one can then establish a mapping between relative treatment effects and potential differences (voltages). 

\subsubsection{Analogy of the NMA problem in electric circuits.} In electrical network theory, graph theoretical methods are used in different ways, for example to construct the electric potentials at the nodes from external currents, or to compute the effective resistance between two nodes from the resistors in the network. R{\"u}cker showed that a similar set of methods can be used in NMA to derive an expression for the network estimates of the relative treatment effects from the observed effects. This leads to the same results as the Aitken estimator in Eq. (\ref{eq:dhat_pair}) \cite{Rucker:2012, Rucker:2014}. 

We do not present details here, but the core of the analogy can be described as follows \cite{Rucker:2012}. The NMA problem consists of finding the network estimates for the relative treatment effects using the effects observed in the trials (for a known network structure and known [inverse-variance] weights of the trials). The observed effects will in general be inconsistent, whereas the estimates resulting from the NMA are consistent by construction. The observed effects translate to `observed' voltages across the resistors in the network (the resistances are determined by the inverse-weights of the trials). As a consequence of the inconsistency, voltages along loops in the network will not add to zero. This means that no electric potentials can be assigned to the nodes from which the voltages would arise as potential differences. The key result is now that the problem of determining the NMA estimates for the treatment effects is equivalent to finding the set of electric potentials so that the resulting (consistent) voltages best approximate the observed (inconsistent) voltages\footnote{These potentials are only unique up to an overall additive constant. This reflects the fact that NMA tries to estimate relative rather than absolute treatment effects.}. Quality of approximation is here measured in terms of the Euclidian norm.

\subsubsection{Reduction of multi-armed trials.} One particularly useful application of the electrical network analogy is in the context of networks with multi-arm trials. Measurements of the different relative treatment effects from a multi-arm trial are correlated, and the presence of such correlations can cause complications for some NMA methodology. One therefore carries out a `reduction' to an equivalent set of two-arm trials. We here briefly describe this, for details see \cite{Rucker:2012, Rucker:2014}.

We focus on a single multi-arm trial with $m$-arms. The corresponding (sub) network involving the $m$ treatments is then fully connected. The idea is to find a network consisting of $m(m-1)/2$ pairwise trials which is `equivalent' to the multi-arm trial in the sense that the variances of the network estimates of relative treatment effects in the pairwise network [given by  Eq.~(\ref{eq:freq_var})] are the same as the variances in the graph describing the multi-arm trial.

As discussed above, variances in NMA combine like resistances in electric networks, i.e., the variances of network estimates are obtained from the individual trial variances in the same way as {\em effective} resistance is obtained from the physical resistors in electric circuits\footnote{The effective resistance between two nodes results as $U/I$ from the current $I$ that flows into the network if a battery of voltage $U$ is attached to the two target nodes. Effective resistance accounts for any direct connection between the target nodes, and for all indirect connections through other nodes in the graph.}. The reduction problem for the multi-armed trial is therefore equivalent to finding the individual resistances $\{R_{ab}\}$ in an electric network given the {\em effective} resistances between pairs of nodes. It is well known (see e.g. \cite{Gutman:2004}) that
\be\label{eq:R_L_plus}
R^{\rm eff}_{ab}=L^+_{aa}+L^+_{bb}-2L^+_{ab},
\ee
where $\mathbf{L}$ is the graph Laplacian describing the electric network, and where ${}^+$ denotes the so-called pseudo-inverse. The graph Laplacian is defined by the individual (physical) resistors via $L_{ab}=-R_{ab}^{-1}$ for $a\neq b$, and $L_{aa}=\sum_b R_{ab}^{-1}$.

The reduction problem therefore maps onto the problem of finding the elements of the graph Laplacian $\mathbf{L}$ in Eq.~(\ref{eq:R_L_plus}) for given effective resistances $\{R_{ab}^{\rm eff}\}$ (i.e., given variances in the multi-armed trial). The individual physical resistors (variances associated with the individual two-armed trials) can then be extracted from the off-diagonal elements of the Laplacian. Further details of the reduction method are given in Appendix \ref{App:BackCalc}. 

We perform this reduction for every multi-arm trial in a meta-analytic network, and use effect estimates from the multi-arm trials to assign estimates to the two-arm trials. This leads to a network of two-arm trials that is equivalent to the original network (i.e., it produces the same network estimates and variances). Methodology that does not allow for correlations can then be used on this new network. 

\medskip

Further details of the analogy between NMA and electrical networks can be found in \cite{Rucker:2012, Rucker:2014}.

\subsection{Random Walks}
\label{NMA-RW}  
Random walks are a familiar concept to statistical physicists. Random hopping processes on networks are of particular interest in a number of areas \cite{Lov:1994,Noh:2004,Masuda:2017}.  In brief, a random walk on a network is a stochastic process describing a series of hops between nodes that are connected by an edge. We focus on discrete time such that each time step is associated with one hop across an edge.

There is a well-known analogy between random walks and electrical networks \cite{Kakutani:1945, Kemeny:1966, Kelly:1979, Doyle:2000}, briefly summarised in the next section. Using the work of R{\"u}cker \cite{Rucker:2012} described in Sec.~\ref{NMA-Electric} we were able to extend this analogy to network meta-analysis.

\subsubsection{Random walks and resistor networks.}

Each edge in an electrical resistor network has an associated resistance. Given such a network, one now constructs a random walk as follows: For a random walker currently at node $a$, the probability with which the walker hops from $a$ to $b$ in the next step is proportional to the inverse-resistance associated with the edge $ab$. More precisely, one defines the transition matrix elements as 
\be\label{eq:RW_R}
P_{ab} = \frac{R_{ab}^{-1}}{\sum_{c\neq a} {R}_{ac}^{-1}},
\ee
where $R_{ab}$ is the resistance of the resistor connecting nodes $a$ and $b$.

Various physical quantities in the electric network then have interpretations in the random-walk picture. A good summary can be found in \cite{Doyle:2000}. For example, consider the following scenario: A battery is attached to two nodes $a$ and $b$ in the resistor network. We assume that the network only has one single connected component. The voltage of the battery is chosen such that one Amp\`ere of current goes into node $a$ from the battery (and consequently one Amp\`ere goes from node $b$ back into the battery). This then induces currents $I_{cd}$ through all edges $cd$ (the resistors) in the network. Now imagine we release a random walker at node $a$, and it performs the random walk defined by Eq.~(\ref{eq:RW_R}). We stop the walk when the walker reaches node $b$ (this will happen eventually given that the network consists of a single component). We can record the net number of times the walker will have crossed edge $cd$ before it reaches node $b$ (hops from $d$ to $c$ contribute negatively to this value). One can then show \cite{Doyle:2000} that the expected net number of crossings from $c$ to $d$ is  given by the current $I_{cd}$ in the electric network with the battery attached to nodes $a$ and $b$. 

\subsubsection{Random walks and flow of evidence in network meta-analysis.} 
Starting from the existing analogies between electrical networks and NMA on the one hand, and electrical networks and random walks on the other, we (along with R\"ucker, Papakonstantinou and Nikolakopoulou) \cite{Davies:2021} proposed an analogy between NMA and random walks. In the following, we briefly summarise the main ideas.

\begin{figure}
    \centering
    \includegraphics[width=1\linewidth]{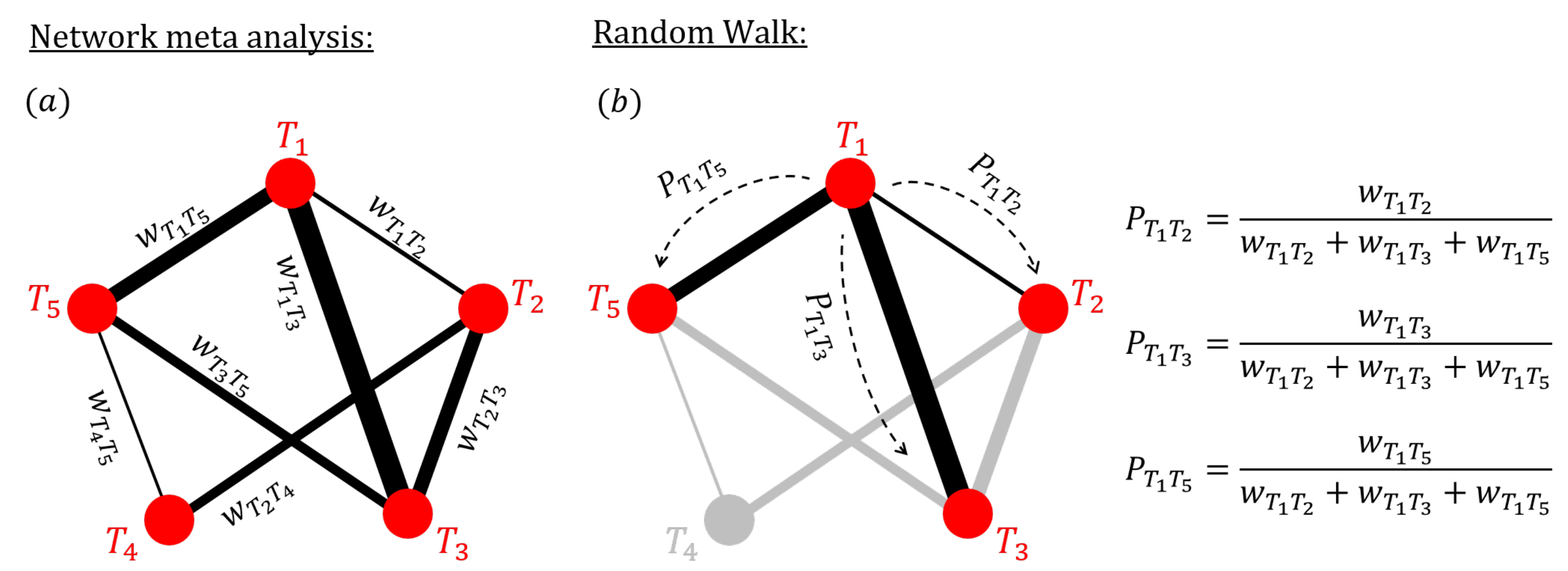}
    \caption{An illustration of the analogy between NMA and random walks. Panel (a) shows an NMA with five treatments $a = T_1, T_2, T_3, T_4, T_5$. Each edge is labelled with the inverse-variance weight associated with that treatment comparison, $w_{ab}$. Panel (b) shows the transition probabilities for a random walker on the network in (a) who is currently at node $T_1$. At the next time step this walker can move to node $T_2$, $T_3$ or $T_5$ with probabilities proportional to the edge weights.
    \label{fig:NMA_RW}}
\end{figure}

We have seen that resistance in an electrical network is analogous to variance in an NMA. Therefore inverse-resistance is associated with inverse-variance weight. Writing $w_{ab}$ for the weight associated with edge $ab$ in a network meta-analysis (see Sec.~\ref{Freq_d_est}), we define the transition matrix of a random walker via
\begin{eqnarray}
    P_{ab} = \frac{{w}_{ab}}{\sum_{c\neq a} {w}_{ac}}.
\end{eqnarray}
This is illustrated in Fig.~\ref{fig:NMA_RW}.

In \cite{Davies:2021} we found that the expected net number of times a walker crosses each edge is equal to the so-called `evidence flow' through that edge. This is a concept introduced by K{\"o}nig et al (2013) \cite{konig:2013}. We do not give full definitions here. Broadly speaking the flow variable $f_{cd}^{(ab)}$ is a coefficient that describes how much the observed relative effect between treatments $c$ and $d$ contributes to the network estimate of the relative effect between $a$ and $b$. The coefficients are related to the entries of the matrix on the right-hand side of Eq.~(\ref{eq:d-freq}). They describe how different pieces of direct evidence in a network meta-analysis combine to give the overall network estimates of relative treatment effects. It turns out \cite{konig:2013} that these coefficients have the properties of a flow. For example, the flow out of a node $c\neq a,b$, $\sum_{x} f_{cx}^{(ab)}$, equals the flow into node $c$, $\sum_{x} f_{xc}^{(ab)}$, indicating that flow is conserved at $c$. The flow variables are defined such that they are non-negative. A value of $f_{cd}^{(ab)}>0$ indicates a positive flow from $c$ to $d$, the transposed variable $f_{dc}^{(ab)}$ is then zero.

For a comparison $ab$, the evidence flow variables $\{f_{cd}^{(ab)}\}$ can be used to define a directed weighted graph. This is for a fixed choice of $a$ and $b$ meaning there is one directed graph for each comparison $ab$. Nodes in this graph again represent the treatment options, but the weights of the (now directed) edges are given by the flow of evidence $\{f_{cd}^{(ab)}\}$. 

\subsubsection{Random walks, streams of evidence and proportion contributions.}
In \cite{Davies:2021} we then defined a second random walk. All walkers start at $a$ and move on the directed graph just described. The construction is such that walkers can never return to a node they have already visited (the graph is acyclic), and all walks end at $b$ (node $b$ is absorbing). We can think of these walkers as collecting evidence along their way. They start at node $a$, hop to intermediate nodes and record differences in treatment effects (similar to differences in height in a mechanical set-up). When a walker arrives at $b$ it reports the total difference in altitude it has experienced. Due to inconsistencies, this reported difference may be different along different paths connecting $a$ and $b$. The average of what the walkers report turns out to be the network estimate of the relative treatment effect between $a$ and $b$ \cite{Davies:2021}. 

In addition, the probability of a walker taking a certain path from $a$ to $b$ is given by the product of the transition probabilities in the edges along that path. This expression can be used to calculate so-called `streams of evidence' \cite{Papakon:2018}, and what is referred to as the `proportion contribution matrix' \cite{Salanti:2014}. In particular, we used the random walk transition probabilities to derive an analytical expression for the contribution each treatment comparison makes to each overall treatment effect estimate. As discussed in more detail in \cite{Davies:2021}, this random walk approach overcomes some limitations of previous algorithms used to construct this quantity. The random-walk method has recently been implemented in the software package {\tt netmeta} \cite{netmeta:2021}.

The random walk analogy is a recent addition to network meta-analysis. As a result, only a small proportion of the random walk literature has been explored in this context. We therefore feel that there is scope to extend the analogy. We hope that  the introduction to NMA we give in this paper will help other statistical physicists with an interest in random walks on networks to join these efforts.

\subsection{System of springs}
Papakonstantinou et al (2021) \cite{Papakon:2021} visualised the meta-analysis problem as a system of springs. In a similar vein to the electrical network analogy, they observe that when connecting systems of springs the inverse of the spring constant combines in the same way as variance in meta-analysis. 

\begin{figure}
    \centering
    \includegraphics[width=0.7\linewidth]{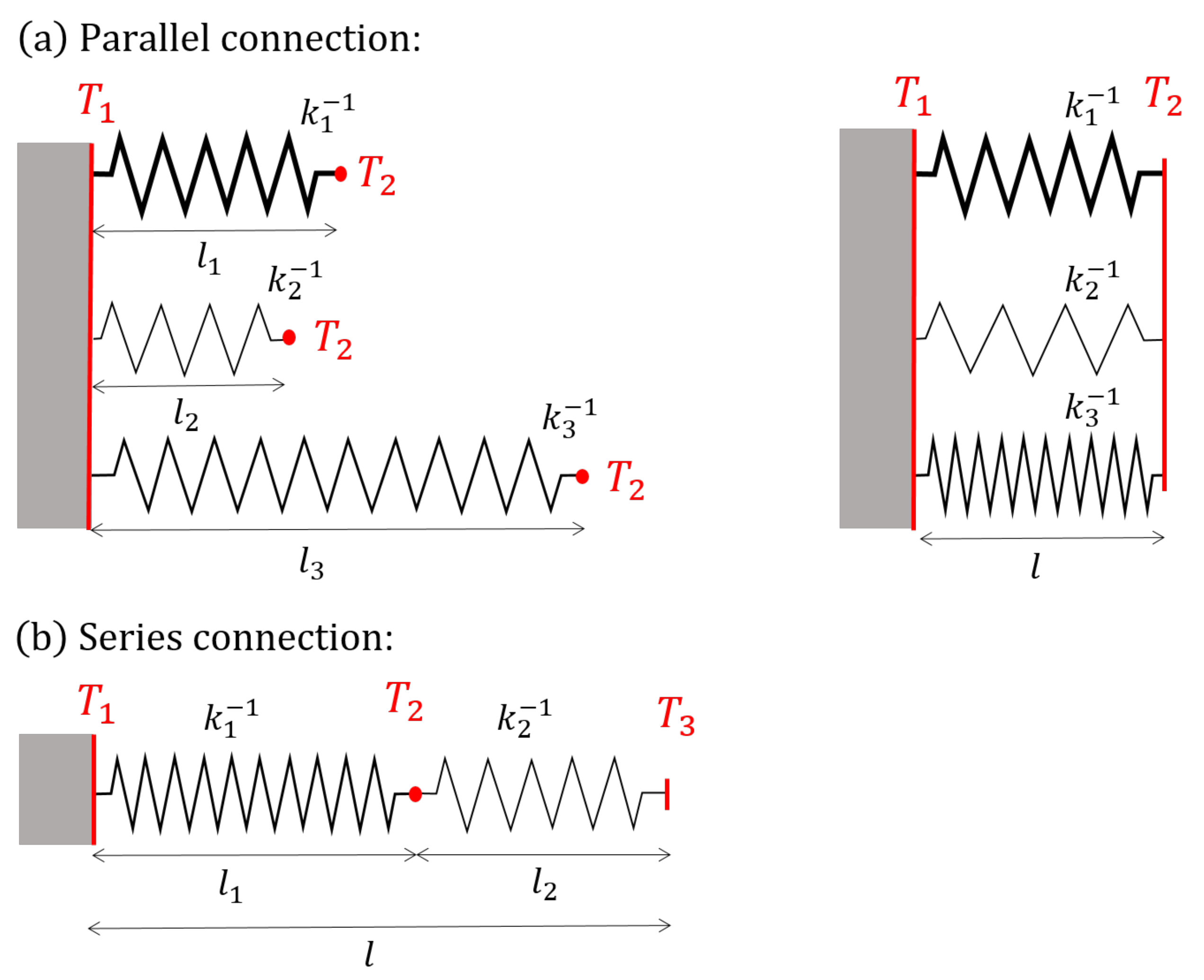}
    \caption{ (a) An illustration of a parallel system of springs. The springs are fixed on one side corresponding to the baseline treatment. The open ends are then forced to the same length so that their natural lengths are displaced by some distance. This is equivalent to a pairwise meta-analysis. (b) A system of springs connected in series. This is equivalent to an indirect comparison in meta-analysis. We label each treatment as $T_{a}$. Each spring is labelled by its natural length, $l_i$, and the inverse of its spring constant, $k_i^{-1}$. $l$ is the effective length of the spring system. The different thicknesses of the springs represent their different spring constants.
    \label{fig:NMAspring}}
\end{figure}

Hooke's law states that the force, $f$, needed to displace a spring by length $x$ is $f = k x$,
where $k$ is a measure of the stiffness of the spring known as the spring constant. The potential energy stored in such a spring is
\begin{eqnarray}
    U = \frac{1}{2} k x^2.
\end{eqnarray}
Consider a set of $M$ springs fixed at one end and arranged in parallel as shown in Fig.~\ref{fig:NMAspring} (a). Each spring has a different spring constant, $k_i$, and a different natural length, $l_i$. The open ends of the spring are then connected so that the springs are forced to assume the same length, called the `effective length' $l$ of the spring. Each spring has therefore been displaced by a different amount, $x_i=l_i-l$. The resulting equilibrium length of the springs, $\hat l$, is such that the total force on $T_2$ vanishes, i.e. $\sum_i k_i x_i=0$. (Equivalently, this length minimises the energy stored in the system, $\hat{l} = \underset{l}{\mathrm{argmin}}\left[ \sum_{i=1}^{M} k_i (l_i-l)^2 \right]$.) One finds,
\begin{eqnarray}
    \hat{l} = \frac{\sum_{i=1}^{M} k_i l_i }{\sum_{i=1}^{M} k_i}.
\end{eqnarray}
This expression is in the form of a weighted mean. Comparing it to the estimate from a pairwise meta-analysis in Eq.~(\ref{eq:dhat_pair}), one observes that we can draw an analogy between the parallel system of springs and a pairwise MA. We interpret the spring constant as inverse-variance weight. The spring constant associated with the `effective spring' is 
\begin{eqnarray}
    k = \sum_{i=1}^{M} k_i. 
\end{eqnarray}
Comparing this to Eq.~(\ref{eq:V_parallel}), it is clear that variance in a pairwise meta-analysis combines in the same way as the inverse of the spring constant for a set of parallel springs. 

One can also make this analogy for springs connected in series. As shown in Fig.~\ref{fig:NMAspring} (b), the effective length of a chain of springs connected together is simply the sum of their natural lengths. The effective spring constant is then
\begin{eqnarray}
    \frac{1}{k} = \sum_{i=1}^{M} \frac{1}{k_i}.
\end{eqnarray}
Comparing this to Eq.~(\ref{eq:V_series}), we observe that the inverse of the spring constants of springs connected in series combine like variances for an indirect estimate in meta-analysis. 

The natural length of each spring can be interpreted as the measurement of the relative treatment effect in each trial. The displacements $x_i$ then relate to the residuals associated with the relative treatment effects (i.e. differences between the relative effects measured in each trial and those predicted by the model). Minimising the energy is analogous to the process of minimising the sum of squared weighted residuals described in Secs.~\ref{OLS} and \ref{GLS}.

This analogy provides a useful visualisation of the meta-analysis process. That is, combining the data from multiple medical trials is like minimising the energy in a system of springs. The energy stored in the final equilibrium system then represents a measure of disagreement between the different pieces of evidence. Though this analogy has been demonstrated for a set of relatively simple configurations, it has not yet been extended to a general network meta-analysis. This would require a much more complex spring system. Further details can be found in \cite{Papakon:2021}.

\subsection{Balance of torques in a mechanical system} Another visualisation of meta-analysis based on a mechanical system was proposed by Bowden and Jackson (2016) \cite{Bowden:2016}. In this analogy, the process of finding the minimum sum of weighted residuals (or, equivalently, maximising the likelihood, see Secs.~\ref{GLS} and \ref{ML}) is equated to balancing torques in a system of weights. This is the same as finding the position of the centre of mass.

\begin{figure}
    \centering
    \includegraphics[width=0.5\linewidth]{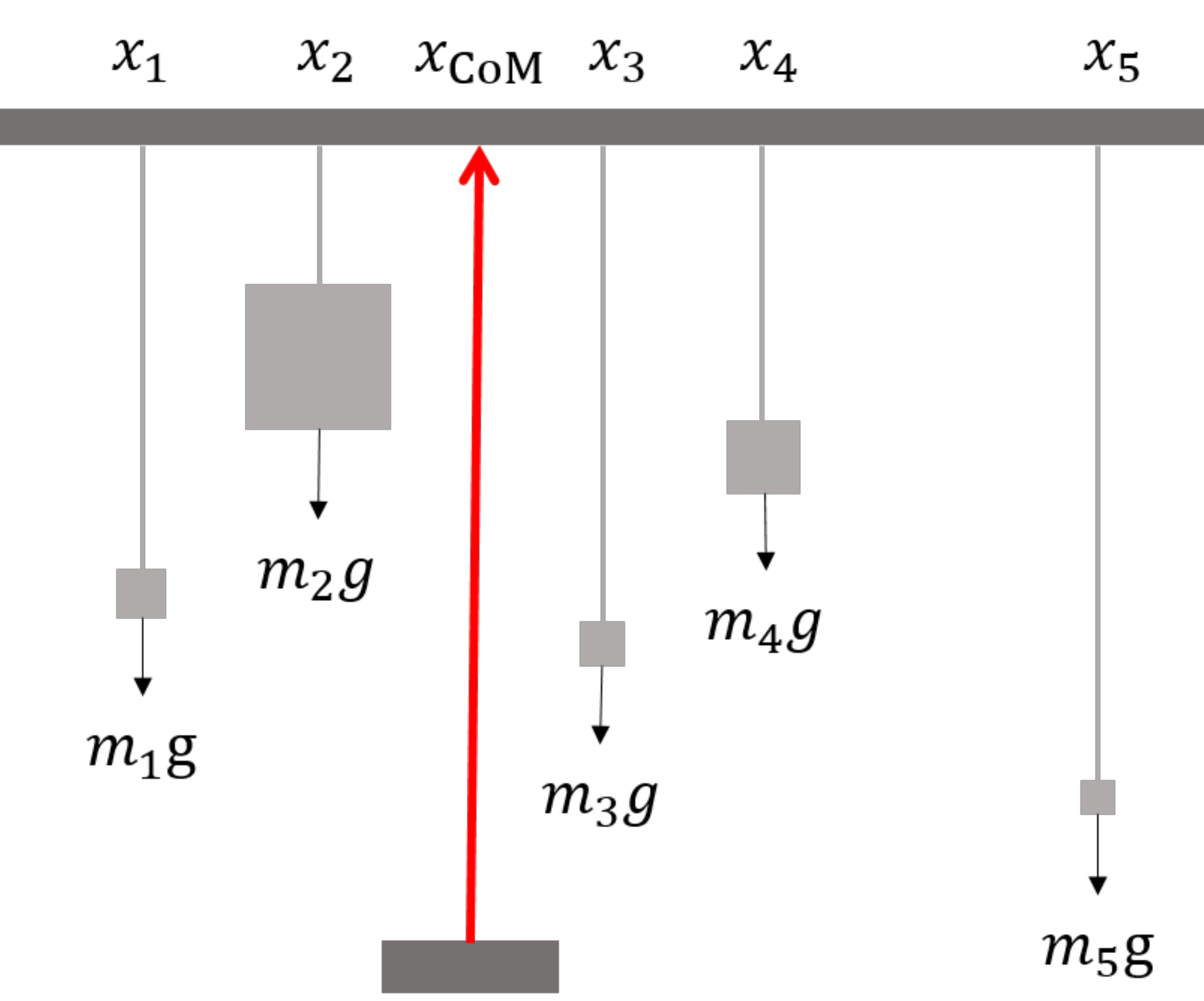}
    \caption{ An illustration of pairwise meta-analysis as a centre of mass problem. Each mass is labelled by its position along the rod, $x_i$, and the force acting on it, $m_i g$, which is equal to its weight. The position of each object represents the measurement of relative treatment effect in each trial. The physical weight of each object represents the inverse-variance weight of each measurement. The centre of mass, $x_{\mathrm{CoM}}$, is the position of the pivot that balances the torques. Finding this position is equivalent to finding the estimate of relative treatment effect in a pairwise meta-analysis.
    \label{fig:NMACoM}}
\end{figure}

Fig.~\ref{fig:NMACoM} shows a mechanical system consisting of a bar with $M$ objects of different masses, $m_i$, hanging at various positions $x_i$ along the bar. The bar is supported by a pivot. The system of masses (or weights) is balanced when the pivot is placed such that the torques exerted by the masses balance, i.e. $\sum_{i=1}^M m_i g(x_i-x_{\rm pivot})=0$, where $g$ is the acceleration due to gravity. Writing the weights of the different masses as $w_i=m_i g$ the balance of torques leads to
\begin{eqnarray}
    x_{\mathrm{pivot}}=x_{\mathrm{CoM}} \equiv \frac{\sum_{i=1}^{M} w_i x_i}{\sum_{i=1}^{M} w_i},
\end{eqnarray}
i.e. the pivot must be located at the centre of mass (in one dimension) defined by the system of weights. 

As before, we compare this to the pairwise meta-analysis estimate in Eq.~(\ref{eq:dhat_pair}) in order to establish the analogy. The position of each mass along the bar then represents the observed relative treatment effect in each trial and the physical weight of each object represents the inverse-variance weight associated with each observation. 

The problem of finding the position of the centre of mass provides a visual representation of finding the best estimate of the relative treatment effect in a pairwise meta-analysis. Bowden and Jackson used this visualisation to create an online visualisation tool. The visualisation shows how different modelling assumptions (for example, fixed effects vs random effects) or the addition/removal of certain studies affect the position of the centre of mass. When the user makes a change to the model or the data, an animation shows the change in balance of the system. This was found to help identify the presence of small-study bias, a phenomenon where smaller studies report a systematically larger relative treatment effect than larger studies \cite{Sterne:2000}. This analogy has not been extended to indirect estimates or network meta-analysis. For details, see \cite{Bowden:2016}.

\section{Ideas for future work: a research programme at the interface of statistical physics and network meta-analysis} \label{sec:future}

In this section we now present a series of broader ideas on how statistical physics might contribute to the field of NMA in the future. Some of these have been formulated in existing literature (e.g. Markov chain methods to rank treatments in Sec.~\ref{sec:mcmc_ranking}), others are more speculative. 

\subsection{Markov chain approaches to ranking}\label{sec:mcmc_ranking}
Markov chain approaches have been used as ranking methods in a variety of fields \cite{Page:1998, Danilowicz:2001, Blanchet:2016}. Most notably, Google's PageRank algorithm \cite{Page:1998} ranks web pages in their search engine results. Broadly summarising, the algorithm is based on a Markov chain that models a `random surfer' who, after being randomly assigned an initial webpage, moves from page to page by randomly clicking links. The algorithm also allows for a damping mechanism that works by assigning a certain probability with which the surfer is re-set to a random page at each step (this ensures that pages with no incoming links are also visited from time to time). The stationary distribution of the Markov chain then informs the ranking of the webpages. Broadly speaking, a page with more incoming links is more likely to be visited and therefore attracts a higher probability at stationarity, resulting in a higher rank.

As discussed in Sec.~\ref{sec:ranking} ranking treatments is an important output of NMA and has received much attention in the literature  \cite{Trinquart:2016, Veroniki:2016, Veroniki:2018, Daly:2019, Chaimani:2019, Chiocchia:2020, Mavridis:2020, Nikolakopoulou:2020b}. Chaimani et al \cite{Chaimani:2019} developed a ranking method for NMA based on the PageRank algorithm.

As in the random-walk framework, each state of the Markov chain is a treatment in the network. At each discrete time step the process moves from one treatment to another. Transitions represent a preference between two treatments: When the process is currently in state $a$, the probability of transitioning to $b$ in the next time step is related to the probability that treatment $b$ is more effective than treatment $a$ \cite{Rucker:2015}. The stationary distribution of the Markov chain can then be used to rank the treatments, where a higher probability of being selected indicates a more effective treatment.

The initial distribution of the Markov chain can be chosen so that it reflects clinically important factors other than the treatment effects, for example the cost or safety of the treatments \cite{Chaimani:2019}. Similar to the PageRank algorithm, Chaimani et al introduce a re-setting mechanism: at every time step there is a non-zero probability of hopping to a state drawn from the initial distribution. This means that the stationary distribution of the Markov chain now depends on this initial distribution as well as the transition probabilities. This allows one to incorporate information about a range of factors that influence decision making.  

\subsection{Using network theory to characterise meta-analytic graphs}
A natural point of contact between NMA and statistical physics is the theory of networks. It is widely recognised that the accuracy and precision of NMA outcomes are likely to be affected by network topology. Indeed, researchers are encouraged to provide graphical and qualitative descriptions of network geometry \cite{PRISMA:2015}. Based on concepts from the theory of networks, researchers have defined topological indices to describe the geometry of meta-analytic networks and related these to the outcomes of NMA \cite{Davies:2020, Tonin:2019, Salanti:2008b}. We give a few examples in this section, recognising that this existing work is only an initial step. We think that there is significant scope to extend these activities.

In network theory, the degree of a vertex is the number of edges the vertex shares. A graph is described as `regular' if all vertices have the same degree. From these definitions we \cite{Davies:2020} defined a measure of `degree irregularity' of the graph of treatment options. This measure is given by 
\begin{equation}
    h^2 = \frac{1}{N} \sum_{a} (k_a - \bar{k})^2,
\end{equation}
and quantifies the variation in the number of studies involving each treatment. We define $k_a$ as the weighted degree of node $a$ (here, the  weight of an edge is given by the number of trials comparing the two treatments connected by the edge). The quantity $\bar{k} = \frac{1}{N} \sum_a k_a$ is the mean degree in the network. Through simulations of NMA, we found that smaller values of $h^2$ were associated with more precise treatment effects and smaller bias on rank probabilities. 

Tonin et al (2019) \cite{Tonin:2019} adapted metrics from graph theory in order to numerically describe network geometry in NMA. They performed a systematic review of 167 published NMAs and used 11 metrics to describe the topology of each network. By performing a sensitivity analysis on each metric and assessing the level of correlation between the metrics, they identified four indicators that were the most useful for describing network geometry. These are (i) Density: A measure of `connectedness' equal to the number of edges in the network divided by the total number of edges possible for $N$ vertices; (ii) Percentage of common comparators: The percentage of vertices with more than one connecting edge relative to the total number of vertices $N$; (iii) Percentage of strong edges: The percentage of edges with more than one study relative to the total number of edges; (iv) Median thickness and dispersion measure: The median number of studies per edge and the IQR (interquartile range). 

Tonin et al recommended that in order to characterise the network of evidence from an NMA these measures, in addition to the number of vertices, the number of edges, and the number of trials per edge, should be reported alongside the network graph.

This existing work shows that concepts used in network theory to characterise the topology of networks can be linked to the outcome of NMA. We think that further work to explore these ideas will be very worthwhile. Significant expertise exists in the complex networks community, and a large number of tools have been developed to characterise, for example, the clustering properties of networks, as well as measures of centrality and assortativity. Techniques are available to detect communities in networks, and more general motifs. It will be exciting to see if and how these concepts can be used to study networks of medical trials. Finding good indicators of network topology for NMA performance also leads to the possibility to suggest targeted additions to an existing meta-analytic network. In other words, these methods from the theory of (complex) networks could be used to propose future trials to be added to a meta-analysis that promise to be the most useful for improving the overall estimate of treatment effects. This could include concepts from chemical graph theory, e.g. topological indices discussed in \cite{Todeschini:2000}.

\subsection{Network meta-analysis, constrained optimisation and statistical mechanics} \label{sec:const-opt-sat}
Counting is very much at the heart of statistical physics. Boltzmann's entropy, $S=k \ln \Omega$, is defined based on the number of accessible microstates $\Omega$ of a system. Establishing the statistical physics of the microcanonical ensemble then boils down to accurately counting the number of microstates associated with a given macrostate. In the canonical and grand-canonical ensembles microstates are `weighted' by the appropriate Boltzmann factors, and the counting is formalised in the canonical and grand-canonical partition functions. Anyone who has taken a course in statistical physics will remember the subtleties associated with determining the number of ways a given number of Fermions or Bosons can be distributed across a set of energy states. 

The Boltzmann weights in the partition functions for the different ensembles can be obtained from the extremisation principles of classical thermodynamics. Depending on circumstances, a system will tend to the state of extremal entropy, Helmholtz free energy or grand potential for example. Using this, the idea of equal a priori probabilities in isolated systems, and the entropy $S=k \ln\,\Omega$ for the combination of the system of interest and the relevant surrounding baths, the canonical and grand-canonical partition functions can be obtained\footnote{We highlight the alternative information-theoretic approach to statistical physics,  beautifully introduced by Jaynes \cite{Jaynes:1957_1, Jaynes:1957_2}.}.

It is no surprise that connections can be established to other areas facing counting and extremisation problems. Examples are the closely related areas of `constrained optimisation' and `constraint satisfaction' in mathematics and computer science. Constrained optimisation problems involve finding the minimum (or maximum) of a function $f(\boldsymbol{x})$, subject to constraints on the variable  $\boldsymbol{x}=(x_1,\dots,x_n)$. These constraints can come as a set of equations connecting the $x_i$, as inequalities or as a combination of equalities and inequalities. One example is graph partitioning, i.e., the problem of partitioning the set of nodes of a network into a given number of subsets while minimising the number of edges between these subsets. Another instance of constrained optimisation is the so-called `knapsack problem' (see e.g. \cite{Pisinger:2005}).

Constraint satisfaction problems are problems in which the variable $\boldsymbol{x}$ must satisfy a set of constraints. Often these are formulated on graphs. A good example is the so-called graph colouring problem (more precisely, we describe the vertex colouring problem). Assume we have a graph consisting of $N$ nodes, connected by a set of links. The problem now consists of assigning a colour to each node so that no two neighbouring nodes have the same colour. The $x_i$, $i=1,\dots,N$ here represent the colours, and the constraints are local (each node  of degree $k\geq 1$ in the network leads to a constraint involving this node and its neighbours). Other constraint satisfaction problems are the celebrated travelling salesman and random $k$-satisfiability problems \cite{Monasson:1997, Mezard:2002, Hartmann:2005}.

The methods used to address such problems include techniques such as message-passing (including belief propagation) and the cavity method. These tools are also relevant for spin glass problems in physics, and it is therefore natural for a statistical physicist to work at the interface of optimisation and computer science. Physicists have also been instrumental in characterising the typical properties of instances of constraint satisfaction or optimisation problems, this will be discussed further below (Sec.~\ref{sec:disorder}).

It is hard to avoid seeing possible connections between NMA and these classes of problems. For example, it is natural to ask if ranking in NMA can be phrased as a constrained optimisation or satisfaction problem. It is perhaps useful to think of the meta-analytic network as a bipartite graph, with one type of node representing treatment options and the other trials [e.g. Fig.~\ref{fig:illustration}~(d)].  If there are $N$ treatment nodes in the graph, then the ranking task consists of assigning the ranks $1,\dots,N$ to those nodes, while satisfying constraints set by the trials or minimising a cost defined by the trials (each trial provides noisy information on the relative ranks of some treatments).  The NMA problem hence falls into a class of broader problems: We start from a set of $N$ objects (the treatment options), who each have an unknown quality (treatment effect). We have noisy estimates of the relative qualities of subsets of these objects (from the trials), and we now wish to estimate the intrinsic property of each object. If this is not possible, what can we say about the relative comparison between pairs of objects, or a ranking of the objects in terms of quality?

While details would have to be thought through carefully we think that it is well possible that parallels between NMA and existing optimisation or satisfaction problems can  be established. There is then potential for statistical physicists to contribute via the methods mentioned above (message passing, cavity method, etc). We think this is an exciting perspective for future work.

\subsection{Network meta-analysis and disordered systems}\label{sec:disorder} 
Related to the previous item there may be interest in looking at the typical properties of the NMA problem and its solution. This is a common approach in constrained optimisation and satisfaction problems. Instead of looking at single instances of these problems, one assumes the network is drawn from an ensemble of graphs, and then asks what the average or typical properties of problems in this ensemble are. For example, how many solutions to the optimisation or satisfaction problem are there, what subspace do they form in the overall state space (e.g. is there one connected manifold of solutions vs fragmented clusters) and what is the typical quality of the solution (i.e., how many of the constraints can be fulfilled)?

To answer these questions, an average over assignments of the graph and constraints needs to be taken. One can then make use of tools from the physics of spin glasses and disordered systems, such as the replica method, dynamic mean field theory or cavity approaches. This allows one to characterise the energy landscape associated with these problems, the geometry of the solution space, and most importantly the performance of algorithms to find solutions to the optimisation or satisfaction problem.

It is conceivable that a similar approach could be taken in NMA. Based on known statistical features of meta-analytic graphs \cite{Niko:2014, Salanti:2008b} one could define an ensemble of random NMA problems (i.e., the configuration of trials, treatment options and their connections is drawn from a distribution). One could then try to assess how features of the ensemble (e.g. connectivity, regularity etc.) affect the outcome of the NMA problems in the ensemble. This would allow one to step away from {\em single} instances, and instead to say more about {\em typical} cases.

\subsection{Machine learning approaches to systematic reviews and Bayesian MCMC} 
The field of machine learning interfaces with statistical physics, and there is increasing interest by physicists in machine learning methods. There are multiple ways in which machine learning can contribute to the field of NMA, and this therefore defines another point of contact with statistical mechanics. 

One way in which machine learning can be used in NMA relates to the Bayesian approach in Sec.~\ref{sec:Bayes}. Methods from machine learning have been proposed as alternatives to MCMC in general. This includes techniques such as expectation propagation \cite{Minka:2001b}, variational Bayesian inference  \cite{Attias:1999} and integrated nested Laplace approximations \cite{Sauter:2015, Rue:2009}. 

A perhaps even stronger link to machine learning presents itself in the process of data acquisition for an NMA. In this paper, we have so far focused on the procedure of carrying out an NMA given a set of data from multiple trials. In reality, the process of compiling the data starts by performing a so-called `systematic review' of the existing trials for a particular medical problem. This is a systematic screening of the literature, using well defined procedures, followed by a process to decide which trials are adopted for the NMA. This decision making again uses well defined protocols and criteria. As part of this process a large number of journal articles must be searched, and appropriate data must be identified and extracted.

This obviously lends itself to automation, which speeds up the process, saves resources, and removes human error and inconsistencies that arise when a team of multiple researchers collects the data. Machine learning methods have indeed been employed to automate or semi-automate this process \cite{Marshall:2019, Marshall:2018, Lange:2021}.

Given these contact points between NMA and machine learning, we think that researchers working at the interface of statistical physics and machine learning may find interest in applying the ideas and methods they are familiar with to the field of evidence synthesis. Their experience may then lead to the development of algorithms that improve on existing methods.

\subsection{Simulation techniques}
Simulation techniques play an important role in NMA. Most evidently, Monte-Carlo sampling is necessary to carry out a Bayesian NMA (Sec.~\ref{sec:Bayes}). Further, so-called `simulation studies' can provide insight into NMA \cite{Morris:2019}. These are studies in which trial data is generated synthetically (simulated). This allows one, for example, to test how NMA fares on different networks, and how different features of the graph affect NMA outcomes \cite{Kibret:2014, Seide:2020, Davies:2020}. The simulation proceeds along the random effects model described in Sec.~\ref{sec:FEM_REM}, we note again the hierarchical structure and the different levels of randomness. In simulation studies one typically has to look at many instances of networks, and average over a large number of realisations of synthetic data. Efficient simulation methods for the generation of data are therefore key.

Statistical physicists are obviously familiar with simulation methods for random processes. Acknowledging that simulation studies are an established part of statistics (and consequently, that there is a significant existing expertise), this defines another point of contact, and  a prospective avenue for statistical physicists to contribute to the field of NMA and evidence synthesis more generally.

\subsection{Meta-analysis in particle physics}
\label{sec:particle}
As a final, more speculative thought, we note that meta-analysis is used in particle physics to obtain the best estimates of particle properties such as masses, widths and lifetimes \cite{Baker:2013}. Expertise developed in this area might also be useful for meta-analysis and network meta-analysis in medical statistics.

\section{Summary}\label{sec:summary}
 Most of the existing NMA literature is naturally written by medical statisticians for medical statisticians, or for researchers actively using NMA tools and software packages in clinical practice. As a consequence, it is not easy to find an account of the essentials of NMA presented in a language physicists would be used to. The objective of this perspective review is to make a first step towards rectifying this. We hope the paper is a useful introduction to network meta-analysis for statistical physicists. It should be noted that the paper was also written by statistical physicists. This means that we work from a limited perspective. The selection of topics and the presentation is subject to personal bias.
 
 Naturally, we could only include what we considered to be the most essential aspects of NMA. Topics which could have been covered in a more extensive review include inconsistency \cite{TSD4}, individual participant data \cite{Riley:2010}, multi-component interventions \cite{Rucker:2020}, multiple outcomes \cite{Riley:2017}, bias adjustment methods \cite{TSD3} and goodness-of-fit assessment \cite{TSD2}. In making our selection of the material for this paper, we aimed to focus on the concepts, ideas and methods, which would best enable the reader to access the wider literature. We tried to write a self-contained systematic introduction which could serve as a starting point for the reader to then explore the field more effectively.
 
  We also hope that the paper highlights the importance of the area of evidence synthesis. Our review is successful if it excites others and if it convinces the members of the community that network meta-analysis is a field in which statistical mechanics can make a difference.  
 
\section*{Acknowledgements}
We thank Gerta R\"ucker, Adriani Nikolakopoulou and Theodoros Papakonstantinou for earlier collaboration and for useful feedback on the current manuscript.  We acknowledge funding by the Engineering and Physical Sciences Research Council (EPSRC UK), grant number EP/R513131/1, and partial financial support by the Maria de Maeztu Programme for Units of Excellence in R\&D, Grant MDM-2017-0711 funded by MCIN/AEI/10.13039/501100011033.

\appendix

\section{Estimating within-study variances and correlations of observed treatment effects}
\label{App:SigCovHat}
In this appendix we discuss how the matrix $\mathbf{V}_i$ in Eq.~(\ref{eq:Vi}) [see Sec.~\ref{sec:freq_intro}] is estimated from trial data. This matrix describes the variance and correlations within a trial due to fluctuations arising from the finite number of subjects in the different arms of the trial (i.e., sampling errors).

\subsection{General setup and estimating the variance of observed treatment effects}
For binomial data, the observed relative treatment effects in trial $i$ are given by the log odds ratios,
\be \label{eq:LOR2}
y_{i,1\ell}= \mathrm{logit}(\hat{p}_{i,\ell})-\mathrm{logit}(\hat{p}_{i,1}) = \ln \frac{\hat{p}_{i,\ell}}{1-\hat{p}_{i,\ell}}- \ln \frac{\hat{p}_{i,1}}{1-\hat{p}_{i,1}},
\ee
where $\hat{p}_{i,\ell}=r_{i,\ell}/n_{i,\ell}$ is the proportion of events in arm $\ell$ of trial $i$. The variances, $\sigma_{i,1\ell}^2$, and covariances, $\mathrm{Cov}(y_{i,1\ell},y_{i,1\ell'})$, associated with these observations define the covariance matrix $\mathbf{V}_i$ in Eq.~(\ref{eq:Vi}). These values can be estimated from the data. To do so we first work out the random sampling variance associated with the values $\hat{p}_{i,\ell}$.

The number of events measured in arm $\ell$ of trial $i$ is a binomial random variable $r_{i,\ell} \sim \mathrm{Bin}(n_{i,\ell},p_{i,\ell})$. It has mean $\mathbb{E}(r_{i,\ell})=n_{i,\ell}p_{i,\ell}$ and variance $\mathrm{Var}(r_{i,\ell})=n_{i,\ell}p_{i,\ell}(1-p_{i,\ell})$. Using $\mathrm{Var}(bx)=b^2\mathrm{Var}(x)$ (for random variable $x$ and constant $b$), we find
\begin{eqnarray}
\label{eq:var_phat}
    \mathrm{Var}(\hat{p}_{i,\ell}) = \frac{p_{i,\ell}(1-p_{i,\ell})}{n_{i,\ell}}.
\end{eqnarray}
Assuming $n_{i,\ell}$ is large, and propagating the errors for the logit function to linear order then leads to 
\begin{eqnarray}
    \mathrm{Var}[\mathrm{logit}(\hat{p}_{i,\ell})] &= \left(\frac{\partial\, \mathrm{logit}(\hat{p}_{i,\ell})}{\partial \hat{p}_{i,\ell}}\right)^2  \mathrm{Var}(\hat{p}_{i,\ell})\nonumber \\
    &=\left(\frac{1}{\hat{p}_{i,\ell}(1-\hat{p}_{i,\ell})}\right)^2 \frac{p_{i,\ell}(1-p_{i,\ell})}{n_{i,\ell}}.
\end{eqnarray}
We get an estimate of this variance by setting $p_{i,\ell}=\hat{p}_{i\ell}$,
\begin{eqnarray}
\label{eq:VarLamHat}
    \widehat{\mathrm{Var}}[\mathrm{logit}(\hat{p}_{i,\ell})] &=\left[n_{i,\ell} \hat{p}_{i,\ell} (1-\hat{p}_{i,\ell})\right]^{-1} = \left[r_{i,\ell} \left(1-\frac{r_{i,\ell}}{n_{i,\ell}}\right)\right]^{-1}.
\end{eqnarray}
Since the values $\hat{p}_{i,\ell}$ for different $\ell$ are independent, the variance of $y_{i,1\ell}$ is $\sigma_{i,1\ell}^2=\mathrm{Var}(y_{i,1\ell}) = \mathrm{Var}[\mathrm{logit}(\hat{p}_{i,\ell})]+\mathrm{Var}[\mathrm{logit}(\hat{p}_{i,1})]$ which we can estimate using Eq.~(\ref{eq:VarLamHat}),
\begin{eqnarray}
\label{eq:SigHat}
    \hat{\sigma}_{i,1\ell}^2 = \left[r_{i,\ell} \left(1-\frac{r_{i,\ell}}{n_{i,\ell}}\right)\right]^{-1} + \left[r_{i,1} \left(1-\frac{r_{i,1}}{n_{i,1}}\right)\right]^{-1}
\end{eqnarray}
\cite{Chang:2001, DerSimonian:1986, Hamza:2008}. The conventional assumption is that the estimates $\hat{\sigma}_{i,1\ell}^2$ can be used in the matrix $\mathbf{V}_i$ in Eq.~(\ref{eq:Vi}) in place of the true variances \cite{Higgins:2009, Chang:2001}. 

\subsection{Estimating correlations between different observed treatment effects in the trial}
The observations $y_{i,1\ell}$ of the relative treatment effects within a trial are correlated because they all involve the same common treatment arm $t_{i,1}$. To calculate the covariance between the observations we start from the general relation
\begin{eqnarray}
    \mathrm{Var}(A-B)=\mathrm{Var}(A)+\mathrm{Var}(B)-2\mathrm{Cov}(A,B),
\end{eqnarray}
for two random variables $A$ and $B$.
Setting $A=y_{i,1\ell}$ and $B=y_{i,1\ell'}$ for $\ell\neq\ell'$ we find
\begin{eqnarray}
\label{eq:Cov}
    \mathrm{Cov}(y_{i,1\ell},y_{i,1\ell'}) = \frac{1}{2}\left[ \mathrm{Var}(y_{i,1\ell}) + \mathrm{Var}(y_{i,1\ell'})-\mathrm{Var}(y_{i,1\ell}-y_{i,1\ell'}) \right].
\end{eqnarray}
We evaluate the final term on the right hand side of Eq.~(\ref{eq:Cov}) using Eq.~(\ref{eq:LOR2}),
\begin{eqnarray}
    \hspace{-15mm}\mathrm{Var}(y_{i,1\ell}-y_{i,1\ell'}) &= \mathrm{Var}\left\{\left[\mathrm{logit}(\hat{p}_{i,\ell})-\mathrm{logit}(\hat{p}_{i,1})\right]-\left[\mathrm{logit}(\hat{p}_{i,\ell'})-\mathrm{logit}(\hat{p}_{i,1})\right]\right\} \nonumber \\
    &=\mathrm{Var}\left[\mathrm{logit}(\hat{p}_{i,\ell})-\mathrm{logit}(\hat{p}_{i,\ell'})\right]\nonumber \\
    &=\mathrm{Var}[\mathrm{logit}(\hat{p}_{i,\ell})]+\mathrm{Var}[\mathrm{logit}(\hat{p}_{i,\ell'})],
\end{eqnarray}
where in the last line we have used the fact that the absolute treatment effects in different arms within a trial are independent. Recalling that $\mathrm{Var}(y_{i,1\ell}) = \mathrm{Var}[\mathrm{logit}(\hat{p}_{i,\ell})]+\mathrm{Var}[\mathrm{logit}(\hat{p}_{i,1})]$ we find \cite{TSD2}
\begin{eqnarray}
    \mathrm{Cov}(y_{i,1\ell},y_{i,1\ell'}) = \mathrm{Var}[\mathrm{logit}(\hat{p}_{i,1})],
\end{eqnarray}
 which we can estimate via Eq.~(\ref{eq:VarLamHat}),
\begin{eqnarray}
\label{eq:CovHat}
    \widehat{\mathrm{Cov}}(y_{i,1\ell},y_{i,1\ell'}) = \left[r_{i,1} \left(1-\frac{r_{i,1}}{n_{i,1}}\right)\right]^{-1}.
\end{eqnarray}
Again, the convention is to assume that the estimate of the covariance in Eq.~(\ref{eq:CovHat}) can be used in place of the true covariance \cite{TSD2}.

\subsection{Limitations of estimating within study variance}
It is evident from Eqs. (\ref{eq:LOR2}) and (\ref{eq:SigHat}) that the values $y_{i,1\ell}$ and $\hat{\sigma}_{i,1\ell}^2$ are correlated (both expressions depend on the variables $(r_{i,\ell}, n_{i,\ell})$ and $(r_{i,1}, n_{i,1})$) \cite{Chang:2001}. This causes a systematic relationship between the magnitude and weight of the observations which leads to a bias on the overall effect estimates \cite{Chang:2001, Stijnen:2010}.

Estimation of $y_{i,1\ell}$ and $\sigma_{i,1\ell}^2$ from the observed data also causes problems when events are observed for either no patients in a trial arm, or for all patients in an arm. By inspection of Eqs.~(\ref{eq:LOR2}) and (\ref{eq:SigHat}), we notice that if $r_{i,\ell}=0$ or $r_{i,\ell}=n_{i,\ell}$ then $y_{i,1\ell}$ and $\hat{\sigma}_{i,1\ell}^2$ will be undefined. A common ad-hoc method to avoid this problem is to add a value of $0.5$ to every $r_{i,\ell}$ and $n_{i,\ell}$. This has been found to produce biased estimates of effect size \cite{Bradburn:2007, Sweeting:2004, Stijnen:2010}. In fact, this limitation, along with the assumption that within-study variances are known, is the main criticism of the frequentist inverse-variance approach to NMA \cite{Hoaglin:2015, Hoaglin:2016}. Alternative methods such as the Mantel-Haenszel method and generalised linear mixed models (GLMM) have been recommended \cite{Stijnen:2010, Efthimiou:2019}.

\section{Expectation of Q under the random effects model}
\subsection{Pairwise meta-analysis}
\label{App:EQ-Pair}
In this section we evaluate the expectation of $Q$,
\begin{eqnarray}
    \mathbb{E}_{\mathrm{RE}}(Q) = \mathbb{E}_{\mathrm{RE}}\left(\sum_{i=1}^{M} a_i (y_i-\hat{y})^2\right), 
\end{eqnarray}
where $\hat{y}$ is defined in Eq.~(\ref{eq:yhat}) and the observations $y_i$ are assumed to follow the random effects (RE) model,
\begin{eqnarray}
\label{eq:RE_MoM}
    y_i \sim \mathcal{N}(d, \sigma_i^2 + \tau^2).
\end{eqnarray}
Using $\mathbb{E}_{\mathrm{RE}}(y_i-\hat{y})=0$, we find
\begin{eqnarray}
    \mathbb{E}_{\mathrm{RE}}\left(\sum_{i=1}^{M} a_i (y_i-\hat{y})^2\right) 
    &=\sum_{i=1}^{M} a_i \mathrm{Var}\left( y_i-\hat{y}\right). \label{eq:E-Var}
\end{eqnarray}
We can then obtain the variance of $y_i-\hat{y}$ using
\begin{eqnarray}
\label{eq:Var_1}
    \mathrm{Var}\left( y_i-\hat{y}\right) = \mathrm{Var}(y_i) + \mathrm{Var}(\hat{y}) - 2\mathrm{Cov}(y_i,\hat{y}).
\end{eqnarray}
From the RE model in Eq.~(\ref{eq:RE_MoM}) we know $\mathrm{Var}(y_i)=\sigma_i^2 + \tau^2$. We now wish to obtain the second and third terms of Eq.~(\ref{eq:Var_1}) in terms of $\mathrm{Var}(y_i)$. To do so we use standard properties of variances and covariances, and the fact that $\mathrm{Cov}(y_i,y_j)=0$ for $i\neq j$ (for a pairwise meta-analysis, the observations $y_i$ are independent).

To calculate $\mathrm{Var}(\hat{y})$ we use Eq.~(\ref{eq:yhat}) in the main paper and find
\begin{eqnarray}
    \mathrm{Var}(\hat{y}) &= \mathrm{Var}\left( \frac{\sum_{i=1}^{M} a_i y_i }{\sum_{i=1}^{M} a_i } \right) = \left(\frac{1}{\sum_{i=1}^{M} a_i}\right)^2 \sum_{i=1}^{M} a_i^2 \mathrm{Var}( y_i).
\end{eqnarray}
For $\mathrm{Cov}(y_i,\hat{y})$ we have
\begin{eqnarray}
    \mathrm{Cov}(y_i,\hat{y}) &= \mathrm{Cov}\left(y_i, \frac{\sum_{i=1}^{M} a_i y_i }{\sum_{i=1}^{M} a_i } \right) \nonumber \\
    &=\frac{1}{\sum_{i=1}^{M} a_i } \left( a_i\mathrm{Cov}(y_i,y_i) + \sum_{j\neq i}^{M} a_j \mathrm{Cov}(y_i,y_j)\right) \nonumber\\
    &=\frac{1}{\sum_{i=1}^{M} a_i }  a_i\mathrm{Var}(y_i).
\end{eqnarray}
Substituting these results into Eq.~(\ref{eq:Var_1}) yields
\begin{eqnarray}
    \mathrm{Var}\left(y_i-\hat{y}\right) = \mathrm{Var}(y_i) + \frac{\sum_{j=1}^{M} a_j^2 \mathrm{Var}( y_j)}{\left(\sum_{j=1}^{M} a_j \right)^2}  - \frac{2a_i\mathrm{Var}(y_i)}{\sum_{j=1}^{M} a_j }.
\end{eqnarray}
Now substituting this into Eq.~(\ref{eq:E-Var}) and using $\mathrm{Var}(y_i)=\sigma_i^2+\tau^2$ we find the expectation of $Q$ under the random effects model to be
\begin{eqnarray}
    \mathbb{E}_{\rm RE}(Q) &= \sum_{i=1}^{M} a_i \mathrm{Var}(y_i) + \sum_{i=1}^{M} a_i \frac{\sum_{j=1}^{M} a_j^2 \mathrm{Var}( y_j)}{\left(\sum_{j=1}^{M} a_j \right)^2}  - \frac{2\sum_{i=1}^{M} a_i^2\mathrm{Var}(y_i)}{\sum_{j=1}^{M} a_j }\nonumber \\
    &=\sum_{i=1}^{M} a_i ( \sigma_i^2+\tau^2 ) -  \frac{\sum_{i=1}^{M} a_i^2 ( \sigma_i^2+\tau^2 )}{\sum_{i=1}^{M} a_i }\nonumber \\
    &= \tau^2 \left(\sum_{i=1}^{M} a_i -  \frac{\sum_{i=1}^{M} a_i^2 }{\sum_{i=1}^{M} a_i }\right) + \left(\sum_{i=1}^{M} a_i \sigma_i^2  - \frac{\sum_{i=1}^{M} a_i^2  \sigma_i^2}{\sum_{i=1}^{M} a_i }\right).
\end{eqnarray}
This is the result quoted in Eq.~(\ref{eq:EQ_pair}) in the main paper.

\subsection{Network meta-analysis}
\label{App:EQ-Net} 
In this section we evaluate the expectation of $Q$ under the random effects model for network meta-analysis. We now have [Eq.~(\ref{eq:Q-NMA})]
\begin{eqnarray}
\label{eq:Q-NMA-App}
    Q = (\boldsymbol{y}-\hat{\boldsymbol{y}})^\top \mathbf{V}^{-1} (\boldsymbol{y}-\hat{\boldsymbol{y}}) 
\end{eqnarray}
and [Eq.~(\ref{eq:yhat-NMA})]
\begin{eqnarray}
    \hat{\boldsymbol{y}} = \mathbf{X}(\mathbf{X}^{\top} \mathbf{V}^{-1} \mathbf{X})^{-1} \mathbf{X}^{\top} \mathbf{V}^{-1} \boldsymbol{y}.
\end{eqnarray} 
To simplify Eq.~(\ref{eq:Q-NMA-App}) we follow Jackson et al (2016) \cite{Jackson:2016} and define the matrix
\begin{eqnarray}
\label{eq:A}
    \mathbf{A} = \mathbf{V}^{-1} - \mathbf{V}^{-1}\mathbf{X}(\mathbf{X}^{\top} \mathbf{V}^{-1} \mathbf{X})^{-1} \mathbf{X}^{\top} \mathbf{V}^{-1}
\end{eqnarray}
such that
\begin{eqnarray}
    \boldsymbol{y}-\hat{\boldsymbol{y}} = \mathbf{V}\mathbf{A}\boldsymbol{y}.
\end{eqnarray}
Therefore,
\begin{eqnarray}
    Q &= (\mathbf{V}\mathbf{A}\boldsymbol{y})^\top \mathbf{V}^{-1} (\mathbf{V}\mathbf{A}\boldsymbol{y}) \nonumber \\
    &= \boldsymbol{y}^\top \mathbf{A} \mathbf{V} \mathbf{A} \boldsymbol{y}
\end{eqnarray}
since both $\mathbf{V}$ and $\mathbf{A}$ are symmetric. The former is symmetric by definition, the latter can be shown to be symmetric by taking the transpose of the right hand side of Eq.~(\ref{eq:A}). By explicit evaluation we find
\begin{eqnarray}
    \hspace*{-65pt}\mathbf{A} \mathbf{V} \mathbf{A} \hspace*{-30pt}&= \mathbf{A}[\mathbf{I} - \mathbf{X}(\mathbf{X}^{\top} \mathbf{V}^{-1} \mathbf{X})^{-1} \mathbf{X}^{\top} \mathbf{V}^{-1}] \nonumber \\
    \hspace*{-30pt}&=\mathbf{A} - [\mathbf{V}^{-1} - \mathbf{V}^{-1}\mathbf{X}(\mathbf{X}^{\top} \mathbf{V}^{-1} \mathbf{X})^{-1} \mathbf{X}^{\top} \mathbf{V}^{-1}]\mathbf{X}(\mathbf{X}^{\top} \mathbf{V}^{-1} \mathbf{X})^{-1} \mathbf{X}^{\top} \mathbf{V}^{-1} \nonumber \\
    \hspace*{-30pt}&=\mathbf{A} - \mathbf{V}^{-1}\mathbf{X}(\mathbf{X}^{\top} \mathbf{V}^{-1} \mathbf{X})^{-1} \mathbf{X}^{\top} \mathbf{V}^{-1} + \mathbf{V}^{-1}\mathbf{X}(\mathbf{X}^{\top} \mathbf{V}^{-1} \mathbf{X})^{-1} \mathbf{X}^{\top} \mathbf{V}^{-1} \nonumber \\
    \hspace*{-30pt}&= \mathbf{A}
\end{eqnarray}
such that
\begin{eqnarray}
    Q=\boldsymbol{y}^\top \mathbf{A}  \boldsymbol{y}.
\end{eqnarray}

As in the pairwise case we take the expectation of $Q$ under the random effects model. Defining $\boldsymbol{\xi}=\boldsymbol{\eta}+\boldsymbol{\epsilon}$, we re-write the RE model [Eq.~(\ref{eq:RE-regress}) in the main paper] as
\begin{eqnarray}
    \boldsymbol{y} = \mathbf{X}\boldsymbol{d}+\boldsymbol{\xi}, \hspace{20 pt} \boldsymbol{\xi}\sim\mathcal{N}(0, \mathbf{V}+\boldsymbol{\Sigma}).
\end{eqnarray}
This leads to 
\begin{eqnarray}
    \mathbb{E}_{\mathrm{RE}}(Q)&=\mathbb{E}_{\mathrm{RE}}[(\mathbf{X}\boldsymbol{d}+\boldsymbol{\xi})^\top \mathbf{A}  (\mathbf{X}\boldsymbol{d}+\boldsymbol{\xi})]\nonumber \\
    &=\mathbb{E}_{\mathrm{RE}}(\boldsymbol{d}^\top\mathbf{X}^\top \mathbf{A} \mathbf{X}\boldsymbol{d} + \boldsymbol{d}^\top\mathbf{X}^\top \mathbf{A}\boldsymbol{\xi} + 
    \boldsymbol{\xi}^\top \mathbf{A}  \mathbf{X}\boldsymbol{d}+\boldsymbol{\xi}^\top \mathbf{A} \boldsymbol{\xi}).
\end{eqnarray}
By explicit evaluation we find
\begin{eqnarray}
    \mathbf{X}^\top \mathbf{A}  &= \mathbf{X}^\top\mathbf{V}^{-1} - \mathbf{X}^\top\mathbf{V}^{-1}\mathbf{X}(\mathbf{X}^{\top} \mathbf{V}^{-1} \mathbf{X})^{-1} \mathbf{X}^{\top} \mathbf{V}^{-1} = 0 \nonumber  \\
    \mathbf{A}\mathbf{X}  &= \mathbf{V}^{-1}\mathbf{X}  - \mathbf{V}^{-1}\mathbf{X}(\mathbf{X}^{\top} \mathbf{V}^{-1} \mathbf{X})^{-1} \mathbf{X}^{\top} \mathbf{V}^{-1}\mathbf{X}  = 0,
\end{eqnarray}
and hence,
\begin{eqnarray}
\label{eq:EQ-NMA1}
     \mathbb{E}_{\mathrm{RE}}(Q) = \mathbb{E}_{\mathrm{RE}}(\boldsymbol{\xi}^\top \mathbf{A} \boldsymbol{\xi}).
\end{eqnarray}
For any vector $\boldsymbol{z}$ with mean $\boldsymbol{\mu}_{z}$ and covariance matrix $\mathbf{V}_{z}$, one has $\mathbb{E}(\boldsymbol{z}^\top \mathbf{B} \boldsymbol{z})=\mathrm{tr}(\mathbf{B}\mathbf{V}_{z})+\boldsymbol{\mu}_z^\top \mathbf{B}\boldsymbol{\mu}_z$ where $\mathbf{B}$ is a square matrix and $\mathrm{tr}(.)$ indicates the trace of a matrix. This identity can be checked directly, see also \cite{Searle:2016} (Theorem 4, pg 75, Chapter 2). 

The vector $\boldsymbol{\xi}$ in Eq.~(\ref{eq:EQ-NMA1}) has mean $\boldsymbol{0}$ and covariance matrix $\mathbf{V}+\boldsymbol\Sigma$, therefore
\begin{eqnarray}
     \mathbb{E}_{\mathrm{RE}}(Q) &= \mathrm{tr}[\mathbf{A}(\mathbf{V}+\boldsymbol\Sigma)]\nonumber \\
     &=\mathrm{tr}(\mathbf{A}\mathbf{V}) + \tau^2\mathrm{tr}(\mathbf{A}\mathbf{P}).
\end{eqnarray}
In the last step we have written  $\boldsymbol{\Sigma}=\tau^2 \mathbf{P}$ where $\mathbf{P}$ is a block diagonal matrix. Each $(m_i-1) \times (m_i-1)$ block (representing trial $i$) has diagonal elements equal to 1 and off-diagonal elements equal to $1/2$. All other elements are zero. Using the fact the trace is invariant under cyclic permutations we find
\begin{eqnarray}
    \mathrm{tr}(\mathbf{A}\mathbf{V}) &= \mathrm{tr}\left[ \mathbf{V}^{-1}\mathbf{V} - \mathbf{V}^{-1}\mathbf{X}(\mathbf{X}^{\top} \mathbf{V}^{-1} \mathbf{X})^{-1} \mathbf{X}^{\top} \mathbf{V}^{-1}\mathbf{V} \right]\nonumber\\
    &=\mathrm{tr}\left[ \mathbf{I}_{\sum_i(m_i-1)}\right]  - \mathrm{tr}\left[ \mathbf{V}^{-1}\mathbf{X}(\mathbf{X}^{\top} \mathbf{V}^{-1} \mathbf{X})^{-1} \mathbf{X}^{\top} \right] \nonumber\\
     &=\sum_{i=1}^{M}(m_i-1) - \mathrm{tr}\left[ (\mathbf{X}^{\top} \mathbf{V}^{-1} \mathbf{X})^{-1} \mathbf{X}^{\top}\mathbf{V}^{-1}  \mathbf{X} \right]\nonumber\\
     &=\sum_{i=1}^{M}(m_i-1) - \mathrm{tr}\left[ \mathbf{I}_{N-1} \right] \nonumber \\
     &=\sum_{i=1}^{M}(m_i-1) - (N-1) \label{eq:dof}
\end{eqnarray}
where $\sum_{i=1}^{M} (m_i-1)$ is the lateral dimension of $\mathbf{V}$ (the number of observations in $\boldsymbol{y}$) and $N-1$ is the lateral dimension of $\mathbf{X}^{\top} \mathbf{V}^{-1} \mathbf{X}$ (the number of mean relative treatment effects we wish to estimate).  The expression in Eq.~(\ref{eq:dof}) is the number of degrees of freedom associated with the regression (that is, the difference between the number of data points and the number of parameters\footnote{This can be understood via a simple example: imagine calculating the mean from $10$ values. Here the number of data points is 10 and the number of parameters is 1. We only need $9$ of those values plus the mean to fully specify the $10$ values. The number of degrees of freedom is then $9$ (which is equal to the number of data points minus the number of parameters).}).

\section{Bias in maximum likelihood variance estimation}
\label{App:Bias_ML_Var}
Consider a random variable $y=(y_1, \dots, y_N)^\top$ with normal distribution, $y\sim\mathcal{N}(\mu,\sigma^2)$. The likelihood function is then
\begin{eqnarray}
    L = \prod_{i=1}^{N} \frac{1}{\sqrt{2\pi \sigma^2}} \mathrm{e}^{-(y_i-\mu)^2/2\sigma^2}.
\end{eqnarray}
Maximising the likelihood function with respect to $\mu$ and $\sigma^2$ leads to the expressions
\begin{eqnarray}
    \hat{y} &= \frac{1}{N}\sum_{i=1}^{N}y_i\\
    \hat{\sigma}^2 &= \frac{1}{N}\sum_{i=1}^{N}(y_i-\mu)^2. 
\end{eqnarray}
The result of the joint maximisation with respect to $\mu$ and $\sigma$ therefore leads to the maximum likelihood estimators,
\begin{eqnarray}
    \hat{y} &= \frac{1}{N}\sum_{i=1}^{N}y_i \nonumber \\
    \hat{\sigma}^2 &= \frac{1}{N}\sum_{i=1}^{N}(y_i-\hat{y})^2, \label{eq:sig_ML} 
\end{eqnarray}
that is, we have to substitute the maximum likelihood estimate $\hat{y}$ into the expression for  $\hat{\sigma}^2$.

To show that the expected value of the variance estimator $\hat{\sigma}^2$ is not equal to the true variance $\sigma^2$ we re-write Eq.~(\ref{eq:sig_ML}) as 
\begin{eqnarray}
    \hat{\sigma}^2 &= \frac{1}{N}\sum_{i=1}^{N}[(y_i-\mu)-(\hat{y}-\mu)]^2,
\end{eqnarray}
which, after some rearranging and using the fact that $\hat{y}-\mu = \frac{1}{N} \sum_{i=1}^N (y_i-\mu)$, gives
\begin{eqnarray}
    \hat{\sigma}^2 &=\left[ \frac{1}{N}\sum_{i=1}^{N} (y_i-\mu)^2\right] - (\hat{y}-\mu)^2.
\end{eqnarray}
The expectation of $\hat{\sigma}^2$ is then
\begin{eqnarray}
    \mathbb{E}\left[\hat{\sigma}^2\right] &= \mathbb{E}\left[\frac{1}{N}\sum_{i=1}^{N} (y_i-\mu)^2\right] - \mathbb{E}\left[(\hat{y}-\mu)^2\right].
\end{eqnarray}
The first term is the true variance $\sigma^2$. The second term is the variance of $\hat{y}$,
\begin{eqnarray}
    \mathbb{E}\left[(\hat{y}-\mu)^2\right] = \mathrm{Var}(\hat{y}) = \mathrm{Var}\left(\frac{1}{N}\sum_{i=1}^{N}y_i\right) = \frac{1}{N^2}\sum_{i=1}^N \mathrm{Var}(y_i) = \frac{\sigma^2}{N}.
\end{eqnarray}
Therefore, 
\begin{eqnarray}
    \mathbb{E}\left[\hat{\sigma}^2\right] &= \sigma^2 - \frac{\sigma^2}{N} = \frac{N-1}{N}\sigma^2,
\end{eqnarray}
indicating that the ML variance estimator is biased downwards by $\frac{\sigma^2}{N}$. For large samples $N \rightarrow \infty$, this bias becomes negligible.

\section{Adjusting multi-arm trial weights using a result from electrical network theory}
\label{App:BackCalc}
Here we explain the method for adjusting variances associated with measurements in a network meta-analysis in order to account for correlations introduced by multi-arm trials \cite{Rucker:2012, Rucker:2014}. This is based on the method described in Gutman and Xiao (2004) \cite{Gutman:2004} for reconstructing the individual resistances in an electrical network from the effective resistances between pairs of nodes.

We focus on a multi-arm trial $i$ which compares $m_i$ treatments.  This trial yields a total of $q_i=\frac{m_i(m_i-1)}{2}$ treatment effect estimates and associated variances $v_{i,ab}$. For a random effects model these variances are $v_{i,ab}=\sigma_{i,ab}^2+\hat{\tau}^2$ where $\hat{\tau}^2$ is an estimate of the between trial heterogeneity. In a fixed effect model $v_{i,ab} = \sigma_{i,ab}^2$. We write these variances in an $m_i \times m_i$ matrix, $\mathbf{\tilde{V}}_i$. We label this matrix $\mathbf{\tilde{V}}_i$ to distinguish it from the $(m_i-1) \times (m_i-1)$ matrix $\mathbf{V}_i$ defined in the main paper. Each row and column of $\mathbf{\tilde{V}}_i$ represents a treatment in trial $i$. The diagonal elements are equal to zero and each off-diagonal element is the variance associated with the comparison of the corresponding pair of treatments. For example, in a multi-arm trial $i$ that compares treatments $\{T_1, T_2, T_3\}$, the variance matrix is
\begin{eqnarray}
    \mathbf{\tilde{V}}_i = \left(\matrix{
    0 & v_{i,T_1T_2} &  v_{i,T_1T_3} \cr
    v_{i,T_1T_2} & 0 & v_{i,T_2T_3} \cr
     v_{i,T_1T_3} & v_{i,T_2T_3} & 0
    }\right).
\end{eqnarray}

The aim now is to reconstruct the (inverse-variance) weights for a set of three two-arm trials that yield network estimates of relative treatment effects whose variances are equal to the variances in $\mathbf{\tilde{V}}_i$. To this end, we use a method from electrical theory for back calculation of edge resistances given a set of effective resistances \cite{Rucker:2012, Gutman:2004}.

We saw in Sec.~\ref{NMA-Electric} that the effective resistances in an electrical network are related to the pseudo-inverse of the Laplacian. In NMA, effective resistances are associated with the variances $\mathbf{\tilde{V}}_i$ observed in the multi-arm trial. A result from electrical network theory is that we can construct the pseudo-inverse of the Laplacian directly from the effective resistances \cite{Gutman:2004}. Using this result for an $m_i$-armed trial with `effective' variances $\mathbf{\tilde{V}}_i$ gives the $m_i \times m_i$ matrix
\begin{eqnarray}
    \mathbf{L}_i^{+} = -\frac{1}{2}\left( \mathbf{\tilde{V}}_i - \frac{1}{m_i}(\mathbf{\tilde{V}}_i\mathbf{O}_{i} + \mathbf{O}_{i}\mathbf{\tilde{V}}_i) +\frac{1}{m_i^2}  \mathbf{O}_{i} \mathbf{\tilde{V}}_i \mathbf{O}_{i} \right)
\end{eqnarray}
where $\mathbf{O}_{i}$ is an $m_i \times m_i$ matrix of ones \cite{Rucker:2012}. An equivalent more compact expression (used in \cite{Rucker:2014}) is 
\begin{eqnarray}
     \mathbf{L}_i^{+} = - \frac{1}{2m_i^2} \mathbf{B}_i^\top \mathbf{B}_i \mathbf{\tilde{V}}_i \mathbf{B}_i^\top \mathbf{B}_i
\end{eqnarray}
where $\mathbf{B}_i$ is the edge-incidence matrix for trial $i$ that describes what edges (treatment comparisons) are present in the trial \cite{Rucker:2014}. $\mathbf{B}_i$ has dimensions $q_i\times m_i$, where we recall $q_i=\frac{m_i(m_i-1)}{2}$. Each row of $\mathbf{B}_i$ represents a pairwise comparison in the trial, and each column represents a treatment. There is a $1$ in the column corresponding to the baseline treatment for that comparison and a $-1$ in the column representing the treatment compared to that baseline. All other entries are zero.

Once we have the pseudo-inverse of the Laplacian we can work out the Laplacian using $\mathbf{L}_i=(\mathbf{L}_i^{+})^{+}$ and \cite{Rucker:2012, Fouss:2007}
\begin{eqnarray}
    \mathbf{L}_i^+ &= (\mathbf{L}_i-\frac{1}{m_i}\mathbf{O}_i)^{-1} + \frac{1}{m_i} \mathbf{O}_i,\\
    (\mathbf{L}_i^+)^{+} &= (\mathbf{L}_i^{+}-\frac{1}{m_i}\mathbf{O}_i)^{-1} + \frac{1}{m_i} \mathbf{O}_i.
\end{eqnarray}

In an electrical network the graph Laplacian is defined by the individual (physical) resistors, $L_{ab}=-R_{ab}^{-1}$ for $a\neq b$, and $L_{aa}=\sum_b R_{ab}^{-1}$. Therefore the values $R_{ab}$ are obtained by inspection of the off diagonal elements of $\mathbf{L}$. Similarly, in the NMA context, the adjusted (inverse-variance) edge weights for multi-arm trial $i$ can be obtained from the off diagonal elements of the Laplacian $\mathbf{L}_i$. For the three-arm trial, 
\begin{eqnarray}
    \mathbf{L}_i = \left(\matrix{
    \tilde{w}_{i,T_1T_2}+\tilde{w}_{i,T_1T_3} & -\tilde{w}_{i,T_1T_2} & -\tilde{w}_{i,T_1T_3} \cr
    -\tilde{w}_{i,T_1T_2} & \tilde{w}_{i,T_1T_2}+\tilde{w}_{i,T_2T_3} & -\tilde{w}_{i,T_2T_3} \cr
    -\tilde{w}_{i,T_1T_3} & -\tilde{w}_{i,T_2T_3} & \tilde{w}_{i,T_1T_3}+\tilde{w}_{i,T_2T_3} 
    }\right),
\end{eqnarray}
where $\tilde{w}_{i,ab}$ are the adjusted edge weights that describe the inverse-variances for a network of two-arm trials which is equivalent to the multi-arm trial.

\section*{References}
 
\bibliographystyle{iopart-num}
\bibliography{Ref_AMA2.bib}

\providecommand{\newblock}{}
\begin{thebibliography}{100}
\expandafter\ifx\csname url\endcsname\relax
  \def\url#1{{\tt #1}}\fi
\expandafter\ifx\csname urlprefix\endcsname\relax\def\urlprefix{URL }\fi
\providecommand{\eprint}[2][]{\url{#2}}

\bibitem{Parisi:1993}
Parisi G 1993 {\em Phys. World.\/} {\bf 6}(9) 42--47

\bibitem{Castellano:2009}
Castellano C, Fortunato S and Loreto V 2009 {\em Rev. Mod. Phys.\/} {\bf 81}(2)
  591--646

\bibitem{Stauffer:2004}
Stauffer D 2004 {\em Physica A\/} {\bf 336}(1) 1--5 {P}roceedings of the XVIII
  Max Born Symposium ``Statistical Physics outside Physics"

\bibitem{Gallegati:2006}
Gallegati M, Keen S, Lux T and Ormerod P 2006 {\em Physica A\/} {\bf 370}(1)
  1--6 {E}conophysics Colloquium

\bibitem{DerSimonian:1986}
DerSimonian R and Laird N 1986 {\em Control. Clin. Trials\/} {\bf 7}(3)
  177--188

\bibitem{Smith:1995}
Smith T~C, Spiegelhalter D~J and Thomas A 1995 {\em Stat. Med.\/} {\bf 14}(24)
  2685--2699

\bibitem{Higgins:2009}
Higgins J~P~T, Thompson S~G and Spiegelhalter D~J 2009 {\em J. R. Stat. Soc.\/}
  {\bf 172}(1) 137--159

\bibitem{Tonin:2017}
Tonin F~S, Rotta I, Mendes A~M and Pontarolo R 2017 {\em Pharm. Pract.
  (Granada)\/} {\bf 15}(1) 943

\bibitem{DIAS:2018}
Dias S, Ades A~E, Welton N~J, Jansen J~P and Sutton A~J 2018 {\em Network
  Meta-Analysis for Decision Making\/} (Oxford, UK: Wiley)

\bibitem{Boland:2003}
Boland A, Dundar Y, Bagust A, Haycox A, Hill R, {Mujica Mota} R, Walley T and
  Dickson R 2003 {\em Health. Technol. Asses.\/} {\bf 7}(15) 1--136

\bibitem{Keeley:2003}
Keeley E~C, Boura J~A and Grines C~L 2003 {\em Lancet\/} {\bf 361}(9351) 13--20

\bibitem{Dias:2010}
Dias S, Welton N~J, Caldwell D~M and Ades A~E 2010 {\em Stat. Med.\/} {\bf
  29}(7-8) 932--944

\bibitem{Spiegelhalter:2003}
Spiegelhalter D, Thomas A, Best N and Lunn D 2003 Win{BUGS} {U}ser {M}anual:
  Version 1.4 MRC Biostatistics Unit, University of Cambridge

\bibitem{Efthimiou:2019}
Efthimiou O, Rücker G, Schwarzer G, Higgins J~P~T, Egger M and Salanti G 2019
  {\em Stat. Med.\/} {\bf 38}(16) 2992--3012

\bibitem{Stijnen:2010}
Stijnen T, Hamza T~H and {\"O}zdemir P 2010 {\em Stat. Med.\/} {\bf 29}(29)
  3046--3067

\bibitem{Davies:2021}
Davies A~L, Papakonstantinou T, Nikolakopoulou A, R{\"u}cker G and Galla T 2022
  {\em Stat. Med.\/}  1--24

\bibitem{TSD2}
Dias S, Welton N~J, Sutton A~J and Ades A~E 2011 {NICE DSU} {T}echnical
  {S}upport {D}ocument 2: {A} generalised linear modelling framework for
  pairwise and network meta analysis of randomised controlled trials Online
  {L}ast updated September 2016; Available from
  \url{http://www.nicedsu.org.uk}. Accessed March 2020

\bibitem{TSD3}
Dias S, Welton N~J, Sutton A~J and Ades A~E 2011 {NICE DSU} {T}echnical
  {S}upport {D}ocument 3: Heterogeneity: subgroups, meta-regression, bias and
  bias-adjustment Online {L}ast updated April 2012; {A}vailable from
  \url{http://www.nicedsu.org.uk}. Accessed December 2021

\bibitem{CochraneBook}
Higgins J~P~T, Thomas J, Chandler J, Cumpston M, Li T, Page M~J and Welch V~A
  (eds) 2019 {\em Cochrane Handbook for Systematic Reviews of Interventions\/}
  2nd ed (Chichester, UK: John Wiley \& Sons)

\bibitem{SALANTI:2012}
Salanti G 2012 {\em Res. Synth. Meth.\/} {\bf 3}(2) 80--97

\bibitem{Greco:2016}
Greco T, Landoni G, Biondi-Zoccai G, D'Ascenzo F and Zangrillo A 2016 {\em
  Stat. Methods. Med. Res.\/} {\bf 25}(5) 1757--1773

\bibitem{Salanti:2008}
Salanti G, Higgins J~P~T, Ades A~E and Ioannidis J~P~A 2008 {\em Stat. Methods.
  Med. Res.\/} {\bf 17}(3) 279--301

\bibitem{Jansen:2008}
Jansen J~P, Crawford B, Bergman G and Stam W 2008 {\em Value. Health\/} {\bf
  11}(5) 956--964

\bibitem{Efthimiou:2016}
Efthimiou O, Debray T~P, {van Valkenhoef} G, Trelle S, Panayidou K, Moons K~G,
  Reitsma J~B, Shang A and Salanti G 2016 {\em Res. Synth. Meth.\/} {\bf 7}(3)
  236--263

\bibitem{NICE-gloss}
NICE Glossary Online available from \url{https://www.nice.org.uk/Glossary}.
  Accessed February 2022

\bibitem{Pisanski:2000}
Pisanski T and Randić M 2000 Bridges between geometry and graph theory {\em
  Geometry at Work: A Collection of Papers Showing Applications of Geometry\/}
  (Washington, DC, USA: Mathematical Association of America) pp 174--194

\bibitem{TSD4}
Dias S, Welton N~J, Sutton A~J, Caldwell D~M, Lu G and Ades A~E 2011 {NICE DSU}
  {T}echnical {S}upport {D}ocument 4: Inconsistency in networks of evidence
  based on randomised controlled trials Online {L}ast updated April 2014;
  {A}vailable from \url{http://www.nicedsu.org.uk}. Accessed December 2021

\bibitem{Hamza:2008}
Hamza T~H, {van Houwelingen} H~C and Stijnen T 2008 {\em J. Clin. Epidemiol.\/}
  {\bf 61}(1) 41--51

\bibitem{Rucker:2009}
Rücker G, Schwarzer G, Carpenter J and Olkin I 2009 {\em Stat. Med.\/} {\bf
  28}(5) 721--738

\bibitem{Deeks:2002}
Deeks J~J 2002 {\em Stat. Med.\/} {\bf 21}(11) 1575--1600

\bibitem{Dias:2013}
Dias S, Sutton A~J, Ades A~E and Welton N~J 2013 {\em Med. Decis. Making.\/}
  {\bf 33}(5) 607--617

\bibitem{Eddy:1992}
Eddy D~M, Hasselblad V and Shachter R 1992 {\em Meta-Analysis by the Confidence
  Profile Method\/} (London, UK: Academic Press)

\bibitem{Lu:Ades:2006}
Lu G and Ades A~E 2006 {\em J. Am. Stat. Assoc.\/} {\bf 101}(474) 447--459

\bibitem{Hong:2015}
Hong H, Fu H, Price K~L and Carlin B~P 2015 {\em Stat. Med.\/} {\bf 34}(20)
  2794--2819

\bibitem{Dias:2016}
Dias S and Ades A~E 2016 {\em Res. Synth. Meth.\/} {\bf 7}(1) 23--28

\bibitem{Hig:White:1996}
Higgins J~P~T and Whitehead A 1996 {\em Stat. Med.\/} {\bf 15}(24) 2733--2749

\bibitem{Lumley:2002}
Lumley T 2002 {\em Stat. Med.\/} {\bf 21}(16) 2313--2324

\bibitem{Lu:Ades:2004}
Lu G and Ades A~E 2004 {\em Stat. Med.\/} {\bf 23}(20) 3105--3124

\bibitem{Seide:2019}
Seide S~E, Jensen K and Kieser M 2019 {\em Stat. Med.\/} {\bf 38}(17)
  3288--3303

\bibitem{Cox:2006}
Cox D~R 2006 {\em Principles of Statistical Inference\/} (Cambridge, UK:
  Cambridge University Press)

\bibitem{Bartholomew:1965}
Bartholomew D~J 1965 {\em Biometrika\/} {\bf 52}(1-2) 19--35

\bibitem{Wagenmakers2008}
Wagenmakers E, Lee M, Lodewyckx T and Iverson G~J 2008 Bayesian versus
  frequentist inference {\em Bayesian Evaluation of Informative Hypotheses\/}
  ed Hoijtink H, Klugkist I and Boelen P~A (New York, NY, USA: Springer) pp
  181--207

\bibitem{Samaniego:2010}
Samaniego F~J 2010 {\em A comparison of the {B}ayesian and frequentist
  approaches to estimation\/} (New York, NY, USA: Springer)

\bibitem{Hespanhol:2019}
Hespanhol L, Vallio C~S and Saragiotto B~T 2019 {\em Braz. J. Phys. Ther.\/}
  {\bf 23}(4) 290--301

\bibitem{Glickman:2007}
Glickman M~E and van Dyk D~A 2007 Basic {B}ayesian methods {\em Topics in
  Biostatistics: Methods in Molecular Biology\/} vol 404 ed Abrosius W~T
  (Totowa, NJ, USA: Humana Press Inc) chap~16, pp 319--338

\bibitem{Greenland:2006}
Greenland S 2006 {\em Int. J. Epidemiol.\/} {\bf 35}(3) 765--775

\bibitem{White:2019}
White I~R, Turner R~M, Karahalios A and Salanti G 2019 {\em Stat. Med.\/} {\bf
  38}(27) 5197--5213

\bibitem{White:2012}
White I~R, Barrett J~K, Jackson D and Higgins J~P~T 2012 {\em Res. Synth.
  Meth.\/} {\bf 3}(2) 111--125

\bibitem{Rucker:2012}
R\"{u}cker G 2012 {\em Res. Synth. Meth.\/} {\bf 3}(4) 312--324

\bibitem{Franchini:2012}
Franchini A~J, Dias S, Ades A~E, Jansen J~P and Welton N~J 2012 {\em Res.
  Synth. Meth.\/} {\bf 3}(2) 142--160

\bibitem{Basu:1977}
Basu D 1977 {\em J. Am. Stat. Assoc.\/} {\bf 72}(358) 355--366

\bibitem{DuMouchel:1994}
DuMouchel W 1994 Hierarchical {B}ayes {L}inear {M}odels for {M}eta {A}nalysis.
  {T}echnical report number 27. National Institute of Statistical Sciences
  {A}vailable from \url{https://www.niss.org/research/technical-reports}.
  Accessed December 2021

\bibitem{Lu:Ades:2009}
Lu G and Ades A~E 2009 {\em Biostatistics\/} {\bf 10}(4) 792--805

\bibitem{Rosenberger:2021}
Rosenberger K~J, Xing A, Murad M, Chu H and Lin L 2021 {\em J. Gen. Intern.
  Med.\/} {\bf 36}(4) 1049--1057

\bibitem{Gelman:2006}
Gelman A 1006 {\em Bayesian Anal.\/} {\bf 1}(3) 515--533

\bibitem{Turner:2012}
Turner R~M, Davey J, Clarke M~J, Thompson S~G and Higgins J~P~T 2012 {\em Int.
  J. Epidemiol.\/} {\bf 41}(3) 818--827

\bibitem{Rhodes:2015}
Rhodes K~M, Turner R~M and Higgins J~P~T 2015 {\em J. Clin. Epidemiol.\/} {\bf
  68}(1) 52--60

\bibitem{Turner:2018}
Turner R~M, Domínguez-Islas C~P, Jackson D, Rhodes K~M and White I~R 2019 {\em
  Stat. Med.\/} {\bf 38}(8) 1321--1335

\bibitem{JAGS}
Plummer M 2017 {JAGS} version 4.3.0 user manual Online {A}vailable from
  \url{https://martynplummer.wordpress.com/}. Accessed December 2021

\bibitem{STAN}
{Stan Development Team} 2019 Stan user’s guide version 2.23 Online
  {A}vailable from \url{https://mc-stan.org/}. Accessed December 2021

\bibitem{Hastings:1970}
Hastings W~K 1970 {\em Biometrika\/} {\bf 57}(1) 97--109

\bibitem{Geyer:2011}
Geyer C~J 2011 {I}ntroduction to {M}arkov chain {M}onte {C}arlo {\em Handbook
  of {M}arkov Chain {M}onte {C}arlo\/} ed Brooks S, Gelman A, Jones G~L and
  Meng X (Boca Raton, FL, USA: CRC Press) chap~1, pp 3--48

\bibitem{Robert:2015}
Robert C~P 2015 The {M}etropolis–{H}astings algorithm {\em Wiley
  {S}tats{R}ef: Statistics Reference Online\/} ed Balakrishnan N, Colton T,
  Everitt B, Piegorsch W, Ruggeri F and Teugels J~L (American Cancer Society)
  pp 1--15

\bibitem{Robert:2004}
Robert C~P and Casella G 2004 {\em Monte Carlo Statistical Methods\/} (New
  York, NY, USA: Springer, Science and Business Media)

\bibitem{Toral:2014}
Toral R and Colet P 2014 {\em Stochastic numerical methods: An introduction for
  students and scientists\/} (Weinheim, Germany: Wiley-VCH Verlag GmbH)

\bibitem{Gelman:1996}
Gelman A, Roberts G~O and Gilks W~R 1996 Efficient {M}etropolis jumping rules
  {\em {B}ayesian Statistics 5\/} ed Bernardo J, Berger J~O, Dawid A~P and
  Smith A~F~M (Oxford, UK: Oxford University Press) pp 599--607

\bibitem{Lynch:2007}
Lynch S~M 2007 {\em Introduction to Applied {B}ayesian Statistics and
  Estimation for Social Scientists\/} (New York, NY, USA: Springer)

\bibitem{Roberts:2001}
Roberts G~O and Rosenthal J~S 2001 {\em Stat. Sci.\/} {\bf 16}(4) 351--367

\bibitem{Gelman:2013}
Gelman A, Carlin J~B, Stern H~S, Dunson D~B, Vehtari A and Rubin D~B 2013 {\em
  Bayesian Data Analysis\/} 3rd ed (Boca Raton, FL, USA: CRC Press)

\bibitem{Gelman:Rubin:1992}
Gelman A and Rubin D 1992 {\em Stat. Sci.\/} {\bf 7}(4) 457--511

\bibitem{Brooks:Gelman:1998}
Brooks S and Gelman A 1998 {\em J. Comput. Graph. Stat.\/} {\bf 7}(4) 434--455

\bibitem{Chang:2001}
Chang B, Waternaux C and Lipsitz S 2001 {\em Stat. Med.\/} {\bf 20}(13)
  1947--1956

\bibitem{Riley:2012}
Riley R~D 2012 {\em J. R. Stat. Soc. A. Stat.\/} {\bf 172}(4) 789--811

\bibitem{Amemiya:1985}
Amemiya T 1985 Generalised least squares theory {\em Advanced Econometrics\/}
  (Cambridge, MA, USA: Havard University Press) chap~6

\bibitem{Charnes:1976}
Charnes A, Frome E~L and Yu P~L 1976 {\em J. Am. Stat. Assoc.\/} {\bf 71}(353)
  169--171

\bibitem{Dolby:1972}
Dolby G~R 1972 {\em J. Roy. Stat. Soc. B. Met.\/} {\bf 34}(3) 393--400

\bibitem{Aitken:1936}
Aitken A~C 1936 {\em P. Roy. Soc. Edinb. B.\/} {\bf 55} 42--48

\bibitem{DerSimonian:2007}
DerSimonian R and Kacker R 2007 {\em Contemp. Clin. Trials.\/} {\bf 28}(2)
  105--114

\bibitem{Hartung:2003}
Hartung J and Makambi K~H 2003 {\em Commun. Stat. Simul. Comput.\/} {\bf 32}(4)
  1179--1190

\bibitem{Sidik:2005}
Sidik K and Jonkman J~N 2005 {\em J. R. Stat. Soc. C-Appl.\/} {\bf 54}(2)
  367--384

\bibitem{Rukhin:2013}
Rukhin A~L 2013 {\em J. R. Stat. Soc. B.\/} {\bf 75}(3) 451--469

\bibitem{Paule:1982}
Paule R~C and Mandel J 1982 {\em J. Res. Natl. Bur. Stand.\/} {\bf 87}(5)
  377--385

\bibitem{Langan:2019}
Langan D, Higgins J~P~T, Jackson D, Bowden J, Veroniki A~A, Kontopantelis E,
  Viechtbauer W and Simmonds M 2019 {\em Res. Synth. Meth.\/} {\bf 10}(1)
  83--98

\bibitem{Veroniki:2016b}
Veroniki A~A, Jackson D, Viechtbauer W, Bender R, Bowden J, Knapp G, Kuss O,
  Higgins J~P~T, Langan D and Salanti G 2016 {\em Res. Synth. Meth.\/} {\bf
  7}(1) 55--79

\bibitem{Petropoulou:2017b}
Petropoulou M and Mavridis D 2017 {\em Stat. Med.\/} {\bf 36}(27) 4266--4280

\bibitem{Jackson:2017}
Jackson D, Veroniki A~A, Law M, Tricco A~C and Baker R 2017 {\em Res. Synth.
  Meth.\/} {\bf 8}(4) 416--434

\bibitem{Jackson:2010}
Jackson D, White I~R and Simon T~G 2010 {\em Stat. Med.\/} {\bf 29}(12)
  1282--1297

\bibitem{Law:2016}
Law M, Jackson D, Turner R, Rhodes K and Viechtbauer W 2016 {\em BMC Med. Res.
  Methodol.\/} {\bf 16}(87) 1--14

\bibitem{Jackson:2016}
Jackson D, Law M, Barrett J~K, Turner R, Higgins J~P~T, Salanti G and White I~R
  2016 {\em Stat. Med.\/} {\bf 35}(6) 819--839

\bibitem{Jackson:2012}
Jackson D, White I~R and Riley R~D 2012 {\em Stat. Med.\/} {\bf 31}(29)
  3805--3820

\bibitem{Kacker:2004}
Kacker R~N 2004 {\em Metrologia\/} {\bf 41}(3) 132--136

\bibitem{Patterson:1971}
Patterson H~D and Thompson R 1971 {\em Biometrika\/} {\bf 58}(3) 545--554

\bibitem{Cochran:1954}
Cochran W~G 1954 {\em Biometrics\/} {\bf 10}(1) 101--129

\bibitem{Harville:1977}
Harville D~A 1977 {\em J. Am. Stat. Assoc.\/} {\bf 72}(358) 320--338

\bibitem{Wakefield:2009}
Wakefield J 2009 Lecture notes {B}io{S}tat 571: Introduction and motivation,
  revision of estimation methods, linear mixed effects models, likelihood
  inference Online {A}vailable from
  \url{http://courses.washington.edu/b571/lectures.html}. Accessed February
  2022

\bibitem{Harville:1974}
Harville D~A 1974 {\em Biometrika\/} {\bf 61}(2) 383--385

\bibitem{Press:1992}
Press W~H, Teukolsky S~A, Vetterling W~T and Flannery B~P 1992 {\em Numerical
  Recipes in {C}. {T}he Art of Scientific Computing\/} 2nd ed (Cambridge, UK:
  Cambridge University Press)

\bibitem{Longford:1987}
Longford N~T 1987 {\em Biometrika\/} {\bf 74}(4) 817--827

\bibitem{Dempster:1977}
Dempster A~P, Laird N~M and Rubin D~B 1977 {\em J. Roy. Stat. Soc. B. Met.\/}
  {\bf 39}(1) 1--38

\bibitem{Viechtbauer:2005}
Viechtbauer W 2005 {\em J. Educ. Behav. Stat.\/} {\bf 30}(3) 261--293

\bibitem{netmeta:2021}
Rücker G, Krahn U, König J, Efthimiou O, Davies A, Papakonstantinou T and
  Schwarzer G 2021 {\em netmeta: Network Meta-Analysis using Frequentist
  Methods\/} R Foundation for Statistical Computing Vienna, Austria {R} package
  version 2.0-0. Available from
  \url{https://CRAN.R-project.org/package=netmeta}

\bibitem{White:2015}
White I~R 2015 {\em Stata J.\/} {\bf 15}(4) 951--985

\bibitem{Hawkins:2009}
Hawkins N, Scott D~A, Woods B~S and Thatcher N 2009 {\em Value. Health.\/} {\bf
  12}(6) 996--1003

\bibitem{Hurley:2020}
Hurley J 2020 {\em J. Clin. Epidemiol.\/} {\bf 121} 110

\bibitem{Salanti:2011}
Salanti G, Ades A~E and Ioannidis J~P~A 2011 {\em J. Clin. Epidemiol.\/} {\bf
  64}(2) 163--171

\bibitem{Mbuagbaw:2017}
Mbuagbaw L, Rochwerg B, Jaeschke R, Heels-Andsell D, Alhazzani W, Thabane L and
  Guyatt G~H 2017 {\em Syst. Rev.\/} {\bf 6}(1) 79

\bibitem{White:2011}
White I~R 2011 {\em Stata J.\/} {\bf 11}(2) 255--70

\bibitem{Rucker:2015}
R{\"u}cker G and Schwarzer G 2015 {\em BMC Med. Res. Methodol.\/} {\bf 15}(1)
  58

\bibitem{Rucker:2014}
Rücker G and Schwarzer G 2014 {\em Stat. Med.\/} {\bf 33}(25) 4353--4369

\bibitem{Gutman:2004}
Gutman I and Xiao W 2004 {\em Bull. Acad. Serbe. Sci. Cl. Sci. Math. Nat. Sci.
  Nat.\/} {\bf 129}(29) 15--23

\bibitem{Lov:1994}
Lov{\'a}sz L 1994 Random walks on graphs: A survey Online {YALE/DCS/TR-1029}.
  Available from
  \url{http://www.cs.yale.edu/publications/techreports/tr1029.pdf}. Accessed
  March 2021

\bibitem{Noh:2004}
Noh J~D and Rieger H 2004 {\em Phys. Rev. Lett.\/} {\bf 92}(11) 118701

\bibitem{Masuda:2017}
Masuda N, Porter M~A and Lambiotte R 2017 {\em Phys. Rep.\/} {\bf 716-717}
  1--58

\bibitem{Kakutani:1945}
Kakutani S 1945 {\em Proc. Jap. Acad.\/} {\bf 21}(4) 227--233

\bibitem{Kemeny:1966}
Kemeny J~G, Snell J~L and Knapp A~W 1966 {\em Markov Chains\/} University
  Series in Higher Mathematics (New York, NY, UK: Van Nostrand)

\bibitem{Kelly:1979}
Kelly F~P 1979 {\em Reversibility and Stochastic Networks\/} Probability and
  Statistics Series (Chichester, UK: Wiley)

\bibitem{Doyle:2000}
Doyle P~G and Snell J~L 2000 Random walks and electric networks
  arXiv:math/0001057

\bibitem{konig:2013}
König J, Krahn U and Binder H 2013 {\em Stat. Med.\/} {\bf 32}(30) 5414--5429

\bibitem{Papakon:2018}
Papakonstantinou T, Nikolakopoulou A, Rücker G, Chaimani A, Schwarzer G, Egger
  M and Salanti G 2018 {\em F1000Res.\/} {\bf 7} 610

\bibitem{Salanti:2014}
Salanti G, {Del Giovane} C, Chaimani A, Caldwell D~M and Higgins J~P~T 2014
  {\em PLOS ONE\/} {\bf 9}(7) e99682

\bibitem{Papakon:2021}
Papakonstantinou T, Nikolakopoulou A, Egger M and Salanti G 2021 {\em Res.
  Synth. Meth.\/} {\bf 12}(1) 20--28

\bibitem{Bowden:2016}
Bowden J and Jackson C 2016 {\em Am. Stat.\/} {\bf 70}(4) 385--394

\bibitem{Sterne:2000}
Sterne J~A~C, Gavaghan D and Egger M 2000 {\em J. Clin. Epidemiol.\/} {\bf
  53}(11) 1119--1129

\bibitem{Page:1998}
Brin S and Page L 1998 {\em Comput. Networks. ISDN.\/} {\bf 30}(1) 107--117

\bibitem{Danilowicz:2001}
Daniłowicz C and Baliński J 2001 {\em Inform. Process. Manag.\/} {\bf 37}(4)
  623--637

\bibitem{Blanchet:2016}
Blanchet J, Gallego G and Goyal V 2016 {\em Oper. Res.\/} {\bf 64}(4) 886--905

\bibitem{Trinquart:2016}
Trinquart L, Attiche N, Bafeta A, Porcher R and Ravaud P 2016 {\em Ann. Intern.
  Med.\/} {\bf 164}(10) 666--673

\bibitem{Veroniki:2016}
Veroniki A~A, Straus S~E, Fyraridis A and Tricco A~C 2016 {\em J. Clin.
  Epidemiol.\/} {\bf 76} 193--199

\bibitem{Veroniki:2018}
Veroniki A~A, Straus S~E, R{\"u}cker G and Tricco A~C 2018 {\em J. Clin.
  Epidemiol.\/} {\bf 100} 122--129

\bibitem{Daly:2019}
Daly C~H, Neupane B, Beyene J, Thabane L, Straus S~E and Hamid J~S 2019 {\em
  {BMJ} Open\/} {\bf 9}(9) e024625

\bibitem{Chaimani:2019}
Chaimani A, Porcher R, Sbidian {\'E} and Mavridis D 2021 {\em Stat. Med.\/}
  {\bf 40}(2) 451--464

\bibitem{Chiocchia:2020}
Chiocchia V, Nikolakopoulou A, Papakonstantinou T, Egger M and Salanti G 2020
  {\em BMJ Open\/} {\bf 10}(8) e037744

\bibitem{Mavridis:2020}
Mavridis D, Porcher R, Nikolakopoulou A and Salanti G 2020 {\em Biometrical
  J.\/} {\bf 62}(2) 375--385

\bibitem{Nikolakopoulou:2020b}
Nikolakopoulou A, Mavridis D, Chiocchia V, Papakonstantinou T, Furukawa T~A and
  Salanti G 2021 {\em Res. Synth. Meth.\/} {\bf 12}(2) 161--175

\bibitem{PRISMA:2015}
Hutton B, Salanti G, Caldwell D~M, Chaimani A, Schmid C~H, Cameron C, Ioannidis
  J~P~A, Straus S~E, Thorlund K, Jansen J~P, Mulrow C, Catalá-López F,
  Gøtzsche P~C, Dickersin K, Boutron I, Altman D~G and Moher D 2015 {\em Ann.
  Intern. Med.\/} {\bf 162}(11) 777--784

\bibitem{Davies:2020}
Davies A~L and Galla T 2021 {\em Res. Synth. Meth.\/} {\bf 12}(3) 316--332

\bibitem{Tonin:2019}
Tonin F~S, Borba H~H, Mendes A~M, Wiens A, Fernandez-Llimos F and Pontarolo R
  2019 {\em PLOS ONE\/} {\bf 14}(2) e0212650

\bibitem{Salanti:2008b}
Salanti G, Kavvoura F~K and Ioannidis J~P~A 2008 {\em Ann. Intern. Med.\/} {\bf
  148}(7) 544--553

\bibitem{Todeschini:2000}
Todeschini R and Consonni V 2000 {\em Handbook of Molecular Descriptors\/}
  Methods and Principles in Medicinal Chemistry (Weinheim, Germany: Wiley-VCH
  Verlag GmbH)

\bibitem{Jaynes:1957_1}
Jaynes E~T 1957 {\em Phys. Rev.\/} {\bf 106}(4) 620--630

\bibitem{Jaynes:1957_2}
Jaynes E~T 1957 {\em Phys. Rev.\/} {\bf 108}(2) 171--190

\bibitem{Pisinger:2005}
Pisinger D 2005 {\em Comput. Oper. Res.\/} {\bf 32}(9) 2271--2284

\bibitem{Monasson:1997}
Monasson R and Zecchina R 1997 {\em Phys. Rev. E\/} {\bf 56}(2) 1357--1370

\bibitem{Mezard:2002}
Mézard M, Parisi G and Zecchina R 2002 {\em Science\/} {\bf 297}(5582)
  812--815

\bibitem{Hartmann:2005}
Hartmann A~K and Weight M 2005 {\em Phase Transitions in Combinatorial
  Optimisation Problems\/} (Weinheim, Germany: Wiley-VCH Verlag GmbH)

\bibitem{Niko:2014}
Nikolakopoulou A, Chaimani A, Veroniki A~A, Vasiliadis H~S, Schmid C~H and
  Salanti G 2014 {\em PLOS ONE\/} {\bf 9}(1) e86754

\bibitem{Minka:2001b}
Minka T~P 2001 Expectation propagation for approximate {B}ayesian inference
  {\em Proceedings of the 17th Conference in Uncertainty in Artificial
  Intelligence\/} ed Breese J and Koller D (San Francisco, CA, USA: Morgan
  Kaufmann Publishers Inc.) pp 362--369

\bibitem{Attias:1999}
Attias H 1999 Inferring parameters and structure of latent variable models by
  variational {B}ayes {\em Proceedings of the 15th Conference in Uncertainty in
  Artificial Intelligence\/} ed Laskey K~B and Prade H (San Francisco, CA, USA:
  Morgan Kaufmann Publishers Inc.) pp 21--30

\bibitem{Sauter:2015}
Sauter R and Held L 2015 {\em Biometrical J.\/} {\bf 57}(6) 1038--1050

\bibitem{Rue:2009}
Rue H, Martino S and Chopin N 2009 {\em J. Roy. Stat. Soc. B.\/} {\bf 71}(2)
  319--392

\bibitem{Marshall:2019}
Marshall I~J and Wallace B~C 2019 {\em Syst. Rev.\/} {\bf 8}(1) 163

\bibitem{Marshall:2018}
Marshall I~J, Noel-Storr A, Kuiper J, Thomas J and Wallace B~C 2018 {\em Res.
  Synth. Meth.\/} {\bf 9}(4) 602--614

\bibitem{Lange:2021}
Lange T, Schwarzer G, Datzmann T and Binder H 2021 {\em Res. Synth. Meth.\/}
  {\bf 12}(4) 506--515

\bibitem{Morris:2019}
Morris T~P, White I~R and Crowther M~J 2019 {\em Stat. Med.\/} {\bf 38}(11)
  2074--2102

\bibitem{Kibret:2014}
Kibret T, Richer D and Bayene J 2014 {\em Clin. Epidemiol.\/} {\bf 6} 451--460

\bibitem{Seide:2020}
Seide S~E, Jensen K and Kieser M 2020 {\em Res. Synth. Meth.\/} {\bf 11}(3)
  363--378

\bibitem{Baker:2013}
Baker R~D and Jackson D 2013 {\em Res. Synth. Meth.\/} {\bf 4} 109--124

\bibitem{Riley:2010}
Riley R~D, Lambert P~C and Abo-Zaid G 2010 {\em BMJ\/} {\bf 340}(7745) c221

\bibitem{Rucker:2020}
Rücker G, Petropoulou M and Schwarzer G 2020 {\em Biometrical J.\/} {\bf
  62}(3) 808--821

\bibitem{Riley:2017}
Riley R~D, Jackson D, Salanti G, Burke D~L, Price M, Kirkham J and White I~R
  2017 {\em BMJ\/} {\bf 358} j3932

\bibitem{Bradburn:2007}
Bradburn M~J, Deeks J~J, Berlin J~A and Localio A~R 2007 {\em Stat. Med.\/}
  {\bf 26}(1) 53--77

\bibitem{Sweeting:2004}
Sweeting A~J, Lambert P~C and Sutton A~J 2004 {\em Stat. Med.\/} {\bf 23}(9)
  1351--1375

\bibitem{Hoaglin:2015}
Hoaglin D~C 2015 {\em Res. Synth. Meth.\/} {\bf 6}(3) 287--289

\bibitem{Hoaglin:2016}
Hoaglin D~C 2016 {\em Stat. Med.\/} {\bf 35}(4) 485--495

\bibitem{Searle:2016}
Searle S~R and Gruber M~H~J 2016 {\em Linear Models\/} 2nd ed (Hoboken, NJ,
  USA: John Wiley \& Sons)

\bibitem{Fouss:2007}
Fouss F, Pirotte A, Renders J and Saerens M 2007 {\em IEEE. T. Knowl. Data.
  En.\/} {\bf 19}(3) 455--369

\end{thebibliography}

\end{document}